\documentclass[12pt]{amsart}
\usepackage{amssymb}
\usepackage{amsmath}
\usepackage{amsfonts}
\usepackage{mathrsfs}
\usepackage{graphicx}
\usepackage{color}
\usepackage{subcaption}
\usepackage{placeins}
\usepackage{xcolor}
\usepackage[onehalfspacing]{setspace}
\usepackage{caption}
\usepackage{multicol}
\usepackage{booktabs}
\usepackage{siunitx}
\usepackage[capposition=top]{floatrow}
\usepackage{natbib}
\usepackage{enumerate}
\usepackage[utf8]{inputenc}

\usepackage{tikz-cd}
\usepackage{dcolumn}
\usepackage[all,2cell]{xy}
\UseAllTwocells
\xyoption{arc}
\xyoption{rotate}
\usepackage[charter,cal=cmcal]{mathdesign}
\usepackage[colorlinks=true,citecolor=blue,urlcolor=blue,pdfpagemode=UseNone,pdfstartview=FitH]{hyperref}

\makeatletter
\def\section{\@startsection{section}{1}
	\z@{1.0\linespacing\@plus\linespacing}{.8\linespacing}{\Large}}

\def\subsection{\@startsection{subsection}{2}
	\z@{.8\linespacing\@plus.7\linespacing}{.7\linespacing}{\large}}

\def\subsubsection{\@startsection{subsubsection}{3}
	\z@{.5\linespacing\@plus.7\linespacing}{-.5em}{\normalfont\bfseries}}
\makeatother

\numberwithin{equation}{section}

\newtheorem{theorem}{Theorem}[section]
\newtheorem{lemma}{Lemma}[section]

\theoremstyle{definition}

\theoremstyle{definition}
\newtheorem{assumption}{Assumption}[section]

\theoremstyle{definition}

\title{}
\begin{document}
	\vspace*{5ex minus 1ex}
	\begin{center}
		\Large \textsc{Estimating Dynamic Spillover Effects along Multiple Networks in a Linear Panel Model}
		\bigskip
	\end{center}
	
	\date{%
		\today%
	}

	\vspace*{7ex minus 1ex}
	\begin{center}
		Clemens Possnig\footnote{Vancouver School of Economics, University of British Columbia}, Andreea Rot\u{a}rescu\footnote{Department of Economics, Wake Forest University}, and Kyungchul Song\footnote{Vancouver School of Economics, University of British Columbia}\\
		\bigskip
		
		\today

		\bigskip
		\bigskip
		
	\end{center}
	
\begin{abstract}
{\footnotesize Spillover of economic outcomes often arises over multiple networks, and distinguishing their separate roles is important in empirical research. For example, the direction of spillover between two groups (such as banks and industrial sectors linked in a bipartite graph) has important economic implications, and a researcher may want to learn which direction is supported in the data. For this, we need to have an empirical methodology that allows for both directions of spillover simultaneously. In this paper, we develop a dynamic linear panel model and asymptotic inference with large $n$ and small $T$, where both directions of spillover are accommodated through multiple networks. Using the methodology developed here, we perform an empirical study of spillovers between bank weakness and zombie-firm congestion in industrial sectors, using firm-bank matched data from Spain between 2005 and 2012. Overall, we find that there is positive spillover in both directions between banks and sectors.}\bigskip

{\footnotesize \noindent \textsc{Key words.} Spillover Effects; Networks; Fixed Effects; Dynamic Linear Panel Models; Cross-Sectional Dependence; Sector-Bank Spillover; Zombie Lending; Capital Misallocation}\bigskip

{\footnotesize \noindent \textsc{JEL Classification: C12, C21, C31, E44, G21, G32}}
\end{abstract}
\thanks{We thank Robert Moffitt and Jonathan Wright for their valuable comments. All errors are ours. Rot\u{a}rescu thanks the European Central Bank for data access. Song acknowledges that this research was supported by Social Sciences and Humanities Research Council of Canada. Corresponding
	address: Kyungchul Song, Vancouver School of Economics, University of
	British Columbia, 6000 Iona Drive, Vancouver, BC, V6T 1L4, Canada. Email address: kysong@mail.ubc.ca.}
\maketitle

\pagebreak

\section{Introduction}

Many economic outcomes, such as profits of firms or performance of students, evolve over time, producing spillover effects on other units over multiple large networks. Each network captures a distinct aspect of the relationship between units, and serves as a channel of spillover effects between them. A researcher may want to investigate the statistical significance of spillover effects separately for each network. One prominent example is a setting with two directed bipartite networks between two groups of cross-sectional units, where one network represents a channel of spillover from one group to the other, and the other network represents its opposite direction. A researcher may want to see which direction of spillover is empirically supported. For this, she needs to find an empirical methodology that allows for both directions of spillovers simultaneously. 

Our paper is motivated by one important area of empirical research which concerns the spillover that arises between banks and firms.\footnote{There is a large and growing empirical literature exploring the real economic effects of bank distress. See, for example, \cite{Peek/Rosengren:00:AER}, \cite{Khwaja/Mian:08:AER} and \cite{Chodorow-Reich:14:QJE} for seminal contributions.} Banks and firms are naturally interconnected through their lending relationships and as such heavily influence one another. Bank health, for one, can be a powerful determinant of firm performance. For example, banks with weak balance sheets may cut lending to their borrowers, thereby depriving them of the funds they need to operate and to invest. Firm performance, in turn, can act as an important driver of bank health: when firms miss payments on their loans, the banks issuing those loans suffer losses which put them in a worse financial position. In this setting, establishing which direction of outcome propagation is significant has important policy implications. If the direction pointing from the banking sector to the real sector is significant, then this makes the case for having prudential regulation in place which ensures that banks are well capitalized and able to absorb shocks to their balance sheets. If, additionally, the spillover from the real sector to the banking sector is significant, then policies that foster a more dynamic business environment might prove beneficial. This includes making it easier for incumbent firms to adjust the scale of their activity, as well as facilitating the exit of inefficient firms. These goals could be achieved, for example, by reducing factor adjustment costs and reforming insolvency regimes.

In this paper, we focus on a dynamic linear panel model where the spillover effects are captured by averages of past outcomes of neighbors across network measurements. For the empirical applications we have in mind, the spillover effects along a network are often detected through the correlation between the outcome of a unit in one period and its neighbors' outcomes in the previous period, after controlling for covariates. However, such correlation can arise simply through time effects that are specific to the cluster that those units jointly belong to. For example, the outcomes of two firms can be correlated because they belong to the same industrial sector and there have been industry-specific demand shocks over time. This correlation between  firms has nothing to do with the spillover effects between firms along a network, and yet contributes to the spillover effects measured through such correlations. To separate out cross-sectional correlation due to clusters, we include cluster-specific time effects and define the spillover effect to be one that is due to the spatio-temporal variations of outcomes \textit{net of} cluster-specific time averages of the outcomes.

Employing the Helmert transform of \cite{Arellano/Bover:95:JOE}, we develop an estimation procedure of the parameters and asymptotic inference on those parameters. Our method of estimation and inference is simple and does not involve any numerical optimization. We also propose a simple multiple testing procedure to detect the direction of spillover effect between two groups while controlling for the familywise error rate (FWER) asymptotically. From an extensive Monte Carlo simulation study, we find that the asymptotic inference performs very well, over a wide range of network configurations.

Using the methodology developed in this paper, we perform an empirical study of spillover between banks and industrial sectors using data from Spain. One of the most striking manifestations of the close interdependence between banks and firms is the existence of so-called zombie firms. These are firms which are known to be in financial distress but are artificially kept alive by weak banks seeking to avoid, or at least postpone, further damaging their balance sheets by realizing the losses caused by these firms. The practice of extending credit to zombie firms, also known as loan evergreening, was first documented in the context of Japan's ``lost decade'' and has attracted renewed attention in light of the weak economic recovery in Europe following the global financial crisis and the subsequent sovereign debt crisis.\footnote{For early evidence from Japan, see \cite{Peek/Rosengren:05:AER}, \cite{Caballero/Hoshi/Kashyap:08:AER} and \cite{Giannetti/Simonov:13:AEJMacro}. For more recent evidence from Europe, see \citet{Acharya/Eisert/Eufinger/Hirsch:19:ReFinStud}, \citet{Blattner/Farinha/Rebelo:19:ECB} and \citet{Schivardi/Sette/Tabellini:22:EJ}. For a recent survey of the theoretical and empirical literature on zombie lending, see \citet{Acharya/Crosignani/Eisert/Steffen:22:NBER}.}

Keeping zombie firms afloat can have adverse spillover effects on the rest of the economy by inhibiting the movement of resources from less productive to more productive uses. Indeed, a number of recent studies provide evidence linking an increase in zombie-firm congestion to a deterioration in sector-level competition, innovation, and productivity-enhancing reallocation.\footnote{For example, \citet{Acharya/Eisert/Eufinger/Hirsch:19:ReFinStud} document a reduction in employment growth and investment of non-zombie firms in industries with a high presence of zombie firms in the euro area. \citet{Adalet/Andrews/Millot:18:EP} additionally find high zombie congestion to be linked with less productivity-enhancing capital reallocation in an international cross-country sample. \citet{Schmidt/Schneider/Steffen/Streitz:20:SSRN} present evidence from Spain which shows capital misallocation from zombie lending to adversely impact output, competition, and patent applications.} These findings are particularly worrisome, given that sclerotic business environments with a high degree of resource misallocation have been shown to contribute to sizable losses in aggregate productivity.\footnote{See \citet{Hsieh/Klenow:09:QJE} and, more recently, \citet{Gopinath/Kalemli-Ozcan/Karababounis/Villegas-Sanchez:17:QJE}.}

Motivated by this discussion, we choose to focus our empirical application on the relationship between bank health and zombie-firm congestion in narrow sectors of activity. Our application is thematically related to a large body of research on the real effects of the bank lending channel, however, we depart from this literature in important methodological ways. Much of the literature has focused on one direction in the causal link, namely the one originating from banks. A common approach in this literature has been to identify exogenous shocks to bank health and trace their effect on bank lending and various types of real economic activity.\footnote{For example, \citet{Peek/Rosengren:00:AER}, \citet{Khwaja/Mian:08:AER}, \citet{Chodorow-Reich:14:QJE}, \citet{Bentolila/Jansen/Jimenez:18:JEEA} provide evidence that a decline in bank health can cause banks to contract lending, raise rates, and/or have an impact on foreign markets or employment.} What makes such an approach challenging is the possibility that, regardless of one's definition of a fragile bank, the composition of the bank's borrower pool contributed to that fragility. 

Our framework allows us to take a more general approach which differs from the previous empirical literature in two crucial ways. First, we are able to separately measure both directions of spillover -- from sectors to banks and from banks to sectors -- and to determine whether the effect is significant in either or both directions. Second, our analysis is not contingent on a singular source of exogenous variation, such as a crisis or an unexpected policy announcement. To our knowledge, ours is the first empirical application to jointly estimate \textit{both} directions of outcome spillover between the banking sector and the real sector and take a stand on their joint significance.

We use firm-bank matched data from Spain between 2005 and 2012 to construct a measure of zombie congestion at the 3-digit sector level based on firms' interest coverage ratios and a measure of bank weakness based on reported loan loss provisions. Using directed networks, which capture the dependence between banks and sectors, we implement our estimator and test for the presence of spillover effects between zombie-firm congestion and bank fragility along these networks. We find sizable and significant positive spillover in both directions: a one standard deviation increase in the sector-level zombie share increases banks' loan loss provisions ratio by about 8.7 p.p. At the same time, weaker banks lead to an increase in the prevalence of zombie firms: an increase of 8.7 p.p. of the loan loss provision ratio leads to a 1.2 p.p. increase in the share of a sector's assets that are sunk in zombie firms.

Our results imply that bank fragility and inefficient resource allocation, as proxied by the degree of zombie congestion, feed into and reinforce each other. From a policy perspective, this means that a combination of prudential bank regulation and policies fostering a dynamic business environment is needed to avoid aggregate productivity losses. 

\medskip

\textbf{Literature Review}\medskip

The literature of dynamic linear panel models with large $n$ and small $T$ has developed various ways to accommodate fixed effects by transforming linear models. (See \cite{Chamberlain:84:Handbook} and \cite{Arellano/Honore:01:Handbook} for an overview of this literature.) Our method relies on the transform suggested by \cite{Arellano/Bover:95:JOE}. See \cite{Hayakawa:09:ET} for a related study of efficient instrumental variables. Our cluster-specific time effects can be viewed as a special case of interactive effects, where we impose equality among the factor loadings in the same cluster. See \cite{Holtz-Eakin/Newey/Rosen:88:Eca} and \cite{Ahn/Lee/Schmidt:01:JOE}, and more recently, \cite{Kuersteiner/Prucha:20:Eca}. The literature of linear panel models with interactive effects has also developed asymptotic theory with large $n$ and large $T$. (See \cite{Fernandez-Val/Weidner:18:ARE} for a review of the related literature. See also \cite{Shi/Lee:17:JOE} for a spatial dynamic panel model with interactive effects.)

Econometric models that allow for spillover effects over cross-sectional units and cross-sectional dependence have been developed in the literature. (The literature is vast. We refer the reader to \cite{Lee:04:Eca}, \cite{Lee/Yu:10:RSUE}, \cite{Lee/Liu/Lin:10:EJ}, \cite{Kuersteiner/Prucha:13:JOE} and \cite{Kuersteiner/Prucha:20:Eca} and references therein.) The perspective of spillover effects realizing along multiple networks has received attention in the literature. \cite{Egami:21:PA} developed methods of sensitivity analysis when there are unobserved networks that capture part of the spillover effects. \cite{Drukker/Egger/Prucha:22:ET} proposed asymptotic analysis of estimation and inference procedures in a linear spatial model, which accommodates multiple networks.

To the best of our knowledge, the proposal of \cite{Kuersteiner/Prucha:20:Eca} covers the most general class of linear panel models accommodating interactive effects, lagged dependent variables and spatial dependence both in the explanatory variables and errors within the large $n$ and small $T$ framework. They proposed a generalized version of the Helmert transform in \cite{Arellano/Bover:95:JOE}, and developed asymptotic theory as $n \rightarrow \infty$ while $T$ is fixed. While the main focus of \cite{Kuersteiner/Prucha:20:Eca} was to develop general asymptotic inference that covers a wide range of linear panel models, our paper pursues a more parsimonious model which is motivated by an area of empirical research on spillover along multiple networks, where each network carries a distinct economic interpretation as a channel of causal effects among cross-sectional units. As such, the estimation and inference procedures and the techniques used for the development of asymptotic theory are different from those of \cite{Kuersteiner/Prucha:20:Eca}.

Our paper is organized as follows. In the next section, we present a linear panel model of spillover along multiple large networks. We also demonstrate how this model can be extended to accommodate the spillover effects between two groups of cross-sectional units. In Section 3, we introduce the estimation and inference procedures, and provide conditions under which the procedures are asymptotically valid. We also develop the method of inference on the direction of spillovers using a multiple testing procedure and show that it controls the FWER asymptotically. In this section, we present results from our Monte Carlo simulation study which show that the inference procedures perform well across a wide class of network structures. In Section 4, we apply our methods to the empirical analysis of spillover effects between firms and banks in Spain. In Section 5, we conclude. The mathematical proofs of the asymptotic inference results and further results on the empirical study are found in the Supplemental Note.

\section{Spillover along Multiple Large Networks}
\subsection{A Linear Panel Model of Spillover along Networks}
Let us first present a dynamic linear panel model that can be useful to study the spillover of outcomes over multiple large networks. Later we extend our model to one that enables us to study the direction of spillover effects between two groups of cross-sectional units. We assume that there are $L$ observed large networks over a set $N$ of cross-sectional units, where each network evolves over time. Each cross-sectional unit may represent a person, a household, or a firm, depending on the application. The networks may represent friendship among people or business relations among firms. There are multiple networks capturing such relations, where each network captures a distinct aspect of the relationships, but the researcher does not know along which set of networks the spillover effects realize.\footnote{This lack of knowledge leads to ambiguity about the networks. Our paper does not necessarily view the networks as multiple proxy measurements of a single underlying network. Instead, each network reflects a distinct aspect of the relationship between cross-sectional units. Estimating the spillover effects with no or partial knowledge of the network structure has received attention in the literature of statistics and econometrics. For example, see \cite{Choi:17:JASA}, \cite{dePaula/Rasul/Souza:20:WP}, \cite{Zhang:20:WP}, \cite{Lewbel/Qu/Tang:21:WP}, and \cite{He/Song:22:WP}.} For each unit $i$, we denote $N_{t-1,\ell}(i)$ to represent the in-neighborhood of unit $i$ in the $\ell$-th network in time $t$. That is, the set $N_{t-1,\ell}(i)$ represents the set of units $j$ whose outcome influences the outcome of unit $i$. We assume that $i \notin N_{t-1,\ell}(i)$ for each $i$ and $\ell$, so that each unit is not its own neighbor.

Separately from the spillover effect through the networks, each unit $i$ belongs to a cluster and is subject to time-varying cluster-specific common shocks. For example, the correlation between two firms' performance may come from network effects due to their direct interdependence, or just from sector-specific macroeconomic common shocks which affect the outcomes of all the firms in the sector. It is important to distinguish between the network effects and sector-specific, time-varying common shocks. 

For each cross-sectional unit $i$, the cluster index is given by $\mu(i) \in \{1,...,\overline c\}$, so that the equality $\mu(i) = c$ indicates that the unit $i$ belongs to cluster $c$. Similarly as in the network notation, we denote the set of units in the same cluster as $i$ by:
\begin{align}
	\label{N_C(i)}
	N_C(i) = \left\{j \in N: \mu(i) = \mu(j) \right\}.
\end{align}
 (The subscript $C$ in $N_C(i)$ is mnemonic for ``cluster''.) Throughout this paper, we regard the networks as random but the cluster structure as non-random.\footnote{We assume a time-invariant cluster structure only for the sake of simplicity. Our framework allows for the case where the cluster structure is time-varying so that $\mu_t(i) = c$ represents that unit $i$ belongs to cluster $c$ in time $t$, by modifying the definition of filtration later appropriately.  The estimation and inference procedures remain the same.}

Suppose that there is a random variable $y_{i,t}$ representing a continuous outcome for cross-sectional unit $i$ in time $t$. We define
\begin{align}
	y_{i,t}^C = \frac{1}{|N_C(i)|}\sum_{j \in N_C(i)} y_{j,t},
\end{align}
which is the cluster average of outcomes. We assume that $y_{i,t}$'s are generated as follows:
\begin{align*}
	y_{i,t} =  \alpha_0 y_{i,t-1} + \overline Y_{i,t-1}^\top \beta_0 + X_{i,t}^\top \gamma_0 + u_{i,t},
\end{align*}
where
\begin{align}
	\label{eq3}
	\overline y_{i,t-1,\ell} = \frac{1}{|N_{t-1,\ell}(i)|}\sum_{j \in N_{t-1,\ell}(i)} (y_{j,t-1} - y_{j,t-1}^C),
\end{align}
and $\overline Y_{i,t-1} = [\overline y_{i,t-1,1},..., \overline y_{i,t-1,L}]^\top$, and $u_{i,t}$ are error terms. (The summation over $N_{t-1,\ell}(i)$ is taken to be zero, if $N_{t-1,\ell}(i)$ is empty.) Hence $\overline y_{i,t-1,\ell}$ is the average of the previous period mean-deviated outcomes of unit $i$'s neighbors according to the $\ell$-th network. The mean deviation is taken to ensure that the parameter $\beta_0$ correctly captures the network effect, after controlling for cluster-specific time effects. We will explain this below. It is worth noting that the covariate vector, $X_{i,t}$, can contain lagged covariates.

We assume that the error term $u_{i,t}$ is decomposed as follows:
\begin{align}
	\label{error term}
	u_{i,t} = v_i + f_{i,t} + \varepsilon_{i,t},
\end{align}
where
\begin{align}
	f_{i,t} = \sum_{c=1}^{\overline c} 1 \{\mu(i) = c\} \pi_{t,c}.
\end{align}
The error term (\ref{error term}) consists of three components: $v_i$, $f_{i,t}$ and $\varepsilon_{i,t}$. The first component $v_i$ represents time-invariant fixed effects. The term $\varepsilon_{i,t}$ represents unobserved idiosyncratic effects which are independent across cross-sectional units and time, conditional on the past values of other random variables and $X_{i,t}$'s and networks. (We will make clear this conditioning on the past values.) The term $f_{i,t}$ represents \textit{cluster-specific time effects}. For example, each cluster represents an industry sector and $\pi_{t,c}$ captures sector-specific common shocks which are time-varying.

In many applications, the main parameter of interest is $\beta_0$. The $\ell$-th entry of the parameter vector $\beta_0$ captures the spillover effect along the $\ell$-th network. The model is useful especially when the researcher does not know which network plays a major role in the spillover of outcomes. The model, once properly extended, is also useful when she is interested in exploring the direction of influence between two groups of cross-sectional units. We will demonstrate this in our empirical application.

The subtraction of $y_{i,t-1}^C$ in (\ref{eq3}) distinguishes our model from existing models in the literature, and requires explanation. The subtraction is made to remove any cumulative cluster-specific time effects reflected in neighbors' outcomes. First, let us write
\begin{align*}
	\frac{1}{|N_{t-1,\ell}(i)|}\sum_{j \in N_{t-1,\ell}(i)} y_{j,t-1} = \frac{1}{|N_{t-1,\ell}(i)|}\sum_{j \in N_{t-1,\ell}(i)} (y_{j,t-1} - y_{i,t-1}^C) + \frac{1}{|N_{t-1,\ell}(i)|}\sum_{j \in N_{t-1,\ell}(i)} y_{i,t-1}^C.
\end{align*}
When $N_{t-1,\ell}(i) \ne \varnothing$ for all $i \in N$, the last term is essentially equal to $y_{i,t-1}^C$ which can be absorbed into the cluster time effect $f_{i,t}$. However, when there are isolated nodes, the last term cannot be absorbed into the cluster-specific time effects $\pi_{t,g}$, and has a cross-sectional variation within the same cluster, because it is the same as
\begin{align*}
	y_{i,t-1}^C 1\{ N_{t-1,\ell}(i) \ne \varnothing\},
\end{align*}
which varies if some units in the same cluster are isolated and others are not. However, this variation has nothing to do with the neighbors' outcomes. Hence without subtracting cluster means, the coefficient $\beta_0$ would capture not only the effect of variations in the neighbors' average actions but also variations in the non-emptiness of neighborhoods, which makes it hard to interpret $\beta_0$ as a measure of spillover effect along the networks. Therefore, we consider the form with subtracting $y_{i,t-1}^C$.\footnote{The mean deviation form is still required even if we focus on a single cluster setting, i.e., including only time effects. In this case, we need to consider subtracting the time average of the outcomes.}

\begin{figure}[t]
	\begin{center}
		\includegraphics[scale=0.55]{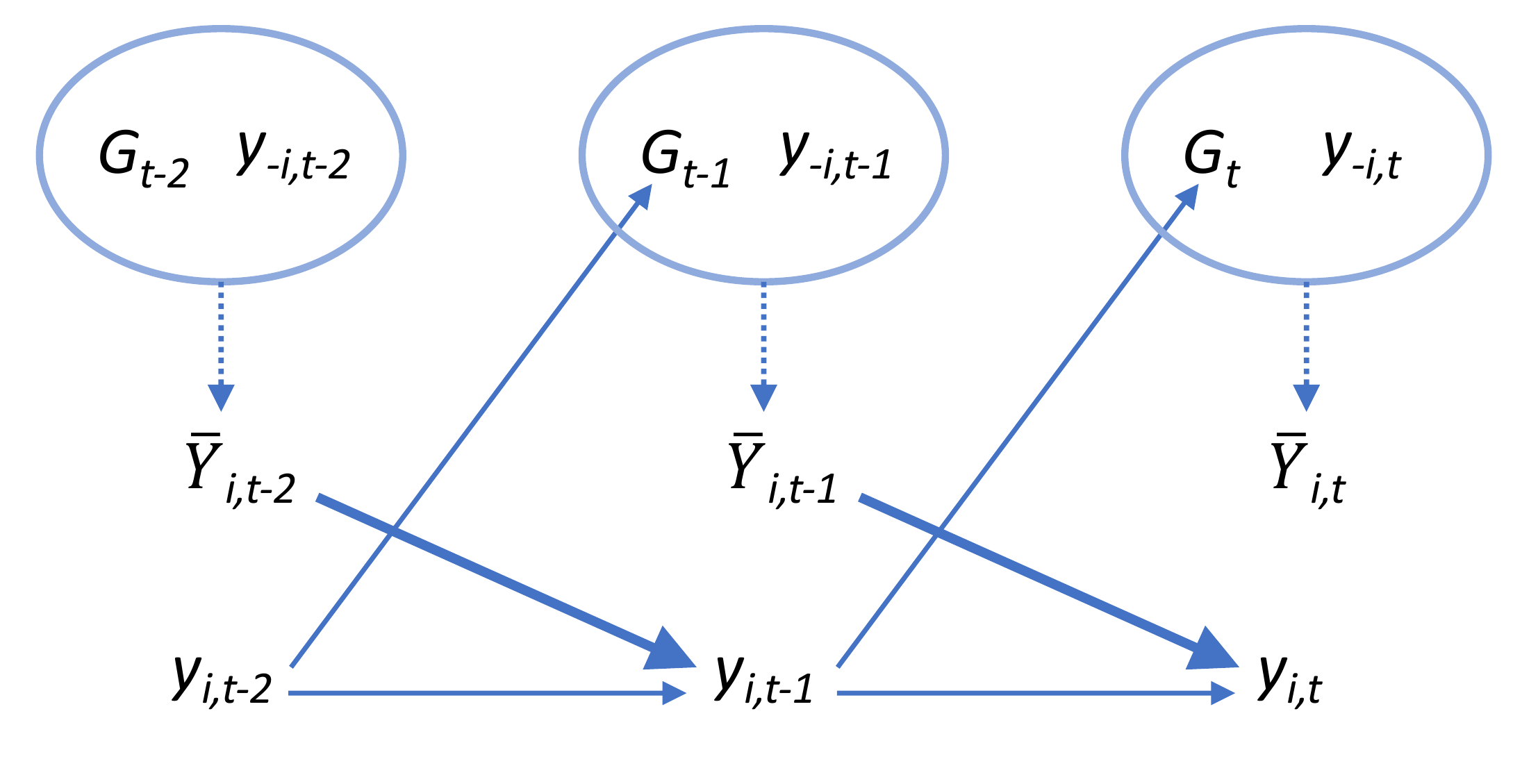}
		\caption{The Causal Diagram of Network Formation and Outcome Realization in the Case of Time-Varying Networks}
		\label{fig:causal diagram}
		\bigskip
		\parbox{6.2in}{\footnotesize Notes: The figure shows the causal diagram of the networks and outcomes. Here each solid arrow indicates the causal direction between variables and networks. The notation $G_t$ denotes the set of networks at time $t$, which is affected by the previous outcome $y_{i,t-1}$. The dotted arrow simply represents that $\overline Y_{i,t}$ is constructed from $G_t$ and $y_{-i,t} = (y_{j,t})_{j \ne i}$. The thick solid arrow from $\overline Y_{i,t-1}$ to $y_{i,t}$ represents the spillover effect. In this case, we assume that the network $G_t$ is formed based on $y_{t-1}$, not on $y_{t}$. Since we do not allow for $y_{t}$ to influence the formation of the network $G_t$ (except through the fixed effects), we allow the endogeneity of network formation to arise only through the fixed effects in this case.}
	\end{center}
\end{figure}

In contrast to the outcome generation, we consider a generic model of evolution of networks over time. First, define for each $i \in N$ and $t$,
\begin{align}
	\label{V}
	V_{i,t-1} = \left[ y_{i,t-2}, X_{i,t-1}^\top, v_i, f_{i,t-1}, \eta_{i,t-1} \right]^\top,
\end{align}
where $\eta_{i,t-1}$ is a time-varying idiosyncratic shock that is independent of $\varepsilon_{i,t-1}$'s. Let $\mathcal{V}$ be the space from which $V_{i,t-1}$ takes values. For the network $G_{t-1,\ell}$, we assume that the link from $j$ to $i$ is formed if and only if $D_{ij,t-1,\ell} = 1$, where
\begin{align*}
	D_{ij,t-1,\ell} = \varphi_{t,\ell}\left(V_{i,t-1}, V_{j,t-1}\right),
\end{align*}
where $\varphi_{t,\ell} : \mathcal{V} \times \mathcal{V} \rightarrow \{0,1\}$ is a nonstochastic, time-varying map. Therefore, the network is formed based on the fixed effects and all the previous-period variables that are observed or unobserved. For example, our model accommodates the situation where a firm to bank relation is revised in every period, based on their previous period's performance.\footnote{Note that while our model can accommodate time-varying networks, the data we employ in our empirical application only allows for static network definitions.} Throughout the paper, we do not make a further specification of the functional form of $\varphi_{t,\ell}$ other than noting that it needs to generate sparse enough networks so that our asymptotic theory works. (The causal relations between outcome variables and networks are depicted in Figure \ref{fig:causal diagram}.) 

\subsection{Extension: Direction of Spillover Between Banks and Firms}
\label{subsec: banks and firms}

Let us explain how our model can be extended to study the direction of spillovers between two groups of cross-sectional units. First, suppose that the set of cross-sectional units $N$ is divided into two groups
\begin{align}
	\label{partition}
	N = N_B  \cup N_F,
\end{align}
where we call $N_B$ the set of ``banks'' and $N_F$ the set of ``firms''. We assume that a bank and a firm cannot be in the same cluster, that is, the cluster structure is a refinement of the partition (\ref{partition}).  

At each time $t$, each node $i$ is associated with an outcome variable $y_{i,t}$. As before, the causal relationship between the variables at each time $t$ is expressed by a set of observed, directed networks whose neighborhoods of a unit $i$ are written as $N_{t-1,\ell}(i)$, $\ell=1,...,L$. For each node $i$ which is either a bank or a firm, we define two in-neighborhoods, one consisting of banks and the other consisting of firms:
\begin{align*}
	N_{t-1,\ell,B}(i) &= \{j \in N_B : j \in N_{t-1,\ell}(i)\}, \text{ and } \\
	N_{t-1,\ell,F}(i) &= \{j \in N_F : j \in N_{t-1,\ell}(i)\}.
\end{align*}
Hence the set $N_{t-1,\ell,B}(i)$ denotes the set of banks whose outcome at time $t-1$ potentially influences the outcome at time $t$ of node $i$ according to network $\ell$, and similarly, the set $N_{t-1,\ell,F}(i)$ the set of firms whose outcome at time $t-1$ potentially influences the outcome at time $t$ of node $i$ according to the $\ell$-th network.

Our primary interest is in the direction of spillover effects between the outcomes of banks and those of firms. To investigate this, let us consider the following linear model. For each $i \in N$, we specify the outcomes to be generated as follows:
\begin{align}
	\label{bank firm model}
	y_{i,t} =  \left\{ \begin{array}{ll}
		y_{i,t-1}\alpha_B + \overline Y_{i,t-1,B}^\top \beta_{BB} + \overline Y_{i,t-1,F}^\top \beta_{FB} + X_{i,t}^\top \gamma_B + u_{i,t}, & \text{ if } i \in N_B, \vspace{5pt}\\  \\
		y_{i,t-1}\alpha_F + \overline Y_{i,t-1,B}^\top \beta_{BF} + \overline Y_{i,t-1,F}^\top \beta_{FF} + X_{i,t}^\top \gamma_F + u_{i,t}, & \text{ if } i \in N_F,
	\end{array}
	\right.
\end{align}
where $\overline Y_{i,t-1,B} = [\overline y_{i,t-1,1,B},..., \overline y_{i,t-1,L,B}]^\top$, and $\overline Y_{i,t-1,F} = [\overline y_{i,t-1,1,F},..., \overline y_{i,t-1,L,F}]^\top$,
\begin{align*}
	\overline y_{i,t-1,\ell,B} &= \frac{1}{|N_{t-1,\ell,B}(i)|}\sum_{j \in N_{t-1,\ell,B}(i)} (y_{j,t-1} - y_{j,t-1}^C), \text{ and } \\
	\overline y_{i,t-1,\ell,F} &= \frac{1}{|N_{t-1,\ell,F}(i)|}\sum_{j \in N_{t-1,\ell,F}(i)} (y_{j,t-1} - y_{j,t-1}^C).
\end{align*}
Here we specify $u_{i,t}$ as in (\ref{error term}). The parameter $\beta_{BF}$ captures the impact of the average outcomes of the banks linked to a firm at time $t-1$ on the firm's outcome at time $t$, similarly $\beta_{FB}$ that of the average outcomes of the firms linked to a bank at time $t-1$ on the bank's outcome. By computing the joint confidence region for $(\beta_{BF}, \beta_{FB})$, we can check whether the spillover works in only one direction and if so, which direction. In a later section, we provide details on this procedure. We apply this procedure to the investigation of the relation between firms and banks in Spain in our empirical application.

It is important to note that the outcomes do not need be of the ``same type'' for banks and firms. This flexibility is crucial when analyzing firms and banks because these are very different types of entities and important characteristics defining one group may not be well defined for the other. For example, for banks, we can take $y_{i,t}$ to be bank health, as measured by loan loss provisions -- a variable which has no natural counterpart on the firm side -- whereas for firms, we can take $y_{i,t}$ to be firm health, as measured by the firm's zombie status -- a concept which solely applies to firms. In this case, the parameter $\beta_{BF}$ captures the spillover effect from the previous period's bank health on the health of its linked firms, where ``health'' has a different meaning for firms than it does for banks.\footnote{Productivity is another prominent example of a concept which is only applicable to firms and which is highly related to bank performance. Our framework can accommodate analyzing the spillover between the health of the banking system and firm productivity and hence shed light on aggregate productivity dynamics.}

\section{Estimation and Inference}

\subsection{Estimation and Inference}

In this section, we focus on the extension in Section \ref{subsec: banks and firms} and explain the estimation and inference procedure.  First, we write the dynamic panel model compactly as
\begin{align}
	\label{y gen}
	y_{i,t} = W_{i,t}^\top \delta_{i} + u_{i,t}, 
\end{align}
where $W_{i,t} = [y_{i,t-1}, \overline Y_{i,t-1,B}^\top, \overline Y_{i,t-1,F}^\top, X_{i,t}^\top]^\top$, and
\begin{align*}
	\delta_{i} = \left\{ \begin{array}{cc}
		\displaystyle \delta_{B} \equiv [\alpha_B, \beta_{BB}^\top, \beta_{FB}^\top, \gamma_B^\top]^\top,& \text{ if } i \in N_B \\  \\
		\displaystyle \delta_{F} \equiv [\alpha_F, \beta_{BF}^\top, \beta_{FF}^\top, \gamma_F^\top]^\top.& \text{ if } i \in N_F.
	\end{array}\right.
\end{align*}
Hence the parameter vector $\delta_i$ takes a different value depending on whether the unit $i$ is a bank or a firm. We construct an estimator of $\delta \equiv [\delta_{B}^\top, \delta_{F}^\top]^\top$ using subgroup units as follows. 

First, we define cluster-specific averages:
\begin{align}
	Y_{i,t}^C = \frac{1}{|N_C(i)|} \sum_{j \in N_C(i)} y_{j,t}, \text{  } W_{i,t}^C = \frac{1}{|N_C(i)|} \sum_{j \in N_C(i)} W_{j,t}, \text{ and } u_{i,t}^C = \frac{1}{|N_C(i)|} \sum_{j \in N_C(i)} u_{j,t}.
\end{align}
We combine the Helmert transform in \cite{Arellano/Bover:95:JOE} with the between-group operation in panel models to transform variables as follows: 
\begin{align*}
	y_{i,t}^H &= \sum_{s=t}^T h_{s,t}\left(y_{i,s} - y_{i,s}^C \right), \\
	W_{i,t}^H &= \sum_{s=t}^T h_{s,t}\left(W_{i,s} - W_{i,s}^C \right), \text{ and }\\
	u_{i,t}^H &= \sum_{s=t}^T h_{s,t}\left(u_{i,s} - u_{i,s}^C \right),
\end{align*}
where $h_{s,t}$'s are constants defined as
\begin{eqnarray*}
	h_{s,t} = \left\{\begin{array}{ll}
		\displaystyle \sqrt{\frac{T-t}{T-t+1}}, & \text{ if } s = t\\
		\displaystyle -\frac{1}{\sqrt{(T-t)(T-t+1)}}, & \text{ if } s = t+1,...,T.
	\end{array} \right.
\end{eqnarray*}
The Helmert transform is useful because it removes any serial correlation in the errors (including fixed effects), and facilitates the analysis of the asymptotic validity of the inference.

Let $Z_{i,t}$ be a $d_Z$-dimensional vector of instrumental variables, $d_Z \ge d_W$, where $d_W$ denotes the dimension of $W_{i,t}$. There are multiple ways to choose instrumental variables in our framework. We will discuss this in more detail later. (Conditions for $Z_{i,t}$ are summarized in Assumption \ref{assump: Zt} below.)

We define $n_B = |N_B|$ and $n_F = |N_F|$, so that $n_B$ denotes the number of banks and $n_F$ the number of firms in the sample. For $K \in \{B,F\}$, we construct
\begin{align*}
	\hat A_{t,K} &= \frac{1}{n_K}\sum_{i \in N_K} \sum_{s=t}^T h_{s,t} (Z_{i,t} - Z_{i,t}^C)(y_{i,s} - y_{i,s}^C), \\
	\hat B_{t,K} &= \frac{1}{n_K}\sum_{i \in N_K} \sum_{s=t}^T h_{s,t} (Z_{i,t} - Z_{i,t}^C) (W_{i,s} - W_{i,s}^C)^\top, \text{ and }\\
	\hat U_{t,K} &= \frac{1}{n_K}\sum_{i \in N_K} \sum_{s=t}^T h_{s,t} (Z_{i,t} - Z_{i,t}^C) (u_{i,s} - u_{i,s}^C),
\end{align*}
where
\begin{align*}
	Z_{i,t}^C = \frac{1}{|N_C(i)|}\sum_{i \in N_C(i)} Z_{i,t}.
\end{align*}
By construction, we have
\begin{eqnarray*}
	\hat U_{t,K} = \frac{1}{n_K}\sum_{i \in N_K} \sum_{s=t}^T h_{s,t} (Z_{i,t} - Z _{i,t}^C) \varepsilon_{i,s}.
\end{eqnarray*}
Hence the Helmert transform eliminates the fixed effects $v_i$. The between-group transformation eliminates the cluster-specific time effects.

We let 
\begin{align}
	\label{AK BK}
	\hat A_K = \sum_{t=1}^{T-1} \hat A_{t,K}, \text{ and } \quad \hat B_K = \sum_{t=1}^{T-1} \hat B_{t,K}
\end{align}
and define our initial estimator of $\delta_K$ to be
\begin{align*}
	\tilde \delta_K = \left( \hat B_K^\top \hat B_K \right)^{-1} \hat B_K^\top \hat A_K.
\end{align*}
Using this, we define $\hat u_{i,t,K}^H = Y_{i,t}^H - W_{i,t}^{H \top} \tilde \delta_K$, and let
\begin{align*}
	\hat \Omega_K =   \frac{1}{n_K}\sum_{i \in N_K}  \left(\sum_{t=1}^{T-1}(Z_{i,t} - Z_{i,t}^C) \hat u_{i,t,K}^H \right)\left(\sum_{t=1}^{T-1}(Z_{i,t} - Z_{i,t}^C)^\top \hat u_{i,t,K}^H\right).
\end{align*}

Then, we take our estimator of $\delta_K$, $K \in \{B,F\}$, as
\begin{align*}
	\hat \delta_K = \left( \hat B_K^\top \hat \Omega_K^{-1} \hat B_K \right)^{-1} \hat B_K^\top \hat \Omega_K^{-1} \hat A_K.
\end{align*}
As we show later, we have
\begin{align*}
	\left[ \begin{array}{c}
		\sqrt{n_B} \hat V_B^{-1/2} (\hat \delta_B - \delta_{B})\\ \\
		\sqrt{n_F} \hat V_F^{-1/2} (\hat \delta_F - \delta_{F})
	\end{array}
	\right]
	\rightarrow_d N\left(0, I\right),
\end{align*}
where for $K \in \{B,F\}$,
\begin{align}
	\label{VK}
	\hat V_K = \left( \hat B_K^\top \hat \Omega_K^{-1} \hat B_K \right)^{-1}.
\end{align}
From the asymptotic normality result, we can obtain the standard error for the $j$-th parameter of $\delta_B$ and $\delta_F$ respectively as follows:
\begin{align*}
	\hat \sigma_{B,j} = \frac{\sqrt{\mathbf{e}_j^\top \hat V_B\mathbf{e}_j}}{\sqrt{n_B}}, \text{ and } \hat \sigma_{F,j} = \frac{\sqrt{\mathbf{e}_j^\top \hat V_F\mathbf{e}_j}}{\sqrt{n_F}},
\end{align*}
where $\mathbf{e}_j$ denotes the column vector of zeros except for its $j$-th entry which is one. Thus, for each $K \in \{B,F\}$ and $\alpha \in (0,1)$, the level $(1- \alpha)$ confidence interval for the $j$-th parameter $\delta_{K,j}$ of $\delta_K$ is given by
\begin{align*}
	\left[ \hat \delta_{K,j} - \frac{c_{1 - \alpha/2} \hat \sigma_{K,j} }{\sqrt{n_K}}, \hat \delta_{K,j} + \frac{c_{1 - \alpha/2} \hat \sigma_{K,j} }{\sqrt{n_K}} \right],
\end{align*}
where $c_{1 - \alpha/2}$ denotes the $(1-\alpha/2)$-percentile of the standard normal distribution, and $\hat \delta_{K,j}$ denotes the $j$-th entry of $\hat \delta_K$.

\subsection{Instrumental Variables}

In this subsection, we discuss various choices of instrumental variables that can be used in practice. (Again, the general conditions for the instrumental variable are given in Assumption \ref{assump: Zt} below.) The first practical choice is a simple one which takes $Z_{i,t} = W_{i,t}$ for all $t=1,...,T$. While this simple choice is valid in our context, we may improve the accuracy of the estimator by choosing alternative instrumental variables.

The following IV is motivated by the optimal IV of \cite{Arellano:03:CEMFI}. We consider the following:
\begin{align*}
	Z_{i,t} = \left( \sum_{j \in N} W_{j,t}^H \varphi_{j,t}^\top \right)\left(  \sum_{j \in N} \varphi_{j,t} \varphi_{j,t}^\top \right)^{-1} \varphi_{i,t},
\end{align*}
where $\varphi_{j,t} = \varphi_t(W_{j,t} - W_{j,t}^C)$, for some function $\varphi_t: \mathbf{R}^{d_W} \rightarrow \mathbf{R}^{d_r}$ for some $d_r \ge d_W$. The choices of $\varphi_t$ in practice can be as follows.\medskip

(A) $\varphi_t(w_{j,t}) = w_{j,t}$.

(B) $\varphi_t(w_{j,t}) = [w_{j,t}^\top, w_{j,t,1}^2, w_{j,t,2}^2,....,w_{j,t,d_W}^2]^\top$.

(C) \begin{align*}
	\varphi_t(w_{j,t}) = \left\{
	     \begin{array}{ll}
	     	&[w_{j,t}^\top, w_{j,t,1}^2, w_{j,t,2}^2,....,w_{j,t,d_Z}^2]^\top,  \text{ if }t=1\\ \\
	     	&[w_{j,t}^\top, w_{j,t,1}^2, w_{j,t,2}^2,....,w_{j,t,d}^2,w_{j,t-1,1}^2, w_{j,t-1,2}^2,....,w_{j,t-1,d_Z}^2]^\top,  \text{ if }t>1.
	     \end{array}
     \right.
\end{align*}
\medskip

One may include higher order polynomial terms in the definition of IV, but this may induce a severe finite sample bias. In the context of a panel model with cross-sectionally independent observations, the optimal choice of IV has been studied in the literature. (See \cite{Donald/Newey:01:Eca} for a general treatment of this problem when there are many choices available for IVs.)

\subsection{Inference on the Direction of Spillover Between Banks and Firms} 
\label{subsec: multiple testing}

Let us turn to the problem of detecting the direction of spillovers between banks and firms. Our parameters of focus are now $\beta_{FB}$ which captures the spillover from firms to banks and $\beta_{BF}$ which captures the spillover from banks to firms. For simplicity, we assume that $\beta_{FB}, \beta_{BF} \in \mathbf{R}$.

We are interested in testing the following multiple individual hypotheses:
\begin{align*}
	&H_{0,FB}: \beta_{FB} = 0, \quad \text{ vs } \quad H_{1,FB}: \beta_{FB} \ne 0, \text{ and }\\
	&H_{0,BF}: \beta_{BF} = 0, \quad \text{ vs } \quad H_{1,BF}: \beta_{BF} \ne 0.
\end{align*}
The null hypothesis $H_{0,FB}$ says that there is no spillover from firms to banks and similarly with $H_{0,BF}$. We follow \cite{Romano/Wolf:05:Eca} and \cite{Romano/Shaikh:10:Eca}, and introduce a simple step-down procedure to determine the rejection of the null hypothesis $H_{0,J}$ for each $J \in \{FB,BF\}$, in a way that controls the familywise error rate asymptotically. For this we proceed as follows.

First, we let $d$ be the dimension of $\delta = [\delta_B^\top, \delta_F^\top]^\top$. Define $\hat \delta = [\hat \delta_B^\top, \hat \delta_F^\top]^\top$. Let us introduce two $d$-dimensional column vectors $\mathbf{e}_{FB}$ and $\mathbf{e}_{BF}$ such that
\begin{align*}
	\mathbf{e}_{FB}^\top \delta = \beta_{FB},  \text{ and } \mathbf{e}_{BF}^\top \delta = \beta_{BF}.
\end{align*} 
Hence the vectors $\mathbf{e}_{FB}$ and $\mathbf{e}_{BF}$ select out parameters $\beta_{FB}$ and $\beta_{BF}$ from $\delta$. We also define
\begin{align*}
	\hat v_{FB}^2 = \mathbf{e}_{FB}^\top \hat V \mathbf{e}_{FB}, \text{ and  } \hat v_{BF}^2 = \mathbf{e}_{BF}^\top \hat V \mathbf{e}_{BF},
\end{align*}
where
\begin{align*}
	\hat V = \left[ \begin{array}{cc}
		\hat V_B & 0\\
		0 & \hat V_F
	\end{array}
	\right],
\end{align*}
and $\hat V_K$, $K \in \{B,F\}$, are defined in (\ref{VK}). Define
\begin{align*}
	\hat Q_{FB} = \frac{n_B \hat \beta_{FB}^2}{\hat v_{FB}^2}, \text{ and } \hat Q_{BF} = \frac{n_F \hat \beta_{BF}^2}{\hat v_{BF}^2}.
\end{align*} 
Note that $\hat Q_{FB}$ and $\hat Q_{BF}$ are squared $t$-test statistics from $\hat \beta_{FB}$ and $\hat \beta_{BF}$.

Now we introduce a procedure to construct a set $\hat S \subset \{FB,BF\}$ such that when we reject all the null hypotheses that are outside of this set $\hat S$, our decisions are under familywise error rate control asymptotically. For any $\tau \in (0,1)$, we denote $c_\tau$ to be the $100 \tau \%$ percentile from the distribution of $\chi^2(1)$.\medskip

\textbf{Step 1: } Suppose that $\max\left\{ \hat Q_{FB}, \hat Q_{BF} \right\} \le c_{\sqrt{1-\alpha}}.$  Then, we set $\hat S = \{FB, BF\}$ and stop.\medskip

\textbf{Step 2: } Suppose that $\hat Q_{FB} \le c_{\sqrt{1-\alpha}}, \text{ but } \hat Q_{BF} > c_{\sqrt{1-\alpha}}$. Then, we set $\hat S = \{FB\}$. On the other hand, suppose that $\hat Q_{FB} > c_{\sqrt{1-\alpha}}, \text{ but } \hat Q_{BF} \le c_{\sqrt{1-\alpha}}$. Then, we set $\hat S = \{BF\}$.\medskip

\textbf{Step 3: } Suppose that $\hat S = \{FB\}$ in Step 2. Then, if $\hat Q_{FB} > c_{1-\alpha}$, we reset $\hat S = \varnothing$ and otherwise we keep $\hat S = \{FB\}$. On the other hand, suppose that $\hat S = \{BF\}$ in Step 2. Then, if $\hat Q_{BF} > c_{1-\alpha}$, we reset $\hat S = \varnothing$ and otherwise we keep $\hat S = \{BF\}$.\medskip

We use the set $\hat S$ to perform multiple hypothesis testing. More specifically, for each $J \in \{FB, BF\}$, we perform the hypothesis testing as follows:
\begin{align}
	\label{dec}
	\text{Reject the null hypothesis }H_{0,J} \text{ if and only if }  J \notin \hat S.
\end{align}
Following the literature of multiple testing (c.f. \cite{Lehmann/Romano:05:TSH}, Chapter 9), let us define the familywise error rate (FWER) as follows:
\begin{align}
	\label{FWER}
	\mathsf{FWER}_P = P\left\{ J \notin \hat S, \text{ for some } J \in S_P \right\},
\end{align}
where $S_P \subset \{FB, BF\}$ is a subset such that $J \in S_P$ if and only if $H_{0,J}$ is true under $P$. Thus the FWER is the probability of rejecting mistakenly at least one individual null hypothesis by following the decision in (\ref{dec}) when the individual null hypothesis holds true under $P$. From the asymptotic results below, we can show that
\begin{align*}
	\limsup_{n_B, n_F \rightarrow \infty } \mathsf{FWER}_P \le \alpha. 
\end{align*}
(See Theorem \ref{thm: asym FWER control} below for a formal statement of this result.) Hence this testing procedure controls the familywise error rate asymptotically.

\subsection{Asymptotic Theory}
In this section, we present the conditions and the results of asymptotic validity of the confidence intervals we have proposed previously. Let us first introduce notation. For each $s=1,...,T$, we define
\begin{align*}
	\tilde Z_{i,s} = \sum_{t=1}^{s \wedge (T-1)} h_{s,t} (Z_{i,t} - Z_{i,t}^C),
\end{align*}
and for $K \in \{B,F\}$, we let
\begin{align}
	\label{xi}
	\xi_{s,K} = \frac{1}{\sqrt{n_K}}\sum_{i \in N_K} \tilde Z_{i,s} \varepsilon_{i,s}.
\end{align}
Define $\hat U_K = \sum_{t=1}^{T-1} \hat U_{t,K}$ and rewrite
\begin{align*}
	\sqrt{n_K}(\hat \delta_K - \delta_K) = \left(\hat B_K^\top \hat \Omega_K^{-1} \hat B_K\right)\hat B_K^\top \hat \Omega_K^{-1} \sqrt{n_K} \hat U_K.
\end{align*}
We observe that (see Lemma \ref{lemm: repres} in the Supplemental Note)
\begin{align*}
	\sqrt{n_K} \hat U_K = \sum_{s=1}^T \xi_{s,K}, \quad K \in \{B,F\}.
\end{align*}
Hence after removing the fixed effects and cluster-specific time effects through the modified Helmert transform, the random vector $\tilde Z_{i,s}$ serves now effectively as our instrumental vector. It is useful to recall that
\begin{align*}
	\tilde Z_{i,s} = \sqrt{\frac{T-s}{T-s+1}} (Z_{i,s} - Z_{i,s}^C) + \sum_{t=1}^{s \wedge (T-1)}\frac{1}{\sqrt{(T-t)(T- t +1)}} (Z_{i,t} - Z_{i,t}^C).
\end{align*}
Hence $\tilde Z_{i,s}$ does not involve $Z_{j,s}$, for $j \ne i$, and has moments bounded as long as $Z_{i,s}$ has moments bounded uniformly over $i$.

For the development of the asymptotic theory, let us define $\mathcal{G}_{t}$ to be the $\sigma$-field generated by the neighborhoods $N_{s,\ell,K}(i)$, $i \in N$, $\ell = 1,...,L$, $K \in \{B,F\}$, and $s=0,1,...,t$. We define
\begin{align*}
	\mathcal{F}_t = \sigma\left(W_t, ..., W_0, \varepsilon_{t-1},...,\varepsilon_0, f_t,f_{t-1},...,f_0, \mathcal{G}_{t} \right) \vee \mathcal{F}_0,
\end{align*}
with $f_t =(f_{i,t})_{i \in N}$, $W_t =(W_{i,t})_{i \in N}$, $Z_t =(Z_{i,t})_{i \in N}$, and $\mathcal{F}_0 = \sigma\left( (Y_{i,0}, v_i)_{i \in N}\right)$. The first set of assumptions is concerned with the error terms $\varepsilon_{i,s}$.

\begin{assumption}\label{assump: errors} For each $s=1,...,T$, the following statements hold.
	
	(i) $\varepsilon_{i,s}$'s are conditionally independent across $i \in N$ given $\mathcal{F}_s$.
	
	(ii) $\mathbf{E}[\varepsilon_{i,s} \mid \mathcal{F}_{s}] = 0$ for all $i \in N$.
	
	(iii) For each $K \in \{B,F\}$, there exists $C>0$ such that for all $n_K \ge 1$,
	\begin{align*}
		\max_{i \in N_K} \mathbf{E}[\varepsilon_{i,s}^4 \mid \mathcal{F}_{s}]  \le C.
	\end{align*}
	
	(iv) For each $K \in \{B,F\}$, $\sigma_{n,s,K}^2 \equiv \mathbf{E}[\varepsilon_{i,s}^2 \mid \mathcal{F}_s ]$ is identical across $i \in N_K$, and as $n_K \rightarrow \infty$, 
	\begin{align*}
		\sigma_{n,s,K}^2 \rightarrow_P \sigma_{s,K}^2,
	\end{align*}
	for some random variable $\sigma_{s,K}^2>0$ which is $\mathcal{F}_0$-measurable.
\end{assumption}

Assumption \ref{assump: errors}(i) says that once we condition on $\mathcal{F}_s$, $\varepsilon_{i,s}$'s do not exhibit any cross-sectional dependence. Hence the errors can still be cross-sectionally correlated through the variables contained in $\mathcal{F}_s$. Assumption \ref{assump: errors}(iii) requires a conditional fourth moment condition for the errors, uniformly over $i \in N$. Assumption \ref{assump: errors}(iv) allows for conditional heteroskedasticity for $\varepsilon_{i,s}$, only if the heteroskedasticity arises through the conditioning variables in $\mathcal{F}_s$.  We allow the conditional variance to be different across groups $K \in \{B,F\}$.

The second assumption below requires that the instrumental variables are approximated by ones that are constructed from the contemporaneous or past values of $W_{i,t}$'s.

\begin{assumption}\label{assump: Zt} 
	\label{assump: IV approx}
	For each $i \in N_K$, $K \in \{B,F\}$ and $t=1,...,T-1$, there exists a random variable $Z_{i,t}^*$ such that
	\begin{align}
		\label{Zit cond}
		\frac{1}{n_K} \sum_{i \in N_K} \left\| Z_{i,t} - Z_{i,t}^* \right\|^4 = o_P(1), \text{ as } n_K \rightarrow \infty,
	\end{align}
    and $Z_{i,t}^*$ is an $\mathcal{F}_{t-1}$-measurable function of $(W_{i,t},W_{i,t-1},....,W_{i,1})$.
\end{assumption}
 
This assumption is satisfied immediately if $Z_{i,t} = W_{i,t}$ or if one chooses $Z_{i,t}$ to be one based on backward orthogonal deviation. For the IV's based on the linear projection, we take
\begin{align*}
	Z_{i,t}^* = \left( \sum_{j \in N} \mathbf{E}\left[ W_{j,t}^H \varphi_{j,t}^\top \mid \mathcal{F}_{t-1}\right] \right)\left(  \sum_{j \in N} \mathbf{E}\left[ \varphi_{j,t} \varphi_{j,t}^\top \mid \mathcal{F}_{t-1}\right] \right)^{-1} \varphi_{i,t}.
\end{align*}
Then Assumption \ref{assump: IV approx} is satisfied, as long as
\begin{align*}
	 &\left( \sum_{j \in N} W_{j,t}^H \varphi_{j,t}^\top \right)\left(  \sum_{j \in N} \varphi_{j,t} \varphi_{j,t}^\top \right)^{-1} \\
	 &=  \left( \sum_{j \in N} \mathbf{E}\left[ W_{j,t}^H \varphi_{j,t}^\top \mid \mathcal{F}_{t-1}\right] \right)\left(  \sum_{j \in N} \mathbf{E}\left[ \varphi_{j,t} \varphi_{j,t}^\top \mid \mathcal{F}_{t-1}\right] \right)^{-1}  + o_P(1),
\end{align*}
and appropriate moment conditions hold.

The third condition below is a condition requiring the networks to be sparse enough.

\begin{assumption}
	\label{assump: networks}
  There exists $\epsilon >0$ such that for $K \in \{B,F\}$, as $n_K \rightarrow \infty$,
   \begin{eqnarray*}
   	  \max_{t=1,...,T} \max_{i \in N_K}  \left| \bigcup_{\ell=1}^L N_{t-1,\ell,K}(i) \right| = O_P\left((\log n_K)^\epsilon \right).
   \end{eqnarray*}
\end{assumption}

A similar set of sparsity assumptions have appeared in the literature involving asymptotic normal inference on network models.\footnote{For example, see \cite{Leung:20:ReStat} and \cite{Kojevnikov/Marmer/Song:21:JOE}.}

The fourth set of assumptions is concerned with the limit of the components in the conditional variance of the estimator $\hat \delta$. Using $Z_{i,t}^*$ in Assumption \ref{assump: IV approx}, we define
\begin{align*}
	\tilde Z_{i,s}^* = \sum_{t=1}^{s \wedge (T-1)} h_{s,t}(Z_{i,t}^* - Z_{i,t}^{*C}), \text{ with } Z_{i,t}^{*C} = \frac{1}{|N_C(i)|}\sum_{i \in N_C(i)} Z_{i,t}^*,
\end{align*}
and for $K \in \{B,F\}$,
\begin{align}
	\label{xi2}
	\xi_{s,K}^* = \frac{1}{\sqrt{n_K}}\sum_{i \in N_K} \tilde Z_{i,s}^* \varepsilon_{i,s}.
\end{align}
Define
\begin{align}
	\label{B O}
	B_{n,K} &= \sum_{s=1}^{T} \frac{1}{n_K}\sum_{i \in N_K} \mathbf{E}\left[ \tilde Z_{i,s}^* W_{i,s}^\top \mid \mathcal{F}_{s-1} \right], \text{ and }\\ \notag
	\Omega_{n,s,K} &= \frac{1}{n_K}\sum_{i \in N_K} \mathbf{E}\left[ \sigma_{n,s,K}^2 \tilde Z_{i,s}^*  \tilde Z_{i,s}^{*\top} \mid \mathcal{F}_{s-1} \right], \text{ for each } s=1,...,T,
\end{align}
where we recall $\sigma_{n,s,K}^2 = \mathbf{E}[ \varepsilon_{i,s}^2 \mid \mathcal{F}_s]$ for $i \in N_K$. As for $B_{n,K}$ and $\Omega_{n,s,K}$, we make the following assumption.

\begin{assumption}
	\label{assump: limit var}
	For $K \in \{B,F\}$, and $s=1,...,T$, the following holds.
	
	(i) There exist $c>0$ and $n_0 \ge 1$ such that for all $n_K \ge n_0$, 
	\begin{align*}
		\lambda_{\min}\left( \frac{1}{n_K}\sum_{i \in N_K} \mathbf{E}\left[\tilde Z_{i,s}^*  \tilde Z_{i,s}^{*\top} \mid \mathcal{F}_{s-1} \right] \right) > c,
	\end{align*}
where $\lambda_{\min}(A)$ for any symmetric matrix $A$ denotes the minimum eigenvalue of $A$.
	
	(ii) As $n_K \rightarrow \infty$,
	\begin{align*}
		B_{n,K} \rightarrow_P B_{K} \text{ and } \Omega_{n,s,K} \rightarrow_p \Omega_{s,K},
	\end{align*}
    where both $B_K$ and $\Omega_{s,K}$ are $\mathcal{F}_0$-measurable, and $B_K^\top B_K$ and $\Omega_{s,K}$ are nonsingular.
\end{assumption}

The assumption (i) requires that the ``effective IVs'', $\tilde Z_{i,s}^*$, are not redundant for large enough sample size. While the convergence assumption (ii) may be replaced by some lower level conditions, it appears that given the heterogeneity of conditional distributions, such convergence seems necessary for the asymptotic validity of the inference procedure, and has often been used in the literature in developing asymptotic theory with heterogeneously distributed random variables. (For example, see Assumption EX of \cite{Kuersteiner/Prucha:20:Eca}.)

Lastly, we introduce conditions that control the conditional moments of the data.
\begin{assumption}
	\label{assump: m_{T,Z}}
	There exist $C>0$ and $n_0 \ge 1$ such that for $K \in \{B,F\}$, and for all $n_K \ge n_0$,
	\begin{align*}
		\sum_{s=1}^T \frac{1}{n_K}\sum_{i \in N_K} \mathbf{E}\left[ \| \tilde Z_{i,s}^* \|^4 \mid \mathcal{F}_{s-1} \right] \le C, \text{ and } \sum_{s=1}^T \frac{1}{n_K}\sum_{i \in N_K} \mathbf{E}\left[ \| W_{i,s} \|^4 \mid \mathcal{F}_{s-1} \right]\le C.
	\end{align*}
\end{assumption}

Under these assumptions, we can obtain the asymptotic normality for the estimators $\hat \delta_B$ and $\hat \delta_F$ as follows:
\begin{theorem}
	\label{thm: asym norm}
	Suppose that Assumptions \ref{assump: errors}-\ref{assump: m_{T,Z}} hold. Then, as $n_B, n_F \rightarrow \infty$,
	\begin{align*}
		\left[ \begin{array}{c}
			\sqrt{n_B} \hat V_B^{-1/2} (\hat \delta_B - \delta_{B})\\ \\
			\sqrt{n_F} \hat V_F^{-1/2} (\hat \delta_F - \delta_{F})
		\end{array}
		\right]
		\rightarrow_d N\left(0, I\right).
	\end{align*}
\end{theorem}

The theorem shows that $\sqrt{n_B} \hat V_B^{-1/2} (\hat \delta_B - \delta_{B})$ and $\sqrt{n_F} \hat V_F^{-1/2} (\hat \delta_F - \delta_{F})$ are asymptotically independent. Our next result shows that the testing procedure based on $\hat S$ controls the FWER asymptotically.

\begin{theorem}
	\label{thm: asym FWER control}
	Suppose that Assumptions \ref{assump: errors}-\ref{assump: m_{T,Z}} hold. Let $\mathsf{FWER}_P$ be as defined in (\ref{FWER}). Then,
	\begin{align*}
		\limsup_{n_B, n_F \rightarrow \infty}\mathsf{FWER}_P \le \alpha.
	\end{align*}
\end{theorem}
\medskip

\subsection{Monte Carlo Simulations}

\subsubsection{Data Generating Process}
We consider a simulation study using the model of spillover between firms and banks described in Section \ref{subsec: banks and firms} above. Let us first explain the data generating process used in this study. In simulations, we choose $n_B=n_F \in \{500, 5000\}$. For the covariates, we simply draw $X_{i,t}$ i.i.d. from $N(\mathbf{1},I)$ across $i$'s and $t$'s, where $\mathbf{1}$ denotes the vector of ones and $X_{i,t}$ is a $p$-dimensional random vector with $p=3$. 

As for the error components, we first generate fixed effects, $v_i$, independently from $N(1,1)$. For the clusters, we consider equal-size clusters for both banks and firms. In particular, we consider the number of the clusters, $\overline c$, to be from $\{10,100\}$ for banks and firms, and let banks have the same number of clusters as firms. Thus when $\overline c = 100$, the banks have 50 clusters and firms have 50 clusters each. Once cluster structures are determined, we generate cluster-specific time effects $\pi_{t,c}$ which are drawn from $N(1,1)$ independently for each cluster $c$ and time $t$. Finally we generate idiosyncratic components $\varepsilon_{i,t}$ from $N(0,1)$ independently for each unit $i$ and period $t$.

For the networks, we use graphs generated based on Barab\'{a}si-Albert (BA) random graphs with varied denseness, where we generate four BA graphs on the entire cross-sectional units, with each graph representing links from firms to firms, firms to banks, banks to banks and banks to firms. See Table \ref{table1: network char} for the network characteristics used in this simulation study. We generate outcomes $y_{i,t}$ according to the model in (\ref{bank firm model}). In the generation, we take $T\in \{5,10\}$, and $\gamma_K=1$ and $\alpha_K,\beta_{KK'} \in \{0,1\}$ for $K,K' \in \{B,F\}$.

\begin{table}[t]
	\caption{Network Characteristics.}
	\label{table1: network char}
	\small
	
	Bank - Network.\medskip
	
	\begin{tabular}{ccccccc}
		\hline
		\hline
		& \multicolumn{2}{c}{BA 1}         & \multicolumn{2}{c}{BA 5}           & \multicolumn{2}{c}{BA 9} \\ \cline{2-7} 
		n                                   & 500  & \multicolumn{1}{c|}{5000} & 500   & \multicolumn{1}{c|}{5000}  & 500         & 5000       \\ \hline
		\multicolumn{1}{c|}{maximum degree} & 63   & \multicolumn{1}{c|}{144}  & 132   & \multicolumn{1}{c|}{452}   & 198         & 644        \\
		\multicolumn{1}{c|}{average degree} & 3.91 & \multicolumn{1}{c|}{3.99} & 19.23 & \multicolumn{1}{c|}{19.92} & 34.51       & 35.85     \\
		\hline
	\end{tabular}\medskip
	\medskip
	
	Firm - Network.\medskip
	
	\begin{tabular}{ccccccc}
		\hline
		\hline
		& \multicolumn{2}{c}{BA 1}         & \multicolumn{2}{c}{BA 5}           & \multicolumn{2}{c}{BA 9} \\ \cline{2-7} 
		n                                   & 500  & \multicolumn{1}{c|}{5000} & 500   & \multicolumn{1}{c|}{5000}  & 500         & 5000       \\ \hline
		\multicolumn{1}{c|}{maximum degree} & 76   & \multicolumn{1}{c|}{117}  & 149   & \multicolumn{1}{c|}{368}   & 253         & 626        \\
		\multicolumn{1}{c|}{average degree} & 3.91 & \multicolumn{1}{c|}{3.99} & 19.22 & \multicolumn{1}{c|}{19.92} & 34.52       & 35.85     \\
		\hline
	\end{tabular}
	\medskip
	\par
	\parbox{6.2in}{\footnotesize	\medskip
		Notes: The tables present the network characteristics of the networks used for the simulation study. Bank-Networks refers to the overall characteristics of all units within a bank's in-neighborhood, that is including both firms and other banks. Analogously for the Firm-Networks. ``BA-1'', ``BA-5'', and ``BA-9'' represent networks generated according to Barab\'{a}si-Albert graphs with increasing denseness.}
\end{table}

\subsubsection{Results}

We first focus on the two-sided testing problem of $H_0: \beta_{FB} = \bar \beta_{FB}$ against $H_1: \beta_{FB} \neq \bar \beta_{FB}$ for some values of $\bar \beta_{FB}$ we choose. For the test we use a two-sided $t$-test based on our results of Theorem \ref{thm: asym norm}. We report the empirical rejection probabilities in Tables \ref{table2: size} and \ref{table3: power}, and empirical familywise error rate (FWER) in Table \ref{table4: FWER}, using $5000$ Monte Caro simulations.

\begin{table}[t]
	\caption{Empirical Rejection Probabilities under the Null Hypothesis}
	\label{table2: size}
	\small
	
	$\bar \beta_{FB}= 0$, $\alpha_B = \alpha_F = 0$, and $\beta_{BF} = \beta_{BB} = \beta_{FF}=0$\medskip
	
	\begin{tabular}{cccccccc}
		\hline
		\hline
		&                                    & \multicolumn{2}{c}{BA 1}            & \multicolumn{2}{c}{BA 5}            & \multicolumn{2}{c}{BA 9} \\ \cline{3-8} 
		n    & $Z$                                & $T=5$ & \multicolumn{1}{c|}{$T=10$} & $T=5$ & \multicolumn{1}{c|}{$T=10$} & $T=5$      & $T=10$      \\ \hline
		& \multicolumn{1}{c|}{(A)} & 0.065 & \multicolumn{1}{c|}{0.070}  & 0.062 & \multicolumn{1}{c|}{0.061}  & 0.063      & 0.063       \\
		500  & \multicolumn{1}{c|}{(B)} & 0.055 & \multicolumn{1}{c|}{0.064}  & 0.066 & \multicolumn{1}{c|}{0.062}  & 0.074      & 0.067       \\
		& \multicolumn{1}{c|}{(C)} & 0.068 & \multicolumn{1}{c|}{0.063}  & 0.065 & \multicolumn{1}{c|}{0.059}  & 0.066      & 0.066       \\ \hline
		& \multicolumn{1}{c|}{(A)} & 0.053 & \multicolumn{1}{c|}{0.057}  & 0.050 & \multicolumn{1}{c|}{0.047}  & 0.054      & 0.050       \\
		5000 & \multicolumn{1}{c|}{(B)} & 0.047 & \multicolumn{1}{c|}{0.053}  & 0.053 & \multicolumn{1}{c|}{0.045}  & 0.048      & 0.050       \\
		& \multicolumn{1}{c|}{(C)} & 0.053 & \multicolumn{1}{c|}{0.050}  & 0.055 & \multicolumn{1}{c|}{0.050}  & 0.049      & 0.049       \\ \hline
	\end{tabular}
	\medskip\medskip
	
	$\bar \beta_{FB}= 1$, $\alpha_B = \alpha_F = 1$, and $\beta_{BF} = \beta_{BB} = \beta_{FF}=1$\medskip

	\begin{tabular}{cccccccc}
		\hline
		\hline
		&                                    & \multicolumn{2}{c}{BA 1}            & \multicolumn{2}{c}{BA 5}            & \multicolumn{2}{c}{BA 9} \\ \cline{3-8} 
		n    & $Z$                                & $T=5$ & \multicolumn{1}{c|}{$T=10$} & $T=5$ & \multicolumn{1}{c|}{$T=10$} & $T=5$      & $T=10$      \\ \hline
		& \multicolumn{1}{c|}{(A)} & 0.063 & \multicolumn{1}{c|}{0.065}  & 0.054 & \multicolumn{1}{c|}{0.069}  & 0.056      & 0.065       \\
		500  & \multicolumn{1}{c|}{(B)} & 0.059 & \multicolumn{1}{c|}{0.068}  & 0.062 & \multicolumn{1}{c|}{0.063}  & 0.069      & 0.065       \\
		& \multicolumn{1}{c|}{(C)} & 0.059 & \multicolumn{1}{c|}{0.058}  & 0.068 & \multicolumn{1}{c|}{0.070}  & 0.076      & 0.067  \\ \hline
		& \multicolumn{1}{c|}{(A)} & 0.051 & \multicolumn{1}{c|}{0.057}  & 0.048 & \multicolumn{1}{c|}{0.052}  & 0.051      & 0.053       \\
		5000 & \multicolumn{1}{c|}{(B)} & 0.050 & \multicolumn{1}{c|}{0.048}  & 0.049 & \multicolumn{1}{c|}{0.055}  & 0.050      & 0.050       \\
		& \multicolumn{1}{c|}{(C)} & 0.049 & \multicolumn{1}{c|}{0.051}  & 0.056 & \multicolumn{1}{c|}{0.050}  & 0.059      & 0.052  \\ \hline
	\end{tabular}
	\medskip\medskip
	
	$\bar \beta_{FB}= 1$, $\bar \beta_{BF}= 0$, with $\alpha_B = \alpha_F = 1$, and $\beta_{BB} = \beta_{FF}=0$\medskip
	
	\begin{tabular}{cccccccc}
		\hline
		\hline
		&                                    & \multicolumn{2}{c}{BA 1}            & \multicolumn{2}{c}{BA 5}            & \multicolumn{2}{c}{BA 9} \\ \cline{3-8} 
		n    & $Z$                                & $T=5$ & \multicolumn{1}{c|}{$T=10$} & $T=5$ & \multicolumn{1}{c|}{$T=10$} & $T=5$      & $T=10$      \\ \hline
		& \multicolumn{1}{c|}{(A)} & 0.053 & \multicolumn{1}{c|}{0.060}  & 0.048 & \multicolumn{1}{c|}{0.058}  & 0.057      & 0.065       \\
		500  & \multicolumn{1}{c|}{(B)} & 0.065 & \multicolumn{1}{c|}{0.069}  & 0.070 & \multicolumn{1}{c|}{0.071}  & 0.069      & 0.074       \\
		& \multicolumn{1}{c|}{(C)} & 0.068 & \multicolumn{1}{c|}{0.067}  & 0.078 & \multicolumn{1}{c|}{0.076}  & 0.091      & 0.079  \\ \hline
		& \multicolumn{1}{c|}{(A)} & 0.042 & \multicolumn{1}{c|}{0.043}  & 0.047 & \multicolumn{1}{c|}{0.057}  & 0.057      & 0.053       \\
		5000 & \multicolumn{1}{c|}{(B)} & 0.056 & \multicolumn{1}{c|}{0.050}  & 0.056 & \multicolumn{1}{c|}{0.047}  & 0.054      & 0.054       \\
		& \multicolumn{1}{c|}{(C)} & 0.049 & \multicolumn{1}{c|}{0.050}  & 0.056 & \multicolumn{1}{c|}{0.057}  & 0.063      & 0.053 \\ \hline
	\end{tabular}\medskip
	\par
	\parbox{6.2in}{\footnotesize	\medskip
		Notes: This table represents the empirical rejection probability under the null hypothesis that $\beta_{FB} = \bar \beta_{FB}$, where we choose $\bar \beta_{FB} \in \{0,1\}$. As before, ``BA-1'', ``BA-5'', and ``BA-9'' represent networks generated according to Barab\'{a}si-Albert graphs with increasing denseness. All tests were run under the nominal level $\alpha = 0.05$. The column $Z$ represents the instrument choice taken as introduced in Section 3.2. The Monte Carlo simulation number was 5,000.}
\end{table}

Table \ref{table2: size} reports the finite sample rejection probabilities under the null hypothesis at levels 0.01, 0.05 and 0.10. First, we note that the finite sample size properties are all reasonably good. As expected, when the sample size increases from $n=500$ to $n=5000$, the empirical sizes get closer to the nominal sizes. When $T$ increases from $5$ to $10$, we have a slight over-rejection with the sample size $n=500$, but this issue is alleviated when the sample size increases. Most interestingly, the size properties do not exhibit much noticeable difference as the network becomes denser. This suggests that our asymptotic inference performs well with this set of networks.

\begin{table}[t]
	\caption{Empirical Rejection Probabilities under the Alternative Hypothesis: $\beta_{FB} = \bar \beta_{FB}+\Delta$ at the Nominal Level 0.05, where $\Delta = 0.1$.}
	\label{table3: power}
	\small
	
	$\bar \beta_{FB}= 0$, $\alpha_B = \alpha_F = 0$, and $\beta_{BF} = \beta_{BB} = \beta_{FF}=0$\medskip
	
	\begin{tabular}{cccccccc}
		\hline
		\hline
		&                                    & \multicolumn{2}{c}{BA 1}            & \multicolumn{2}{c}{BA 5}            & \multicolumn{2}{c}{BA 9} \\ \cline{3-8} 
		n    & $Z$                                & $T=5$ & \multicolumn{1}{c|}{$T=10$} & $T=5$ & \multicolumn{1}{c|}{$T=10$} & $T=5$      & $T=10$      \\ \hline
		& \multicolumn{1}{c|}{(A)} & 0.999 & \multicolumn{1}{c|}{1.000}  & 0.642 & \multicolumn{1}{c|}{0.957}  & 0.425      & 0.778       \\
		500  & \multicolumn{1}{c|}{(B)} & 0.999 & \multicolumn{1}{c|}{1.000}  & 0.638 & \multicolumn{1}{c|}{0.956}  & 0.433      & 0.788       \\
		& \multicolumn{1}{c|}{(C)} & 1.000 & \multicolumn{1}{c|}{1.000}  & 0.646 & \multicolumn{1}{c|}{0.963}  & 0.428      & 0.790 \\ \hline
		& \multicolumn{1}{c|}{(A)} & 1.000 & \multicolumn{1}{c|}{1.000}  & 1.000 & \multicolumn{1}{c|}{1.000}  & 1.000      & 1.000       \\
		5000 & \multicolumn{1}{c|}{(B)} & 1.000 & \multicolumn{1}{c|}{1.000}  & 1.000 & \multicolumn{1}{c|}{1.000}  & 1.000      & 1.000       \\
		& \multicolumn{1}{c|}{(C)} & 1.000 & \multicolumn{1}{c|}{1.000}  & 1.000 & \multicolumn{1}{c|}{1.000}  & 1.000      & 1.000 \\ \hline
	\end{tabular}
	\medskip\medskip
	
	$\bar \beta_{FB}= 1$, $\alpha_B = \alpha_F = 1$, and $\beta_{BF} = \beta_{BB} = \beta_{FF}=1$\medskip
	
	\begin{tabular}{cccccccc}
		\hline
		\hline
		&                                    & \multicolumn{2}{c}{BA 1}            & \multicolumn{2}{c}{BA 5}            & \multicolumn{2}{c}{BA 9} \\ \cline{3-8} 
		n    & $Z$                                & $T=5$ & \multicolumn{1}{c|}{$T=10$} & $T=5$ & \multicolumn{1}{c|}{$T=10$} & $T=5$      & $T=10$      \\ \hline
		& \multicolumn{1}{c|}{(A)} & 1.000 & \multicolumn{1}{c|}{1.000}  & 0.258 & \multicolumn{1}{c|}{1.000}  & 0.111      & 0.992       \\
		500  & \multicolumn{1}{c|}{(B)} & 1.000 & \multicolumn{1}{c|}{1.000}  & 0.276 & \multicolumn{1}{c|}{1.000}  & 0.111      & 0.993       \\
		& \multicolumn{1}{c|}{(C)} & 1.000 & \multicolumn{1}{c|}{1.000}  & 0.296 & \multicolumn{1}{c|}{1.000}  & 0.110      & 0.997 \\ \hline
		& \multicolumn{1}{c|}{(A)} & 1.000 & \multicolumn{1}{c|}{1.000}  & 0.991 & \multicolumn{1}{c|}{1.000}  & 0.697      & 1.000       \\
		5000 & \multicolumn{1}{c|}{(B)} & 1.000 & \multicolumn{1}{c|}{1.000}  & 0.991 & \multicolumn{1}{c|}{1.000}  & 0.709      & 1.000       \\
		& \multicolumn{1}{c|}{(C)} & 1.000 & \multicolumn{1}{c|}{1.000}  & 0.989 & \multicolumn{1}{c|}{1.000}  & 0.716      & 1.000 \\ \hline
	\end{tabular}
	\medskip\medskip
	
	$\bar \beta_{FB}= 1$, $\bar \beta_{BF}= 0$, with $\alpha_B = \alpha_F = 1$, and $\beta_{BB} = \beta_{FF}=0$\medskip
	
	\begin{tabular}{cccccccc}
		\hline
		\hline
		&                                    & \multicolumn{2}{c}{BA 1}            & \multicolumn{2}{c}{BA 5}            & \multicolumn{2}{c}{BA 9} \\ \cline{3-8} 
		n    & $Z$                                & $T=5$ & \multicolumn{1}{c|}{$T=10$} & $T=5$ & \multicolumn{1}{c|}{$T=10$} & $T=5$      & $T=10$      \\ \hline
		& \multicolumn{1}{c|}{(A)} & 0.200 & \multicolumn{1}{c|}{0.761}  & 0.065 & \multicolumn{1}{c|}{0.323}  & 0.064      & 0.251       \\
		500  & \multicolumn{1}{c|}{(B)} & 0.310 & \multicolumn{1}{c|}{0.897}  & 0.081 & \multicolumn{1}{c|}{0.330}  & 0.071      & 0.261       \\
		& \multicolumn{1}{c|}{(C)} & 0.372 & \multicolumn{1}{c|}{0.955}  & 0.074 & \multicolumn{1}{c|}{0.367}  & 0.068      & 0.268  \\ \hline
		& \multicolumn{1}{c|}{(A)} & 0.556 & \multicolumn{1}{c|}{1.000}  & 0.380 & \multicolumn{1}{c|}{0.991}  & 0.339      & 0.983       \\
		5000 & \multicolumn{1}{c|}{(B)} & 0.666 & \multicolumn{1}{c|}{1.000}  & 0.380 & \multicolumn{1}{c|}{0.990}  & 0.331      & 0.977       \\
		& \multicolumn{1}{c|}{(C)} & 0.718 & \multicolumn{1}{c|}{1.000}  & 0.369 & \multicolumn{1}{c|}{0.991}  & 0.311      & 0.981 \\ \hline
	\end{tabular}
	\par
	\parbox{6.2in}{\footnotesize	\medskip
		Notes: The table shows the empirical rejection probability under the alternative hypothesis that $\beta_{FB} = \bar \beta_{FB} + \Delta$ at the nominal level 0.05, where we choose $\bar \beta_{FB} = 0$ or $\bar \beta_{FB} = 1$, and $\Delta = 0.1$. As before, ``BA-1'', ``BA-5'', and ``BA-9'' represent networks generated according to Barab\'{a}si-Albert graphs with increasing denseness. The Monte Carlo simulation number was 5,000. As expected, the power of the test increases with the sample size. It is also worth noting that it also increases substantially as the time $T$ increases from $T=5$ to $T=10$. As the network becomes denser, the power decreases. It appears that instrument choices (B) and (C) both fare best among our suggested instruments. }
\end{table}
\FloatBarrier

In Table \ref{table3: power}, we present the empirical rejection probabilities under the alternative hypothesis of the following form:
\begin{align*}
	H_1: \beta_{FB} = \bar \beta_{FB}+\Delta,
\end{align*}
for numbers $\Delta>0$. We choose $\Delta$ to be from $\{0.1, 0.5\}$ and use the test at the level 0.05. For brevity, we report only the results with $\Delta = 0.1$. As expected, when the sample size increases from $n=500$ to $n=5000$, the power of the test increases substantially. Also, it is interesting to note that the power increases with the number of the time periods.

However, the power decreases with the denseness of the networks. Such a reduction in power with denser networks has been observed in the literature. (See \cite{Kojevnikov/Marmer/Song:21:JOE}.) In many situations, the reduction in power happens due to an increase in the cross-sectional dependence of the observations. This increase is reflected in the HAC (Heteroskedasticity-Autocorrelation Consistent) estimation of the asymptotic variance, and this estimator usually depends on the network. However, in our case, our asymptotic variance estimator does not involve networks. Furthermore, even when all the parameters are set to be zero, that is, there is no cross-sectional dependence created along the network, the power of the test decreases with the denseness of the network. This suggests that the reduction in power in this case is not due to the increase in the cross-sectional dependence of the observations. In fact, this is due to the decrease in the variations of $\overline y_{i,t}$ as the network becomes denser. Recall that $\overline y_{i,t}$ is the average of the outcomes of the neighbors of unit $i$. When the size of the neighborhood increases, and the outcomes are weakly dependent and distributionally similar, as in our simulation design, the law of large numbers can reduce the variations of $\overline y_{i,t}$, as the network becomes denser, eventually, leading to the weak power. We expect that this reduction in power will be alleviated when the heterogeneity in distribution of the outcomes is large.

Finally, we investigate the finite sample performance of the multiple testing procedure explained in Section \ref{subsec: multiple testing}. In particular we consider two settings. The first setting focuses on the case with $\beta_{FB} = 0.0$ and $\beta_{BF} = 0.5$, so that we have $S_P = \{FB\}$, and the second setting focuses on the case with $\beta_{FB} = 0$ and $\beta_{BF} = 0.0$, so that we have $S_P = \{FB,BF\}$. In Table \ref{table4: FWER}, we report the finite sample FWER of the multiple testing procedure, as the empirical average of the incidence of $\hat S$ failing to contain $S_P$ as a subset. The results show that the finite sample performance of the FWER is reasonable. When the sample sizes are large, the finite sample FWER gets closer to its nominal counterparts.

Let us summarize our findings from the Monte Carlo simulation study. First, the finite sample size properties of the tests are quite stable, regardless of the denseness of the networks used in this study. However, when the network becomes denser, the power of the tests gets weaker, and this is mostly due to the reduced variation of the prior average outcomes over the neighborhood, as the neighborhood gets larger. Second, the power of the tests increases with the sample size $n$ and the number of the time periods $T$. Third, the finite sample FWER performs reasonably well across the network configurations and time periods.

\begin{table}[t]
	\caption{Empirical Familywise Error Rate}
	\label{table4: FWER}
	\small
	
	$\bar \beta_{FB}= 0$, $\bar \beta_{BF}= 0$, with $\alpha_B = \alpha_F = 0$, and $\beta_{BB} = \beta_{FF}=0$\medskip
	
	\begin{tabular}{cccccccc}
		\hline
		\hline
		&                                    & \multicolumn{2}{c}{BA 1}            & \multicolumn{2}{c}{BA 5}            & \multicolumn{2}{c}{BA 9} \\ \cline{3-8} 
		n    & $Z$                                & $T=5$ & \multicolumn{1}{c|}{$T=10$} & $T=5$ & \multicolumn{1}{c|}{$T=10$} & $T=5$      & $T=10$      \\ \hline
		& \multicolumn{1}{c|}{(A)} & 0.067 & \multicolumn{1}{c|}{0.074}  & 0.064 & \multicolumn{1}{c|}{0.064}  & 0.066      & 0.069       \\
		500  & \multicolumn{1}{c|}{(B)} & 0.065 & \multicolumn{1}{c|}{0.072}  & 0.069 & \multicolumn{1}{c|}{0.070}  & 0.071      & 0.069       \\
		& \multicolumn{1}{c|}{(C)} & 0.075 & \multicolumn{1}{c|}{0.063}  & 0.070 & \multicolumn{1}{c|}{0.066}  & 0.071      & 0.067  \\ \hline
		& \multicolumn{1}{c|}{(A)} & 0.049 & \multicolumn{1}{c|}{0.049}  & 0.052 & \multicolumn{1}{c|}{0.051}  & 0.049      & 0.049       \\
		5000 & \multicolumn{1}{c|}{(B)} & 0.047 & \multicolumn{1}{c|}{0.055}  & 0.052 & \multicolumn{1}{c|}{0.052}  & 0.047      & 0.052       \\
		& \multicolumn{1}{c|}{(C)} & 0.050 & \multicolumn{1}{c|}{0.055}  & 0.053 & \multicolumn{1}{c|}{0.051}  & 0.047      & 0.049  \\ \hline
	\end{tabular}
	\medskip\medskip
	
	$\bar \beta_{FB}= 1$, $\bar \beta_{BF}= 0$, with $\alpha_B = \alpha_F = 1$, and $\beta_{BB} = \beta_{FF}=0$\medskip
	
	\begin{tabular}{cccccccc}
		\hline
		\hline
		&                                    & \multicolumn{2}{c}{BA 1}            & \multicolumn{2}{c}{BA 5}            & \multicolumn{2}{c}{BA 9} \\ \cline{3-8} 
		n    & $Z$                                & $T=5$ & \multicolumn{1}{c|}{$T=10$} & $T=5$ & \multicolumn{1}{c|}{$T=10$} & $T=5$      & $T=10$      \\ \hline
		& \multicolumn{1}{c|}{(A)} & 0.053 & \multicolumn{1}{c|}{0.067}  & 0.059 & \multicolumn{1}{c|}{0.071}  & 0.059      & 0.064       \\
		500  & \multicolumn{1}{c|}{(B)} & 0.057 & \multicolumn{1}{c|}{0.060}  & 0.062 & \multicolumn{1}{c|}{0.060}  & 0.059      & 0.066       \\
		& \multicolumn{1}{c|}{(C)} & 0.067 & \multicolumn{1}{c|}{0.069}  & 0.056 & \multicolumn{1}{c|}{0.066}  & 0.058      & 0.061  \\ \hline
		& \multicolumn{1}{c|}{(A)} & 0.055 & \multicolumn{1}{c|}{0.044}  & 0.055 & \multicolumn{1}{c|}{0.051}  & 0.049      & 0.055       \\
		5000 & \multicolumn{1}{c|}{(B)} & 0.053 & \multicolumn{1}{c|}{0.050}  & 0.055 & \multicolumn{1}{c|}{0.054}  & 0.050      & 0.050       \\
		& \multicolumn{1}{c|}{(C)} & 0.055 & \multicolumn{1}{c|}{0.052}  & 0.049 & \multicolumn{1}{c|}{0.047}  & 0.050      & 0.052  \\ \hline
	\end{tabular}\medskip
	\par
	\parbox{6.2in}{\footnotesize	\medskip
		Notes: The table presents the empirical familywise error rate (FWER) for the step down procedure explained in Section \ref{subsec: multiple testing}. The nominal FWER is set to be $\alpha =  0.05$. As the sample size $n$ increases, the FWER comes closer to the nominal rate.} 
\end{table}

\section{Empirical Analysis of Spillover between Industrial Sector Outcomes and Bank Weakness}

\subsection{Data}

In order to put our methodology to the test, we collect detailed data from Spain between 2005 and 2012 which links narrow sectors of real economic activity to the credit institutions providing them with external funding.

Our sector-level data is derived from annual firm-level balance sheets from Bureau van Dijk's (BvD) Orbis dataset. The coverage of the firm-level data is comprehensive: firms in the sample account for 69-82\% of Spanish gross output in the period 2005-2012 and the share of activity accounted for by small, medium and large firms closely resembles that observed in aggregate data.\footnote{See \citet{Kalemli-Ozcan/Sorensen/Villegas-Sanchez/Volosovych/Yesiltas:15:NBER}. 2006 gross output shares for small ($<19$ employees), medium ($20-249$ employees) and large ($>250$ employees) firms are, respectively, $(0.22, 0.39, 0.40)$ in the Orbis data and $(0.21, 0.38, 0.41)$ if the aggregate data from Eurostat.} We drop firm-year observations with non-positive values for total assets, tangible fixed assets, and number of employees as well as entries with negative liabilities and net worth. We drop firms in the financial sector (NACE Rev. 2 codes 64-66) and only keep observations for which basic accounting identities are satisfied.\footnote{The criteria are as in \citet{Gopinath/Kalemli-Ozcan/Karababounis/Villegas-Sanchez:17:QJE}.} Nominal quantities are deflated using sector-specific GDP deflators from Eurostat.

In order to match firms to their banks, we exploit the banker variable in Orbis, which reports the names of up to ten credit institutions with which the firm has a relationship. We take the fact that a firm reports the name of a bank to also mean that the bank lends to the firm, an assumption commonly made in the literature on firm-bank relationships.\footnote{See, for example, \citet{Kalemli-Ozcan/Laeven/Moreno:22:JEEA} and \citet{Laeven/McAdam/Popov:18:ECB} who also infer a lending relationship from the same data source.} The banker variable does not include a time stamp, meaning that we cannot determine when a lending relationship started or whether it changed over time. This shortcoming is mitigated by evidence that lender-borrower relationships tend to be sticky over the business cycle.\footnote{\citet{Giannetti/Ongena:12:JIE} look at different vintages of the banker variable and find it to be very persistent. \citet{Chodorow-Reich:14:QJE} finds consistent evidence using different data from the U.S.} 

We match the bank names reported in the firm-level data with bank financial statements from BvD's Bankscope dataset.\footnote{The matching is done based on names for lack of a common identifier in the two data sources.} We exclude credit institutions specializing in consumer credit, such as credit card and leasing companies, as well as private security and asset management companies. We construct our yearly panel of banks so as to maximize both the number of banks and the number of time periods, as both dimensions are important for our methodology. To do so, we prioritize unconsolidated accounts over consolidated ones where possible, while making sure to avoid double-counting issues.\footnote{We follow the steps outlined in \citet{Duprey/Le:16:SSRN} to create consistent time series.} By using unconsolidated accounts we also avoid the possibility that variation at the individual bank level is lost at the consolidated level.

After cleaning the data according to the procedure outlined above, we are left with $188,923$ unique firms matched to $97$ unique banks and operating in $600$ three-digit NACE Rev 2. sectors. The number of cross-sectional units underlying the main results in Section \ref{sec:results} are lower because our estimation procedure requires strongly balanced panels. To mitigate the loss of observations, we interpolate gaps in our variables of interest of up to three years.

The following section details the definitions of the variables of interest for our spillover analysis and presents some basic stylized facts.

\subsection{Definitions and Stylized Facts}

Zombie lending is the practice of extending credit to distressed firms that would exit the market under normal conditions. The incentive to evergreen loans to ailing firms is especially strong for weak banks seeking to avoid further damaging their balance sheet by reporting their losses.

\subsubsection*{Zombie Firms}
Several ways of identifying zombie firms have been proposed in the literature, and they are all designed to  capture persistent financial weakness of the firm and/or the extent to which the firm is receiving subsidized credit. In this paper we follow \citet{Adalet/Andrews/Millot:18:EP} and use a definition based on the interest coverage ratio defined as the ratio of profits (EBIT) to interest payments. According to this definition, a firm is considered a zombie in a given year if it has reported an interest coverage ratio below one for three consecutive years. This definition, which we will refer to as the baseline definition, states that a zombie firm is not profitable enough to make its debt payments. We interpret the fact that such firms are kept afloat as evidence of bank forbearance.


It is worth pointing out that our baseline measure, unlike the one proposed by \citet{Caballero/Hoshi/Kashyap:08:AER} or \citet{Acharya/Eisert/Eufinger/Hirsch:19:ReFinStud}, does not seek to identify firms receiving subsidized credit, that is, firms paying lower interest rates than their most creditworthy counterparts. The reasons behind this are twofold. First, we do not have information on the interest rates on individual loans, so we would have to infer the ``average'' interest rate paid by a firm from the ratio of interest payments to total debt. Second, we would need an appropriate benchmark to compare our interest rate proxy to. Other papers use the interest rate paid by AAA-rated publicly listed firms for this purpose, however, this benchmark is hardly relevant for our sample which mainly consists of unlisted, small- to medium-sized enterprises which are very different from large public corporations.\footnote{To get a sense of how exceptional listed firms are in Spain, the average value-added share of listed firms in manufacturing was approximately 14\% in 2006 (\cite{Garcia-Macia:17:IMF}).}

\begin{table}[t]\centering
	\caption{Firm characteristics by zombie status}
	\small
	\begin{tabular}{l*{2}{SS}}
		\toprule
		&\multicolumn{2}{c}{Non-zombie}&\multicolumn{2}{c}{Zombie}\\\cmidrule(lr){2-3}\cmidrule(lr){4-5}
		&      {Mean}&    {Median}&      {Mean}&    {Median}\\
		\hline
		Firm age            &      15.172&      14.000&      17.293&      16.000\\
		Total assets (million euro)&       3.040&       0.757&       4.782&       1.003\\
		Sales (million euro)&       2.700&       0.697&       1.987&       0.362\\
		Number of employees &      16.707&       7.000&      16.238&       5.000\\
		Return on assets    &       0.020&       0.014&      -0.060&      -0.031\\
		Investment ratio    &       0.429&      -0.045&       0.224&      -0.039\\
		Leverage ratio      &       0.270&       0.223&       0.315&       0.273\\
		Interest coverage ratio&     210.656&       2.119&    -385.585&      -2.764\\
		Debt service capacity&      39.113&       0.324&     -15.119&      -0.001\\
		Exit                &       0.089&       0.000&       0.113&       0.000\\
		\midrule
		Observations  & \multicolumn{2}{c}{1,309,561}  & \multicolumn{2}{c}{  135,573}\\
		\midrule
	\end{tabular}
	\label{tab:firm_zombies}
	\parbox{6in}{\footnotesize	\medskip
		
		Notes: This table shows a break-down of firm characteristics by zombie status according to the baseline definition. The sample period is 2005-2012. The return on assets is the ratio of net income to total assets; the investment ratio is the percent change in real fixed assets from the year before; the leverage ratio is the ratio of total debt to total assets; the interest coverage ratio is the ratio of profits to the amount of interest paid; the debt service capacity is the ratio of profits to total debt; exit is an indicator equal to one if a firm is dissolved or declares bankruptcy in the future. }
\end{table}

Table \ref{tab:firm_zombies} summarizes some firm characteristics by zombie status. We can see that zombie firms, on average, exhibit markedly lower sales, profitability, and investment while at the same time being more leveraged and more likely to exit. Zombie firms are also older.\footnote{Some zombie firm definitions also impose a firm age of over ten years to avoid misidentifying potentially successful start-ups as zombies. However, less than 25\% of observations attributed to zombies come from firms under ten years of age so we omit this criterion.}

\subsubsection*{Weak Banks}
Our choice of a variable to capture bank weakness is informed by the previous literature on zombie lending and on the bank lending channel more generally. Some papers, such as \citet{Acharya/Eisert/Eufinger/Hirsch:19:ReFinStud}, define weak banks to be those with a low ratio of regulatory capital, while others build on that to create an index which includes additional balance sheet items such as non-performing loans (\cite{Storz/Koetter/Setzer/Westphal:17:ECB}, \cite{Andrews/Petroulakis:19:ECB}). A different approach relies on the receipt of government aid (\cite{Laeven/McAdam/Popov:18:ECB}, \cite{Acharya/Birchert/Jager/Steffen:21:ReFinStud}), while yet another is to exploit the decline in the creditworthiness of some European sovereigns during 2010-2012 to measure bank weakness as the exposure to risky government debt (\cite{Kalemli-Ozcan/Laeven/Moreno:22:JEEA}). 

The ideal measure of bank weakness in the context of our methodology is one that, in addition to capturing balance sheet vulnerability, also displays substantial within-bank time variation and is available for many banks, especially for smaller ones. We therefore settle for the ratio of loan loss provisions to gross loans as our preferred bank weakness measure. Loan loss provisions are an expense set aside by banks to cover different kinds of loan losses, including non-performing loans. The loan loss provisions are then added to the loan loss reserves, which banks are required to maintain as a buffer against potential future losses. 

\begin{table}[t]\centering
	\caption{Bank outcome: Correlation with other characteristics}
	\small
	\begin{tabular}{l*{3}{S}}
		\toprule
		&\multicolumn{3}{c}{Loan loss provisions}\\
		&     {Corr.}& {$p$-value}&      {Obs.}\\
		\hline
		Return on avg. assets&      -0.104&       0.000&        1246\\
		Return on avg. equity&      -0.065&       0.021&        1244\\
		Capital ratio (equity to assets)&       0.179&       0.000&        1246\\
		Loan loss reserves (share of gross loans)&       0.316&       0.000&         513\\
		Sovereign exposure (share of total assets)&       0.114&       0.023&         396\\
		Non-performing loans (share of gross loans)&       0.136&       0.008&         378\\
		\midrule
	\end{tabular}
	\label{tab:corr_banks}
	\parbox{6in}{\footnotesize	\medskip
		
		Notes: This table shows the correlation between our baseline measure of bank weakness, given by the share of loan loss provisions in gross loans, and other bank characertistics. The sample period is 2005-2012. }
\end{table}

Table \ref{tab:corr_banks} shows how our bank weakness measure is related to other bank characteristics. Bank weakness is negatively correlated with profitability, as proxied by the return on average assets and the return on average equity. Weaker banks also have larger sovereign bond holdings, which is not surprising given that our sample period contains the European sovereign debt crisis.\footnote{The sovereign debt variable in Bankscope includes the debt of all sovereigns, not just that of the home country. However, in the case of Spain, most of the sovereign debt held by banks was that of the Spanish government.} Somewhat surprisingly, banks reporting higher loan loss provisions also seem to be better capitalized. While a higher capital ratio is generally considered a sign of a healthy balance sheet, \citet{Andrews/Petroulakis:19:ECB} point out that it could also be the result of low risk-taking and little lending activity. Conversely, banks possessing high-quality assets which generate a steady flow of income with limited risk may afford to have relatively lower levels of capital adequacy.\footnote{A better capital-based measure would be one based on risk-weighted capital or Tier 1 capital ratios, however, both these variables are riddled with data availability issues in our sample.}

It should be noted that, ultimately, any balance sheet item is subject to manipulation on the part of the bank, as documented by \citet{Blattner/Farinha/Rebelo:19:ECB} using granular loan-level data from Portugal. This is why, in a future extension of the paper, we will explore measuring bank weakness through government aid. More work needs to be done to make such a variable continuous and time-varying.

\begin{figure}[t]\centering
	\caption{Evolution of outcomes over time}
	\begin{subfigure}{.49\textwidth}
		\caption{Bank outcome}
		\includegraphics[width=\textwidth]{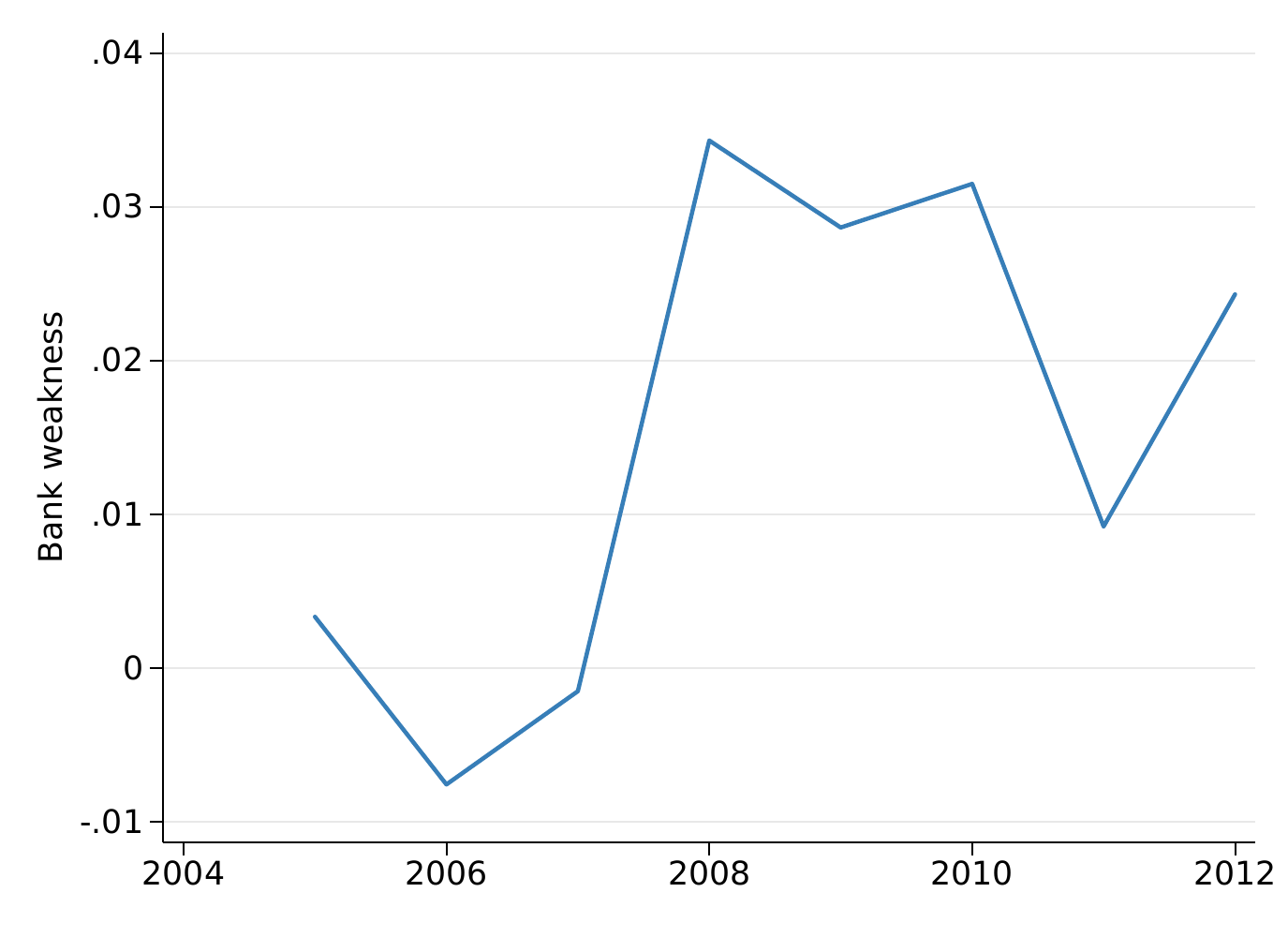}
	\end{subfigure}
	\hfill
	\begin{subfigure}{.49\textwidth}
		\caption{Sector outcome}
		\includegraphics[width=\textwidth]{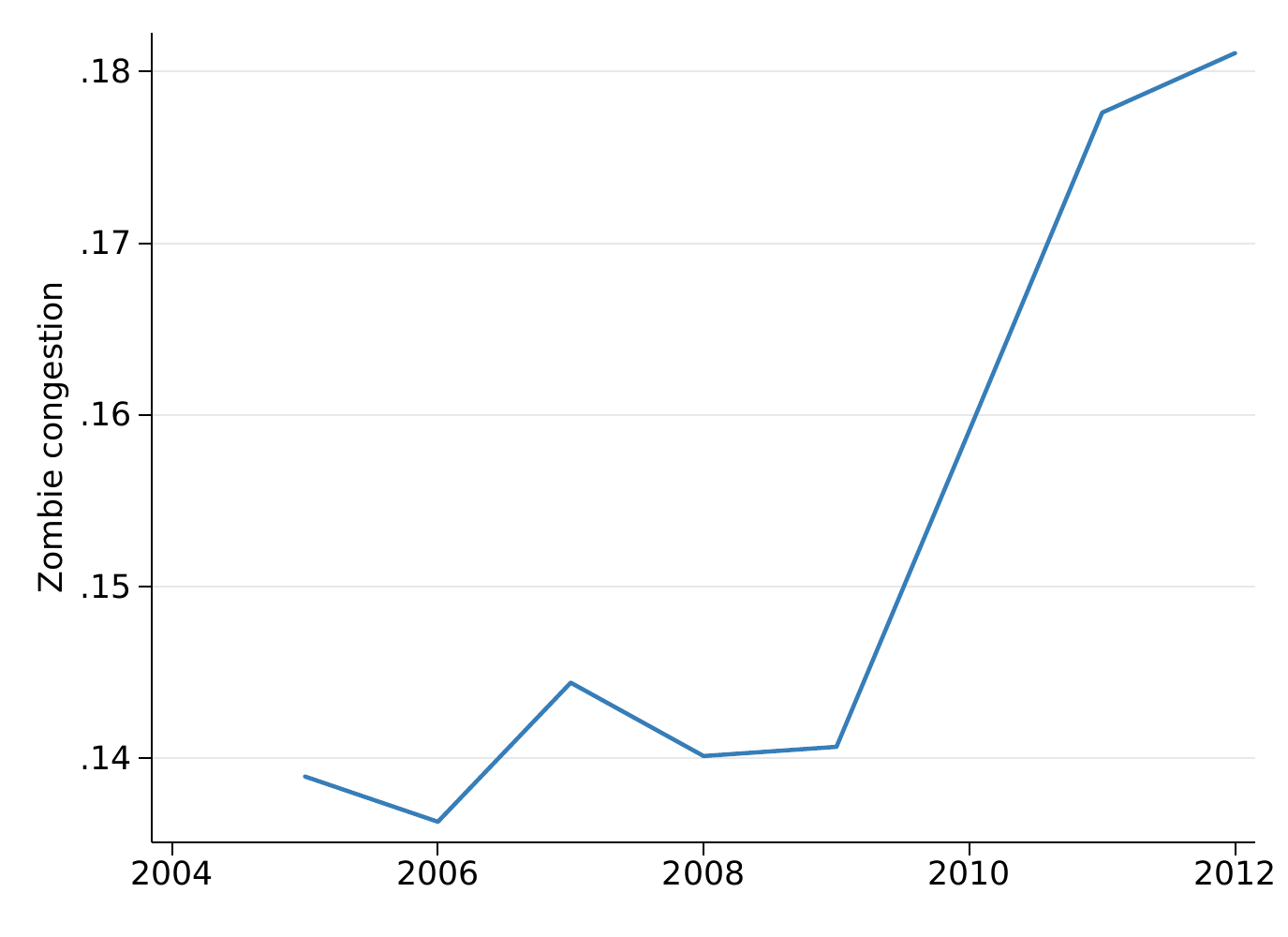}
	\end{subfigure}
	\parbox{6in}{\footnotesize	\medskip
		
		Notes: This figure plots the outcomes of interest over time. Panel (a) plots bank weakness as measured by the average loan loss provision ratio, while Panel (b) plots sector-level zombie congestion as measured by the asset-weighted share of zombie firms in each sector according to the baseline definition.}
	\label{fig:tvar_stats}
\end{figure}

Figure \ref{fig:tvar_stats} plots the time averages of the our main outcome variables: Panel (a) refers to the bank outcome measured by the loan loss provision share, while Panel (b) shows the asset-weighted average of zombie firms based on the interest-coverage definition. We can see that both measures vary significantly over our sample period and that a surge in bank weakness around 2008 is followed by an increase in sectoral zombie-firm congestion two years later.

\subsection{Networks}

Having fixed our outcomes of interest at both the bank and the sector level, we now move on to defining the network structure governing the spillover between these outcomes.

Identifying the networks along which we expect sector-level outcomes to have an effect on bank performance and vice versa is a crucial part of our analysis. This section is devoted to a careful discussion of how to define the networks which best capture the causal channels going from sectors to banks and from banks to sectors. 

At its core, the interdependence between the real sector and the financial sector comes from the former borrowing from the latter. It makes sense, then, to assume that a firm is related to a bank if the firm borrows from that bank. However, when we move past the level of the individual firm and look at the sector in which the firm is active, the exact definition of what it means for that sector to be tied to the bank becomes less clear-cut. In what follows, we take each spillover direction in turn and discuss the associated network definition.

\subsubsection{Sector to Bank} For the sector-to-bank spillover, we want to explain a bank's health using the share of distressed firms in the sectors that the bank caters to. Some sectors will be more important for a bank than others, so we should only expect spillover to occur if a bank is sufficiently exposed to a given sector. A natural measure of exposure to a given sector can be obtained by looking at what share in the total amount of loans originating from a bank goes to that sector. A bank's solvency relies on the ability of its borrowers to pay back their debt, so if a bank is heavily invested in a sector where many firms are experiencing financial distress, then we should expect that bank to suffer losses.

This definition of exposure makes sense in light of the bank and sector outcomes we are interested in. On the bank side, we are interested in bank health as measured by the provisions set aside by the bank to absorb losses from missed loan payments. On the sector side, we are interested in the prevalence of zombie firms as measured by a chronic inability to make interest payments on their debt.

Ideally, we would derive our exposure measure from loan-level data which would contain the identity of the bank, the identity of the firm and the firm's sector of activity. Unfortunately, our data does not possess this level of granularity since we do not observe a bank's individual loans or their breakdown by sector. We do, however, observe firms' total amount of debt and sector of activity as well as the identity of the banks they are related to.\footnote{Around 36\% of firms in our sample report more than one bank. We use all bank relationships for our baseline specification, but we will also explore the robustness of our results to using only the first bank mentioned by each firm.}

\begin{figure}[t]\centering
	\caption{Network characteristics (25th prc.)}
	\begin{subfigure}{.49\textwidth}
		\caption{Sector to bank}
		\includegraphics[width=\textwidth]{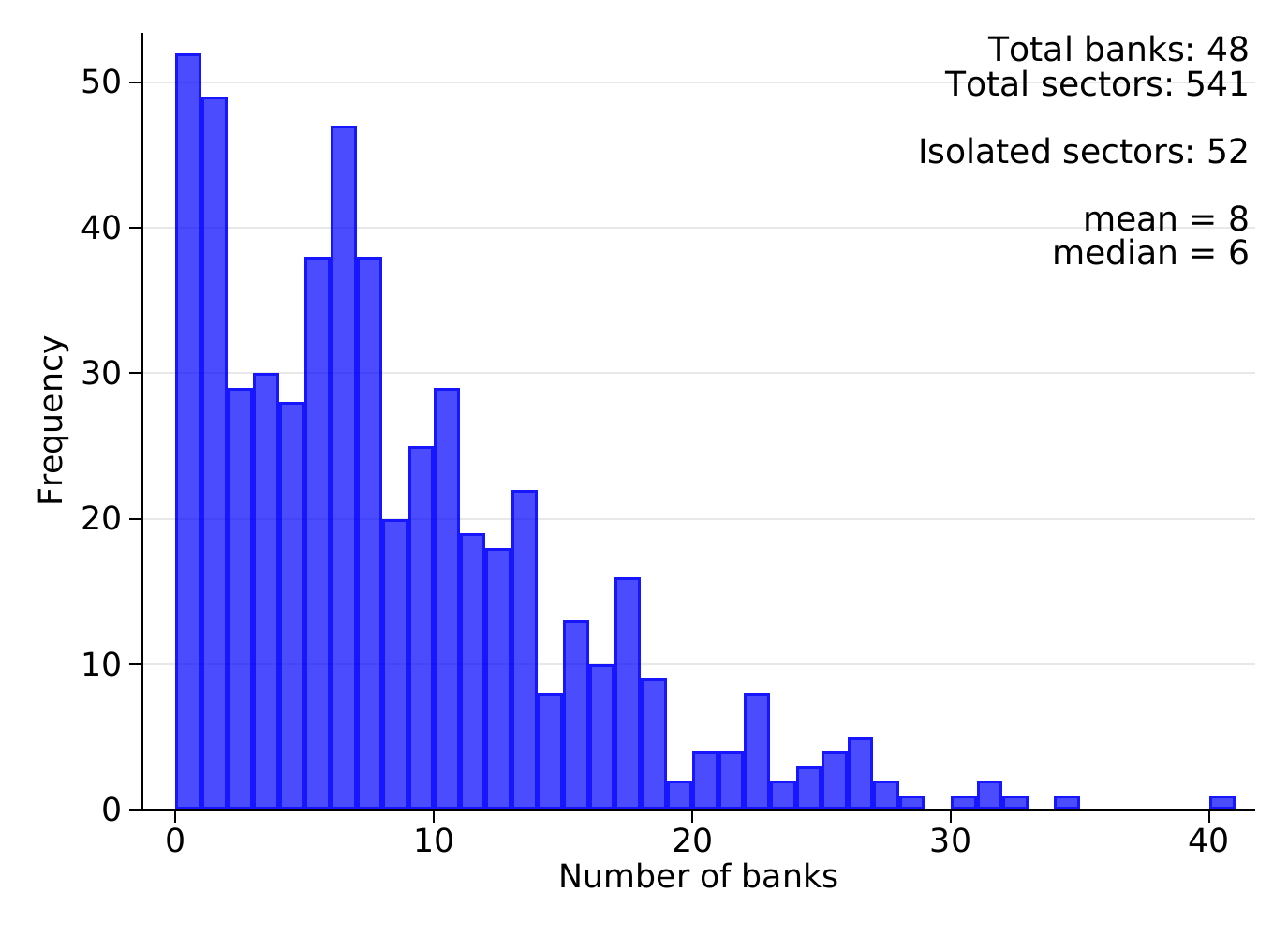}
	\end{subfigure}
	\hfill
	\begin{subfigure}{.49\textwidth}
		\caption{Bank to sector}
		\includegraphics[width=\textwidth]{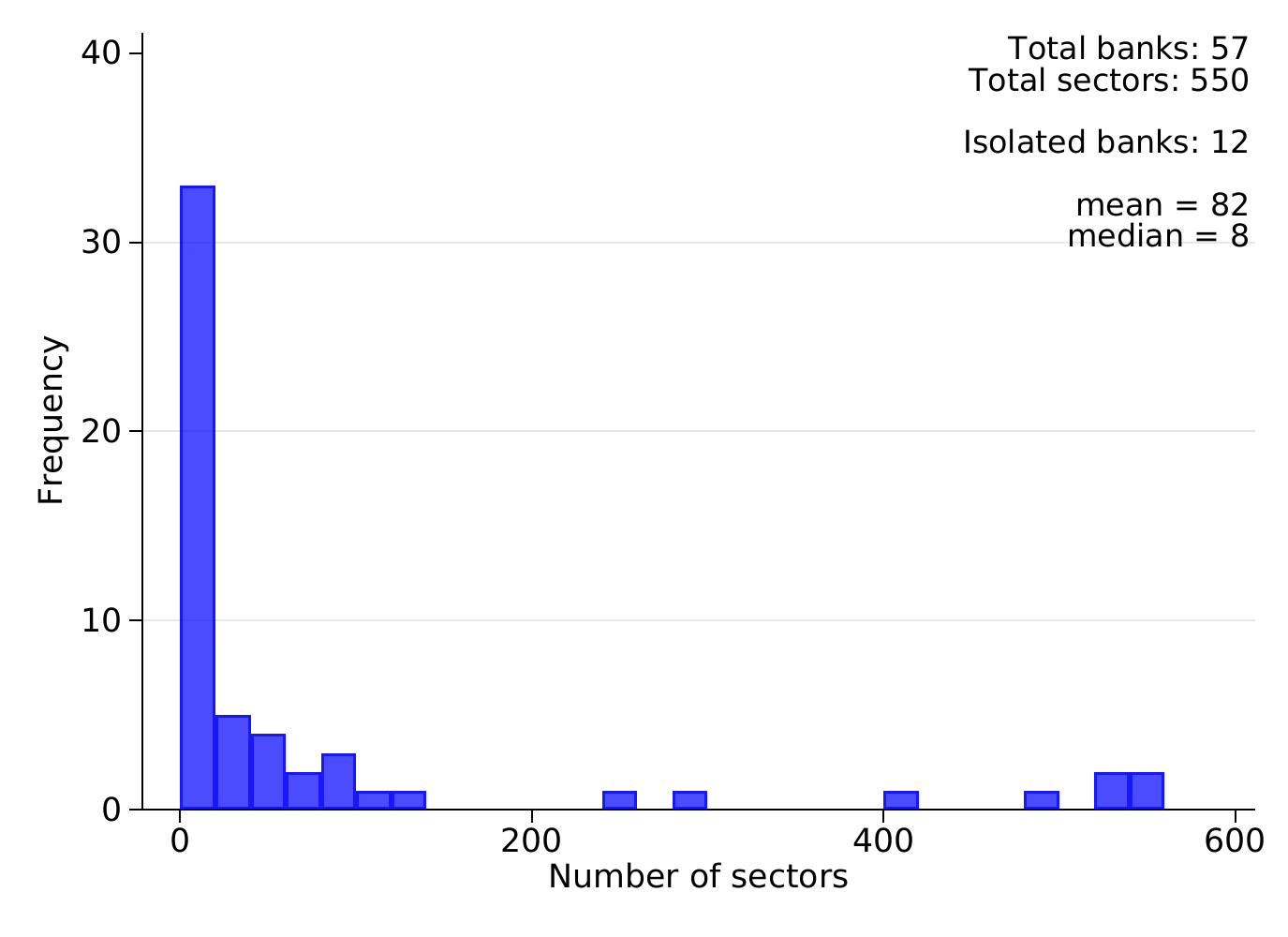}
	\end{subfigure}
	\parbox{6in}{\footnotesize	\medskip
		
		Notes: This figure shows the network characteristics for the baseline specification. Panel (a) shows the distribution of the number of sectors per bank, while Panel (b) shows the distribution of the number of banks per sector. The network thresholds in Panels (a) and (b) are set to the 25th percentile of the relevant debt share and asset share, respectively.}
	\label{fig:net25}
\end{figure}

We define a sector-to-bank relationship based on the share of a sector's debt in the total debt associated with a bank, measured in the year before our sample starts. If that share is higher than the 25th percentile of all bank-sector pairs, we assume that the sector potentially affects the bank.\medskip

\subsubsection{Bank to Sector} For the bank-to-sector spillover, we want to explain the degree of a sector's zombie congestion using the health of its main lenders. Following a similar logic as above, we would expect a bank to matter for a sector if the bank lends to a sizable segment of that sector. Given that we are interested in measuring a bank's impact on a sector, and that our sector-level outcome is the asset-weighted share of zombie firms, it makes sense to look at the asset share in that sector that is accounted for by that bank.

We define a bank-to-sector relationship based on the share of a sector's assets belonging to the firms which borrow from a bank, measured in the year before our sample starts. If that share is higher than the 25th percentile of all bank-sector pairs, we assume that the bank potentially affects the sector.

The choice of threshold for both spillover directions is faced with the following trade-off: the higher the threshold, the more sparse our network structure will be and the more statistical power our estimation will have; however, a higher threshold also means dropping smaller banks or sectors from the analysis, which could introduce a bias. Figure \ref{fig:net25} summarizes the network characteristics for both directions for a threshold set to equal the 25th percentile of the relevant debt share (in the sector-to-bank direction) and asset share (in the bank-to-sector direction), which our empirical analysis has chosen to focus on. In the Supplemental Note, we report results from different choices of the threshold.

\subsection{Specification Details}

We are now equipped with a sector- and a bank-level outcome, as well as with a set of directed networks along which to analyze the spillover between the two outcomes. It remains to discuss the control variables for our regressions. 

Both regressions of the bank-to-sector and the sector-to-bank equations include the lagged outcome as well as time- and individual-specific fixed effects. Here we assume that the cluster structure on the cross-sectional units coincides with their partition into banks and sectors. Hence cluster-specific time effects consist of two kinds of time effects, one that is common among banks and the other that is common among sectors. These time effects absorb any variation that is common across all banks or all sectors in a given year, such as adverse aggregate demand shocks. Similarly, individual sector (bank) effects control for any time-invariant characteristics that may influence sector (bank) performance. In the case of sectors, such characteristics could include the extent of government regulation, while for banks they could be management practices or overall institutional quality. 

In the bank-level equation, we opt for the most parsimonious specification which only includes the lagged outcome and the time- and bank effects. The reason behind this is that most bank assets and liabilities are not marked to market, meaning that these balance sheet variables are very stable and do not register large enough movements over time to distinguish them from the time-invariant effects. Including additional controls would also reduce our sample size while providing little in terms of explanatory power.

\begin{table}[t]\centering
	\caption{Summary statistics}
	\small
	\begin{tabular}{l*{5}{S}}
		\toprule
		&      {Mean}& {Std. dev.}&  {5th prc.}& {50th prc.}& {95th prc.}\\
		\hline
		Bank weakness       &       0.008&       0.040&       0.001&       0.005&       0.026\\
		Zombie congestion   &       0.142&       0.125&       0.019&       0.103&       0.404\\
		Market concentration&       0.040&       0.088&       0.001&       0.013&       0.153\\
		Sales growth        &      -0.020&       0.175&      -0.282&       0.005&       0.162\\
		Capital intensity   &       0.729&       1.230&       0.133&       0.356&       2.555\\
		\midrule
		Observations & \multicolumn{5}{l}{   35,144}\\
		\midrule
	\end{tabular}
	\label{tab:summary_allvars}
	\parbox{6in}{\footnotesize	\medskip
		
		Notes: This table shows summary statistics of the variables included in the baseline estimation. The sample consists of bank-to-sector pairs observed between 2005 and 2012. Bank weakness is given by the ratio of loan loss provsions to total assets; zombie congestion is the asset-weighted share of zombie firms in a sector according to the baseline definition; market concentration is the HHI index of sector-level sales; sales growth is computed as the log difference in total sector sales from the year before; capital intensity is the share of a sector's fixed assets in total sales. }
\end{table}

In the sector-level equation, the richness of our data permits the inclusion of additional time-varying controls. We control for previous-period market concentration (Herfindahl index), sales growth, and capital intensity (share of fixed assets). A priori, a higher market concentration could affect the zombie share in a sector either positively -- by distorting competition -- or negatively -- by favoring large firms with high enough profit margins that make them more resilient to shocks. As for sales growth, we would expect it to have a negative impact on zombie congestion, since a boost in sales can provide firms with a buffer that insulates them against falling behind on loan payments. Finally, capital intensity is meant to capture the degree to which a sector is dependent on banks. We include this control to make sure that it is the characteristics of the banks driving the zombie share, and not the sector's intrinsic reliance on banks. Table \ref{tab:summary_allvars} provides summary statistics of our outcome and control variables.

\subsection{Results}\label{sec:results}

\begin{table}[t]\centering
	\caption{Baseline specification (using IV based on LP with option B)}
	\small
	\begin{tabular}{l*{5}{SS}}
		\toprule
		&\multicolumn{1}{c}{Coef.}&\multicolumn{1}{c}{Std. err.}&\multicolumn{1}{c}{$p$-value}&\multicolumn{2}{c}{[95\% Conf. interval]}\\
		\hline \multicolumn{5}{l}{ \linebreak \textbf{\textit{Panel A: Sector to bank}}} \\
		Bank weakness $ _{t-1} $&       0.330&       0.002&       0.000&       0.326&       0.334\\
		Zombie congestion $ _{t-1} $&       0.964&       0.322&       0.003&       0.333&       1.596\\
		\hline
		Number of banks&          48&            &            &            &            \\
		Number of sectors&         541&            &            &            &            \\
		Observations&       35144&            &            &            &            \\
		\midrule
		\multicolumn{5}{l}{\linebreak \textbf{\textit{Panel B: Bank to sector}}} \\
		Zombie congestion $ _{t-1} $&       0.698&       0.049&       0.000&       0.603&       0.793\\
		Bank weakness $ _{t-1} $&       0.138&       0.047&       0.003&       0.046&       0.230\\
		Market concentration $ _{t-1} $&      -0.025&       0.099&       0.803&      -0.218&       0.169\\
		Sales growth $ _{t-1} $&      -0.001&       0.008&       0.908&      -0.016&       0.014\\
		Capital intensity $ _{t-1} $&      -0.015&       0.009&       0.092&      -0.031&       0.002\\
		\hline
		Number of banks&          57&            &            &            &            \\
		Number of sectors&         550&            &            &            &            \\
		Observations&       37448&            &            &            &            \\
		\midrule
		Direction of spillover & & & & &\\
		\quad with FWER control & $S \leftrightarrow B$ & $\text{ at 1\%}$ & & & \\
		\midrule
	\end{tabular}
	\label{tab:results_all_2005_2012_asymm_optB_net25_spec12}
	\parbox{6in}{\footnotesize	\medskip
		
		Notes: This table shows the results from estimating equation (\ref{bank firm model}). Panel A refers to the direction from sectors to banks while Panel B refers to the direction from banks to sectors. The network thresholds in Panels A and B are set to the 25th percentile of the relevant debt share and asset share, respectively. The sample contains all reported firm-bank relationships and covers the period between 2005 and 2012. Results are obtained using using IVs based on Linear Projection with Option B. Bank weakness is given by the ratio of loan loss provsions to total assets; zombie congestion is the asset-weighted share of zombie firms in a sector according to the baseline definition; market concentration is the HHI index of sector-level sales; sales growth is computed as the log difference in total sector sales from the year before; capital intensity is the share of a sector's fixed assets in total sales. }
\end{table}

Our main results are shown in Table \ref{tab:results_all_2005_2012_asymm_optB_net25_spec12}. Panel A refers to the bank-level equation where we are interested in how the sector-level zombie share affects bank weakness. Panel B refers to the sector-level equation where we are interested in the opposite spillover direction. In both panels, the first row represents the persistence of the outcome and the second row is our main coefficient of interest. 

Focusing first on the sector-to-bank direction (Panel A), we see that a higher level of zombie congestion in a bank's neighboring sectors leads to a significant increase in bank weakness. To get a sense of the magnitude of the effect, note that the average zombie share increased by about 4 percentage points from peak to trough (see Figure \ref{fig:tvar_stats}). Such an increase in the zombie share translates to a an increase in the bank's loan loss provisions of $4 \times .96 = 3.8$ percent of gross loans. This means that banks tied to sectors where resources are tied up in distressed firms suffer bigger losses.

Turning now to Panel B, we see that the spillover effect is still highly significant when looking at the bank-to-sector direction. Weaker banks lead to a significant increase in the prevalence of zombie firms in their neighboring sectors. An increase of 4 percentage points in the loan loss provision share of a bank (as was observed on average from peak to trough) leads to a $4 \times .14 = .6$ percentage point increase in a sector's share of resources sunk in zombie firms. Interestingly, the control variables do not appear to play a significant role explaining the next period zombie share of the sectors.

Our evidence suggests that there is significant spillover originating from banks to sectors, as well as from sectors to banks.

\subsection{Robustness}
\label{subsec: robustness}

Here we explore how our results vary as we modify specifications of networks and instruments. To save space, we provide a summary of the results here, and present details of the results in the Supplemental Note.

\subsubsection{Varying Networks}

As a robustness check, we first varied the denseness of the networks, by taking a threshold of 10 percent to generate a denser network, and a threshold of 50 percent for a sparser network. (See Figures \ref{fig:net10} and \ref{fig:net50} for the summary of network statistics and Tables \ref{tab:results_all_2005_2012_asymm_optB_net50_spec12} and \ref{tab:results_all_2005_2012_asymm_optB_net10_spec12} for results in the Supplemental Note.) When we used the denser network, the results remain robust; both directions of the spillover between banks and sectors are positive with statistical significance at 1\%. Interestingly, when we used the sparser network, the spillover from banks to sectors becomes statistically insignificant whereas the spillover from sectors to banks remain significant. This change mostly stems from the links eliminated from the network rather than from a reduction of the set of banks and sectors, which suggests that by using the sparser network, one loses information on the spillover from banks to sectors.

Second, we considered an alternative definition of links from banks to sectors, which is based on debt, rather than assets. More specifically, bank $i$ is linked to sector $j$ if the firms in sector $j$ with which bank $i$ has a relationship account for more than the 25\% percentile of the shares for all firms in sector $j$. We believe that this alternative definition is less reasonable than our previous choice; for example, if there is a large zombie firm (in terms of assets) in sector $j$ and that firm is tied to a weak bank, but at the same time, the debt of that firm is not too high, then it would appear that the zombie congestion in sector $j$ (which will be very large because the firm is very large) is unrelated to the bank's weakness when in fact it is. Nevertheless, we followed our estimation procedure using this specification for a robustness check. The result is reported in Table \ref{tab:results_all_2005_2012_asymm_optB_net25_spec12_ltdebt} in the Supplemental Note. While the spillover from banks to sectors remains positive, the statistical significance is reduced from a p-value of 0.003 to 0.146. This appears to suggest that the spillover from banks to sectors is better captured using asset-based links rather than debt-based links.

\subsubsection{Varying Instruments}

In the baseline specification, we considered the IVs based on Linear Projection with Option B. We also obtained estimates using Options A and B. Recall that Option A uses just-identification, taking $Z_{i,t} = W_{i,t}$ and Option C uses overidentification like the baseline specification but with more IVs based on the past values of $W_{i,t}$. The results are reported in Tables \ref{tab:results_all_2005_2012_asymm_optA_net25_spec12} (with Option A) and \ref{tab:results_all_2005_2012_asymm_optC_net25_spec12} (with Option C). 

When we use Option A, the result shows that while the spillover from banks to sectors remains almost the same, the positive spillover from sectors to banks exhibits reduced statistical significance from a p-value of 0.003 to 0.133. This suggests that the quadratic transformation of $W_{i,t}$ as IVs is strongly relevant for explaining the spillover from sectors to banks. On the other hand, when we used Option C, the statistical significance of positive spillovers between the banks and the sectors is restored at 5\% in both directions.

\subsubsection{Alternative Definition of Bank Relationships}

As mentioned previously, we have estimated the same model but using the sector-bank networks resulting from only using the first reported bank relationship for each firm. Other papers in the literature restrict their attention to the first bank named by each firm based on the assumption that the ranking of the firm's banks represents the strength of the relationship (see \cite{Ferrando/Popov/Udell:18:JMCB} for example). The results are shown in Table \ref{tab:results_first_2005_2012_asymm_optB_net25_spec12}. The spillover effect from banks to sectors remains significant at the 1\% level. In the opposite direction, however, the coefficient is still positive but no longer significant. This could be due to the resulting smaller number of linked sector-bank pairs which leads to a loss of statistical power.

\section{Conclusion}

In many empirical settings, a researcher is interested in the influence of the outcomes of one group of units on those of another group. However, in many such settings, it is also plausible that the influence in the other direction also exists. In such a situation, we would like to have an empirical model where the directions are assigned a separate role in the model. In this paper, we introduce a new approach of empirical modeling through the use of multiple networks in the context of dynamic linear panel models, and develop asymptotic inference on those spillover effects. Our method is quite simple to use, and shown to perform very well in finite samples in Monte Carlo simulations.

Using bank-firm data from Spain, we demonstrate how our methodology can be harnessed to measure the spillover effects between banks and sectors, with the direction of spillover explicitly distinguished in the model. From the analysis, we find that there is positive spillover between banks and sectors in both directions.

While we believe that our empirical study has its own value as a contribution to the literature that studies the relation between financial sectors and real sectors, our methodology is quite generally applicable in various contexts of group-wise spillovers, with both directions of spillover allowed in the same model. Furthermore, our model allows for within group spillover along a network. Such flexibility can be quite useful, for example, in studying peer effects between two groups of students, where each group of students has their own friendship network.

\bibliographystyle{econometrica}
\bibliography{linear_diffusion}

\newpage

\begin{center}
	\Large \textsc{Supplemental Note to ``Estimating Dynamic Spillover Effects along Multiple Networks in a Linear Panel Model''}
	\bigskip
\end{center}

\date{%
	\today%
}

\vspace*{7ex minus 1ex}
\begin{center}
	Clemens Possnig\footnote{Vancouver School of Economics, University of British Columbia}, Andreea Rot\u{a}rescu\footnote{Department of Economics, Wake Forest University}, and Kyungchul Song\footnote{Vancouver School of Economics, University of British Columbia}\\
	
	\bigskip
	\bigskip
\end{center}

\setcounter{section}{0}

The Supplemental Note consists of two sections. The first section provides the mathematical proofs of the results in the main paper. The second section gives details on further results from the empirical study of bank-sector spillover in Spain. 

\section{Mathematical Proofs}

In this section, we provide mathematical proofs of Theorems \ref{thm: asym norm}-\ref{thm: asym FWER control}. Throughout this section, we assume that Assumptions \ref{assump: errors}-\ref{assump: m_{T,Z}} hold. The following representation shows the advantage of the Helmert transform: the error term $\hat U_K$ after the transform can be written as a sum of martingale difference arrays with filtration $\{\mathcal{F}_s\}_{s=1}^T$.

\begin{lemma}
	\label{lemm: repres}
	For $K \in \{B,F\}$,
	\begin{align*}
		\sqrt{n_K} \hat U_K = \sum_{s=1}^T \xi_{s,K}^* + o_P(1), \text{ as } n_K \rightarrow \infty,
	\end{align*}
	where $\xi_{s,K}^*$ is defined in (\ref{xi2}).
\end{lemma}

\noindent \textbf{Proof: } Let
\begin{align}
	\varepsilon_{i,s}^C = \frac{1}{|N_C(i)|} \sum_{i \in N_C(i)} \varepsilon_{i,s}, \text{ and } v_{i}^C = \frac{1}{|N_C(i)|} \sum_{i \in N_C(i)} v_{i}.
\end{align}
We write
\begin{align*}
	\sqrt{n_K} \hat U_K = \sqrt{n_K}\sum_{t=1}^{T-1}  \hat U_{t,K} &= \frac{1}{\sqrt{n_K}}\sum_{t=1}^{T-1}  \sum_{s=t}^{T} h_{s,t}    \sum_{i \in N_K} (Z_{i,t} - Z_{i,t}^C)(u_{i,s} - u_{i,s}^C)\\
	&= \frac{1}{\sqrt{n_K}} \sum_{t=1}^{T-1}  \sum_{s=t}^{T} h_{s,t}  \sum_{i \in N_K} (Z_{i,t} - Z_{i,t}^C) (\varepsilon_{i,s} + v_i - (\varepsilon_{i,s}^C + v_i^C))\\
	&= \frac{1}{\sqrt{n_K}} \sum_{t=1}^{T-1}  \sum_{s=t}^{T} h_{s,t}  \sum_{i \in N_K} (Z_{i,t} - Z_{i,t}^C) (\varepsilon_{i,s} + v_i).
\end{align*}
The last equality follows due to the mean deviation $Z_{i,t} - Z_{i,t}^C$. By the definition of $h_{s,t}$, we have $\sum_{s=t}^T h_{s,t} = 0$, and hence the last sum is equal to
\begin{align*}
	\frac{1}{\sqrt{n_K}} \sum_{i \in N_K} \sum_{t=1}^{T-1}  (Z_{i,t} - Z_{i,t}^C) \sum_{s=t}^{T} h_{s,t}\varepsilon_{i,s}
	= \sum_{s=1}^{T} \frac{1}{\sqrt{n_K}}\sum_{i \in N_K}  \sum_{t=1}^{s \wedge (T-1)}  h_{s,t} (Z_{i,t} - Z_{i,t}^C)\varepsilon_{i,s}.
\end{align*}
Now, by Assumptions \ref{assump: errors}-\ref{assump: IV approx}, we have
\begin{align}
	\label{ineq34}
	\mathbf{E}\left[\left\|\frac{1}{\sqrt{n_K}} \sum_{i \in N_K} (Z_{i,t} - Z_{i,t}^*) \varepsilon_{i,s} \right\|^2 \mid \mathcal{F}_{s} \right] &= \frac{1}{n_K} \sum_{i \in N_K} \left\| Z_{i,t} - Z_{i,t}^* \right\|^2 \mathbf{E}\left[\varepsilon_{i,s}^2 \mid \mathcal{F}_s \right] \\ \notag
	&\le \max_{i \in N_K} \mathbf{E}\left[\varepsilon_{i,s}^2 \mid \mathcal{F}_s \right]  \frac{1}{n_K} \sum_{i \in N_K} \left\| Z_{i,t} - Z_{i,t}^* \right\|^2  = o_P(1),
\end{align}
and
\begin{align*}
	\sqrt{n_K} \hat U_K =  \sum_{s=1}^{T} \frac{1}{\sqrt{n_K}}\sum_{i \in N_K}  \sum_{t=1}^{s \wedge (T-1)}  h_{s,t} (Z_{i,t}^* - Z_{i,t}^{*C})\varepsilon_{i,s} +o_P(1).
\end{align*}
$\blacksquare$\medskip

Recall the definition: for $K \in \{B,F\}$,
\begin{align*}
	N_{t-1,\ell,K}(i) = \{j \in N_K: j \in N_{t-1,\ell}(i)\}.
\end{align*}
For each $t=1,...,T$ and $i \in N$, define
\begin{align*}
	N_{t-1}(i) = \left\{j \in N: j \in N_{t-1,\ell,K}(i), \text{ for some } \ell = 1,...,L, \text{ and some } K \in \{B,F\}\right\},
\end{align*}
and $\overline N_{t-1}(i) = N_{t-1}(i) \cup \{i\}$. Thus $N_{t-1}(i) $ represents the union of the in-neighborhoods $N_{t-1,\ell,K}(i)$, $\ell=1,...,L$ and $K \in \{B,F\}$.
\begin{lemma}
	\label{lemm: LLN}
	For $K \in \{B,F\}$, as $n_K \rightarrow \infty$,
	\begin{align*}
		\frac{1}{n_K}\sum_{i \in N_K} \left( \tilde Z_{i,s} W_{i,s}^\top - \mathbf{E}\left[\tilde Z_{i,s}^* W_{i,s}^\top \mid \mathcal{F}_{s-1}\right]\right) \rightarrow_P 0.
	\end{align*}
\end{lemma}

\noindent \textbf{Proof: } By Assumptions \ref{assump: IV approx} and \ref{assump: m_{T,Z}}, it is not hard to see that
\begin{align*}
	\frac{1}{n_K}\sum_{i \in N_K} \left( \tilde Z_{i,s} - \tilde Z_{i,s}^*\right) W_{i,s}^\top = o_P(1).
\end{align*}
For $k=1,...,d_Z$ and $m=1,...,d_W$, we let $\tilde Z_{i,s,k}^*$ be the $k$-th entry of $\tilde Z_{i,s}^*$ and $W_{i,s,m}$ be the $m$-th entry of $W_{i,s}$. For any such $k,m$, once we condition on $\mathcal{F}_{s-1}$, the randomness of $\tilde Z_{i,s,k}^* W_{i,s,m}$ solely comes from $(\varepsilon_{j,s-1})_{j \in \overline N_{s-1}(i)}$. Furthermore, $\overline N_{s-1}(i)$ is already $\mathcal{F}_{s-1}$-measurable. By Assumption \ref{assump: errors}(i), we have
\begin{align*}
	\text{Cov}\left( \tilde Z_{i,s,k}^*W_{i,s,m}, \tilde Z_{j,s,k}^* W_{j,s,m} \mid \mathcal{F}_{s-1}\right) = 0,
\end{align*}
whenever $\overline N_{s-1}(i) \cap \overline N_{s-1}(j) = \varnothing.$ Therefore,
\begin{align*}
	\text{Var}\left( \frac{1}{n_K}\sum_{i \in N_K} \tilde Z_{i,s,k}^* W_{i,s,m}  \mid \mathcal{F}_{s-1}\right)
	=\frac{1}{n_K^2} \sum_{i,j \in N_K: \overline N_{s-1}(i) \cap \overline N_{s-1}(j) \ne \varnothing} \text{Cov}\left( \tilde Z_{i,s,k}^* W_{i,s,m}, \tilde Z_{j,s,k}^* W_{j,s,m} \mid \mathcal{F}_{s-1}\right).
\end{align*}
By Assumptions \ref{assump: networks} and \ref{assump: m_{T,Z}}, we find that the last term is $o_P(1)$. $\blacksquare$

\begin{lemma}
	\label{lemm: B}
	For each $K \in \{B,F\}$, $\hat B_K - B_{n,K} \rightarrow_P 0$, as $n_K \rightarrow \infty$, where $\hat B_K$ and $B_{n,K}$ are defined in (\ref{AK BK}) and (\ref{B O}).
\end{lemma}

\noindent \textbf{Proof: } First we write
\begin{align*}
	\sum_{t=1}^{T-1} \frac{1}{n_K}\sum_{i \in N_K} \left(Z_{i,t} - Z_{i,t}^C \right) W_{i,t}^{H \top}
	&= \sum_{s=1}^{T} \frac{1}{n_K}\sum_{i \in N_K} \tilde Z_{i,s} \left(W_{i,s} - W_{i,s}^C\right)^\top = \sum_{s=1}^{T} \frac{1}{n_K}\sum_{i \in N_K} \tilde Z_{i,s} W_{i,s}^\top.
\end{align*}
Then we write that for each $s=1,...,T$,
\begin{align*}
	\frac{1}{n_K}\sum_{i \in N_K} \tilde Z_{i,s}W_{i,s}^\top = \frac{1}{n_K}\sum_{i \in N_K} \mathbf{E}\left[ \tilde Z_{i,s}^*W_{i,s}^\top \mid \mathcal{F}_{s-1} \right] + o_P(1),
\end{align*}
by Lemma \ref{lemm: LLN}. This gives the desired result. $\blacksquare$\medskip

\begin{lemma}
	\label{lemm: tilde delta}
	For $K \in \{B,F\}$, as $n_K \rightarrow \infty$, $\tilde \delta_K - \delta_K = o_P(1)$.
\end{lemma}
\noindent \textbf{Proof: } We first show that $\hat U_K = o_P(1)$. By Lemma \ref{lemm: repres},
\begin{align*}
	\sqrt{n_K} \hat U_K = \frac{1}{\sqrt{n_K}} \sum_{s=1}^T \xi_{s,K}^* +o_P(1).
\end{align*}
Note that
\begin{align*}
	\sum_{s=1}^T \text{Var}(\xi_{s,K}^* \mid \mathcal{F}_s) &= \sum_{s=1}^T \frac{1}{n_K}\sum_{i \in N_K} \sigma_{n,s,K}^2 \tilde Z_{i,s}^* \tilde Z_{i,s}^{* \top}\\
	&= \sum_{s=1}^T \frac{1}{n_K}\sum_{i \in N_K} \mathbf{E}\left[ \sigma_{n,s,K}^2 \tilde Z_{i,s}^* \tilde Z_{i,s}^{* \top} \mid \mathcal{F}_{s-1} \right] +o_P(1) = \Omega_{K} + o_P(1),
\end{align*}
by following the same arguments as in the proof of Lemma \ref{lemm: LLN} and Assumptions \ref{assump: errors} and \ref{assump: limit var}. Since $\mathbf{E}\left[ \xi_{s,K}^*\right] = 0$, by Assumption \ref{assump: m_{T,Z}}, we have
\begin{align*}
	\mathbf{E}\left[\sum_{s=1}^T \text{Var}(\xi_{s,K}^* \mid \mathcal{F}_s) \right] = O(1),
\end{align*}
as $n_K \rightarrow \infty$. Hence, for $K \in \{B,F\}$, $\xi_{s,K}^* = O_P(1)$, as $n_K \rightarrow \infty$. We conclude that
\begin{align*}
	\hat U_K = O_P(n_K^{-1/2}) = o_P(1).
\end{align*}
Hence
\begin{align}
	\tilde \delta_K - \delta_K &= \left( \hat B_K^\top \hat B_K\right)^{-1}\hat B_K^\top \hat U_K \\ \notag
	&=\left( B_{K}^\top B_{K}\right)^{-1} B_{K}^\top \hat U_K + o_P(1) = o_P(1),
\end{align}
by Lemma \ref{lemm: B} and Assumption \ref{assump: limit var}. $\blacksquare$\medskip

Define
\begin{align*}
	\Omega_{n,K} = \sum_{s=1}^T \Omega_{n,s,K},
\end{align*}
and $\Omega_{n,s,K}$ is defined in (\ref{B O}).

\begin{lemma}
	\label{lemm: Omega}
	For $K \in \{B,F\}$, $\hat \Omega_K - \Omega_{n,K} \rightarrow_P 0$, as $n_K \rightarrow \infty$.
\end{lemma}

\noindent \textbf{Proof: } Since $\hat u_{i,t,K}^H - u_{i,t}^H = W_{i,t}^{H \top}(\delta_K - \tilde{\delta}_K)$, we write
\begin{align*}
	\hat \Omega_K &= \sum_{t=1}^{T-1} \sum_{t'=1}^{T-1} \frac{1}{n_K}\sum_{i \in N_K} \left(Z_{i,t} - Z_{i,t}^C \right)\left(Z_{i,t'} - Z_{i,t'}^C \right)^\top \hat u_{i,t,K}^H \hat u_{i,t',K}^H\\
	&= \sum_{t=1}^{T-1} \sum_{t'=1}^{T-1} \frac{1}{n_K}\sum_{i \in N_K} \left(Z_{i,t} - Z_{i,t}^C \right)\left(Z_{i,t'} - Z_{i,t'}^C \right)^\top u_{i,t}^H u_{i,t'}^H + R_{n},
\end{align*}
where
\begin{align*}
	R_{n} &= \sum_{t=1}^{T-1} \sum_{t'=1}^{T-1} \sum_{j=1}^{d_W} \sum_{j'=1}^{d_W} (\delta_{K,j}- \tilde \delta_{K,j}) \left(\frac{1}{n_K}\sum_{i \in N_K} W_{i,t,j}^H \left(Z_{i,t} - Z_{i,t}^C \right)\left(Z_{i,t'} - Z_{i,t'}^C \right)^\top W_{i,t',j'}^H \right) (\delta_{K,j'}- \tilde \delta_{K,j'})\\
	&\quad + 2 \sum_{t=1}^{T-1} \sum_{t'=1}^{T-1} \sum_{j'=1}^{d_W} u_{i,t}^H \left(\frac{1}{n_K}\sum_{i \in N_K} \left(Z_{i,t} - Z_{i,t}^C \right) \left(Z_{i,t'} - Z_{i,t'}^C \right)^\top W_{i,t',j'}^H \right) (\delta_{K,j'}- \tilde \delta_{K,j'}),
\end{align*}
with $W_{i,t,j}^H$ denoting the $j$-th entry of $W_{i,t}^H$ and similarly with $\delta_{K,j}$ and $\tilde \delta_{K,j}$. Note that
\begin{align*}
	&\sum_{t=1}^{T-1} \sum_{t'=1}^{T-1} \frac{1}{n_K}\sum_{i \in N_K} W_{i,t,j}^H \left(Z_{i,t} - Z_{i,t}^C \right)\left(Z_{i,t'} - Z_{i,t'}^C \right)^\top W_{i,t',j'}^H\\
	&=\sum_{s=1}^T \sum_{s'=1}^T \frac{1}{n_K}\sum_{i \in N_K} W_{i,s,j} \tilde Z_{i,s} \tilde Z_{i,s'}^\top W_{i,s',j'}\\
	&=\frac{1}{n_K}\sum_{i \in N_K}\left(\sum_{s=1}^T W_{i,s,j} \tilde Z_{i,s} \right)\left(\sum_{s=1}^T W_{i,s,j'} \tilde Z_{i,s} \right)^\top.
\end{align*}
By Cauchy-Schwarz inequality, we bound the $(m,m')$-th entry of the last term by
\begin{align*}
	\sqrt{\frac{1}{n_K}\sum_{i \in N_K}\left(\sum_{s=1}^T W_{i,s,j} \tilde Z_{i,s,m} \right)^2}
	\times \sqrt{\frac{1}{n_K}\sum_{i \in N_K}\left(\sum_{s=1}^T W_{i,s,j} \tilde Z_{i,s,m'} \right)^2},
\end{align*}
where $\tilde Z_{i,s,m}$ denotes the $m$-th entry of $\tilde Z_{i,s}$. We write
\begin{align*}
	\frac{1}{n_K}\sum_{i \in N_K}\left(\sum_{s=1}^T W_{i,s,j} \tilde Z_{i,s,m} \right)^2 &= \sum_{s=1}^T \sum_{s'=1}^T \frac{1}{n_K}\sum_{i \in N_K} W_{i,s,j}W_{i,s',j} \tilde Z_{i,s,m}\tilde Z_{i,s',m}\\
	&\le \sum_{s=1}^T \sum_{s'=1}^T  \sqrt{\frac{1}{n_K}\sum_{i \in N_K} \tilde Z_{i,s,m}^2W_{i,s,j}^2}\sqrt{\frac{1}{n_K}\sum_{i \in N_K} \tilde Z_{i,s',m}^2W_{i,s',j}^2}.
\end{align*}
Using the same arguments as in the proof of Lemma \ref{lemm: LLN},
\begin{align*}
	\frac{1}{n_K}\sum_{i \in N_K} \tilde Z_{i,s,m}^2W_{i,s,j}^2&= \frac{1}{n_K}\sum_{i \in N_K} \mathbf{E}\left[\tilde Z_{i,s,m}^{*2}W_{i,s,j}^2 \mid \mathcal{F}_{s-1} \right] + o_P(1)\\
	&\le \sqrt{\frac{1}{n_K}\sum_{i \in N_K} \mathbf{E}\left[\|W_{i,s}\|^4 \mid \mathcal{F}_{s-1} \right]}
	\times \sqrt{\frac{1}{n_K}\sum_{i \in N_K} \mathbf{E}\left[\| \tilde Z_{i,s}^* \|^4 \mid \mathcal{F}_{s-1} \right]} + o_P(1).
\end{align*}
By Assumptions \ref{assump: IV approx} and \ref{assump: m_{T,Z}}, we find that
\begin{align*}
	\frac{1}{n_K}\sum_{i \in N_K} W_{i,t,j}^H \left(Z_{i,t} - Z_{i,t}^C \right)\left(Z_{i,t'} - Z_{i,t'}^C \right)^\top W_{i,t',j'}^H = O_P(1).
\end{align*}
Since $\tilde \delta_K = \delta_K + o_P(1)$ by Lemma \ref{lemm: tilde delta}, the leading term in the definition of $R_n$ is $o_P(1)$. We can deal with the second term similarly to show that it is $o_P(1).$ Hence, we have
\begin{align*}
	&\frac{1}{n_K}\sum_{i \in N_K} \left(Z_{i,t} - Z_{i,t}^C \right)\left(Z_{i,s} - Z_{i,s}^C \right)^\top \hat u_{i,t,K}^H \hat u_{i,s,K}^H\\
	&= \frac{1}{n_K}\sum_{i \in N_K} \left(Z_{i,t} - Z_{i,t}^C \right)\left(Z_{i,s} - Z_{i,s}^C \right)^\top u_{i,t}^H u_{i,s}^H + o_P(1).
\end{align*}
This gives us
\begin{align*}
	\hat \Omega_K &= \frac{1}{n_K}\sum_{i \in N_K} \left( \sum_{t=1}^{T-1} \left(Z_{i,t} - Z_{i,t}^C \right) u_{i,t}^H \right)\left( \sum_{t=1}^{T-1} \left(Z_{i,t} - Z_{i,t}^C \right) u_{i,t}^H \right)^\top + o_P(1)\\
	&= \frac{1}{n_K}\sum_{i \in N_K} \left( \sum_{s=1}^{T} \tilde Z_{i,s} \varepsilon_{i,s} \right) \left( \sum_{s=1}^{T} \tilde Z_{i,s} \varepsilon_{i,s} \right)^\top + o_P(1)\\
	&= \frac{1}{n_K}\sum_{i \in N_K} \left( \sum_{s=1}^{T} \tilde Z_{i,s}^* \varepsilon_{i,s} \right) \left( \sum_{s=1}^{T} \tilde Z_{i,s}^* \varepsilon_{i,s} \right)^\top + o_P(1),
\end{align*}
where the first equality is due to our derivation above, and the third equality due to (\ref{ineq34}).

For $s \ne s'$, we have for all $j,j'=1,...,d_Z$,
\begin{align*}
	\mathbf{E}\left[ \left(\tilde Z_{i,s,j}^* \tilde Z_{i,s',j'}^* \right) \varepsilon_{i,s} \varepsilon_{i,s'} \mid \mathcal{F}_0\right] = 0,
\end{align*}
by Assumption \ref{assump: errors}(i). Hence for $s \ne s'$,
\begin{align*}
	&\text{Var}\left( \frac{1}{n_K}\sum_{i \in N_K} \tilde Z_{i,s,j}^* \tilde Z_{i,s',j'}^* \varepsilon_{i,s} \varepsilon_{i,s'} \mid \mathcal{F}_0\right)\\
	&= \frac{1}{n_K^2} \sum_{i \in N_K} \mathbf{E}\left[ \left(\tilde Z_{i,s,j}^* \tilde Z_{i,s,j'}^*\right)^2 \varepsilon_{i,s}^2 \varepsilon_{i,s'}^2 \mid \mathcal{F}_0\right]\\
	&\le \frac{1}{n_K}\sqrt{\frac{1}{n_K}\sum_{i \in N_K} \mathbf{E}\left[ \left(\tilde Z_{i,s,j}^* \varepsilon_{i,s} \right)^4 \mid \mathcal{F}_0 \right] } \sqrt{\frac{1}{n_K}\sum_{i \in N_K} \mathbf{E}\left[ \left(\tilde Z_{i,s,j'}^* \varepsilon_{i,s} \right)^4 \mid \mathcal{F}_0 \right] }.
\end{align*}
Since
\begin{align*}
	\frac{1}{n_K}\sum_{i \in N_K} \mathbf{E}\left[ \tilde Z_{i,s,j}^{*4} \varepsilon_{i,s}^4 \mid \mathcal{F}_s \right] \le \frac{1}{n_K} \sum_{i \in N_K} \tilde Z_{i,s,j}^{*4} \max_{i \in N} \mathbf{E}\left[ \varepsilon_{i,s}^4 \mid \mathcal{F}_s \right] =O_P(1),
\end{align*}
we find that whenever $s \ne s'$, for all $j,j'=1,...,d_Z$,
\begin{align*}
	\text{Var}\left( \frac{1}{n_K}\sum_{i \in N_K} \tilde Z_{i,s,j}^* \tilde Z_{i,s',j'}^* \varepsilon_{i,s} \varepsilon_{i,s'} \mid \mathcal{F}_0\right) = O_P\left(n_K^{-1}\right).
\end{align*}
Hence 
\begin{align*}
	\hat \Omega_K &= \frac{1}{n_K}\sum_{i \in N_K} \sum_{s=1}^{T} \tilde Z_{i,s}^* \tilde Z_{i,s}^{*\top} \varepsilon_{i,s}^2 + o_P(1)\\
	&= \frac{1}{n_K}\sum_{i \in N_K} \sum_{s=1}^{T} \tilde Z_{i,s}^* \tilde Z_{i,s}^{*\top} \sigma_{n,s,K}^2 + o_P(1)\\
	&= \frac{1}{n_K}\sum_{i \in N_K} \sum_{s=1}^{T} \mathbf{E}\left[\sigma_{n,s,K}^2 \tilde Z_{i,s}^* \tilde Z_{i,s}^{*\top} \mid \mathcal{F}_{s-1}\right] + o_P(1) = \Omega_{n,K} +o_P(1).
\end{align*}
$\blacksquare$\medskip

For each $s=1,...,T$, recall the definition of $\xi_{s,K}^*, K \in \{B,F\}$ in (\ref{xi2}). Let $\xi_{s}^* = [\xi_{s,B}^{*\top}, \xi_{s,F}^{*\top}]^\top$, and
\begin{align}
	\label{Omega circ}
	\tilde \Omega_{n,s} = \text{Var}\left( \xi_s^* \mid \mathcal{F}_s\right).
\end{align}
We can write
\begin{align*}
	\xi_s^* = \frac{1}{\sqrt{n}}\sum_{i \in N} \Lambda_{i,s} \varepsilon_{i,s},
\end{align*}
where $\Lambda_{i,s} = [ \tilde Z_{i,s}^{*\top} 1_{i,B}, \tilde Z_{i,s}^{*\top} 1_{i,F} ]^\top$, and 
\begin{align*}
	1_{i,B} = \frac{1\{i \in N_B\} \sqrt{n}}{\sqrt{n_B}}, \text{ and }  1_{i,F} = \frac{1\{i \in N_F\} \sqrt{n}}{\sqrt{n_F}}.
\end{align*}
After some algebra, it is not hard to see that for each $s=1,...,T$,
\begin{align}
	\label{Omega circ2}
	\tilde \Omega_{n,s} = \left[ \begin{array}{cc}
		\displaystyle \frac{1}{n_B}\sum_{i \in N_B} \sigma_{n,s,B}^2 \tilde Z_{i,s}^* \tilde Z_{i,s}^{*\top}  & 0\\
		0 & \displaystyle \frac{1}{n_F}\sum_{i \in N_F} \sigma_{n,s,F}^2 \tilde Z_{i,s}^* \tilde Z_{i,s}^{*\top} 
	\end{array}
	\right].
\end{align}
For $K \in \{B,F\}$, recall the definition of $\Omega_{n,s,K}$ in (\ref{B O}) and let $\Omega_{s,K}$ be the matrix in Assumption \ref{assump: limit var}. For each $s=1,...,T$, let
\begin{align}
	\label{Omega 3}
	\Omega = \left[ \begin{array}{cc}
		\Omega_B& 0 \\ 0& \Omega_F
	\end{array}
	\right], \quad \Omega_s = \left[ \begin{array}{cc}
		\Omega_{s,B}& 0 \\ 0& \Omega_{s,F}
	\end{array}
	\right] \text{ and }
	\Omega_{n,s} = \left[ \begin{array}{cc}
		\Omega_{n,s,B} & 0 \\ 0& \Omega_{n,s,F}
	\end{array}
	\right],
\end{align}
where for $K \in \{B,F\}$,
\begin{align*}
	\Omega_K = \sum_{s=1}^T \Omega_{s,K}.
\end{align*}
For each $b \in \mathbf{R}^{2d_Z}$ with $b^\top b = 1$, we define
\begin{align*}
	q_{n,s}(b) = \frac{\displaystyle \frac{1}{n}\sum_{i \in N} |b^\top \Lambda_{i,s}|^3 \mathbf{E}\left[|\varepsilon_{i,s}|^3 \mid \mathcal{F}_s\right]}{\displaystyle \left(b^\top \left(\frac{1}{n} \sum_{K \in \{B,F\}} \sum_{i \in N_K} \Lambda_{i,s} \Lambda_{i,s} ^\top \sigma_{n,s,K}^2 \right) b\right)^{3/2}}.
\end{align*}
\begin{lemma}
	\label{lemm: q_n}
	For each $s=1,...,T$, and each $b \in \mathbf{R}^{2d_Z}$ with $b^\top b = 1$,
	\begin{align*}
		q_{n,s}(b) = O_P(1),
	\end{align*}
   as $n_B, n_F \rightarrow \infty$.
\end{lemma}

\noindent \textbf{Proof: } By the same arguments in the proof of Lemma \ref{lemm: LLN}, 
\begin{align*}
	\frac{1}{n}\sum_{i \in N} |b^\top  \Lambda_{i,s}|^3\mathbf{E}\left[|\varepsilon_{i,s}|^3 \mid \mathcal{F}_s\right]
	&\le \max_{i \in N} \mathbf{E}\left[|\varepsilon_{i,s}|^3 \mid \mathcal{F}_s\right]  \frac{1}{n}\sum_{i \in N} |b^\top  \Lambda_{i,s}|^3\\\
	&= \max_{i \in N} \mathbf{E}\left[|\varepsilon_{i,s}|^3 \mid \mathcal{F}_s\right] \frac{1}{n}\sum_{i \in N} \mathbf{E}\left[ |b^\top  \Lambda_{i,s}|^3 \mid \mathcal{F}_{s-1} \right] + o_P(1).
\end{align*}
Thus, by Assumptions \ref{assump: errors} and \ref{assump: m_{T,Z}}, we have
\begin{align*}
	\max_{i \in N} \mathbf{E}\left[|\varepsilon_{i,s}|^3 \mid \mathcal{F}_s\right] \frac{1}{n}\sum_{i \in N} \mathbf{E}\left[ |b^\top  \Lambda_{i,s}|^3 \mid \mathcal{F}_{s-1} \right] = O_P(1),
\end{align*}
as $n_B,n_F \rightarrow \infty$. 

There exist $n_0 \ge 1$ and $c>0$ such that for $K \in \{B,F\}$,  for all $n_K \ge n_0$, $\sigma_{n,s,K}^2 > c$ and
\begin{align*}
	\lambda_{\min}\left(\frac{1}{n} \sum_{i \in N}\mathbf{E}\left[ \Lambda_{i,s} \Lambda_{i,s}^\top \mid \mathcal{F}_{s-1} \right] \right) \ge c,
\end{align*}
by Assumption \ref{assump: limit var}(i). Hence we obtain the desired result. $\blacksquare$

\begin{lemma}
	\label{lemm: approx}
	For any vector $b \in \mathbf{R}^{2d_Z}$ such that $b^\top b = 1$, and for each $s=1,...,T$,
	\begin{align*}
		\sup_{\tilde c \in \mathbf{R}} \left| P\left\{ b^\top \xi_{s}^* \le \tilde c \mid \mathcal{F}_s \right\} - P\left\{ b^\top \tilde \xi_{s}^\infty \le \tilde c \mid \mathcal{F}_s \right\} \right| = o_P(1),
	\end{align*}
	as $n_B,n_F \rightarrow \infty$, where $\tilde \xi_{s}^\infty = \Omega_s^{1/2} \mathbb{Z}_s$, $\mathbb{Z}_s \in \mathbf{R}^{2d_Z}$, are i.i.d. standard normal random vectors independent of all other random variables and $\Omega_s$ is defined in (\ref{Omega 3}).
\end{lemma}

\noindent \textbf{Proof: } Define $\tilde \xi_s = \tilde \Omega_{n,s}^{1/2} \mathbb{Z}_s$, where $\tilde \Omega_{n,s}$ is as defined in (\ref{Omega 3}). Since $\varepsilon_{i,s}$, $i=1,...,n$, are conditionally independent given $\mathcal{F}_s$ by Assumption \ref{assump: errors}(i), we use the Berry-Esseen bound (Theorem 3 of \cite{Chow/Teicher:88:ProbTheory}, p.304) to deduce that\footnote{The theorem itself is concerned with the sum of independent random variables. However, with appropriate modifications, the same bound with replacing the moments by the conditional moments given the common shock applies to a sum of conditionally independent random variables given the common shocks.}
\begin{align*}
	\sup_{\tilde c \in \mathbf{R}} \left| P\left\{ b^\top \xi_{s}^* \le \tilde c \mid \mathcal{F}_s \right\} - P\left\{ b^\top \tilde \xi_{s} \le \tilde c \mid \mathcal{F}_s \right\} \right| \le \frac{7.5 q_{n,s}(b)}{\sqrt{n}}.
\end{align*}
The last bound is $o_P(1)$ by Lemma \ref{lemm: q_n}. Furthermore, by Assumption \ref{assump: limit var} and as we saw in the proof of Lemma \ref{lemm: Omega},
\begin{align*}
	\tilde \Omega_{n,s} = \Omega_s + o_P(1),
\end{align*}
as $n_B, n_F \rightarrow \infty$. Hence
\begin{align*}
	& \sup_{\tilde c \in \mathbf{R}} \left| P\left\{ b^\top \tilde \xi_{s}^\infty \le \tilde c \mid \mathcal{F}_s \right\} - P\left\{ b^\top \tilde \xi_{s} \le \tilde c \mid \mathcal{F}_s \right\} \right| \\
	&= \sup_{\tilde c \in \mathbf{R}} \left| P\left\{ b^\top \tilde \xi_{s}^\infty \le \tilde c \mid \mathcal{F}_s \right\} - P\left\{ b^\top \tilde \xi_{s}^\infty \le \tilde c + b^\top (\Omega_{s}^{1/2} - \tilde \Omega_{n,s}^{1/2}) \mathbb{Z}_s \mid \mathcal{F}_s \right\} \right|\\
	&= \sup_{\tilde c \in \mathbf{R}} \left| P\left\{ b^\top \tilde \xi_{s}^\infty \le \tilde c \mid \mathcal{F}_s \right\} - P\left\{ b^\top \tilde \xi_{s}^\infty \le \tilde c + o_P(1) \mid \mathcal{F}_s \right\} \right| = o_P(1),
\end{align*}
as $n_B, n_F \rightarrow \infty$, because $b^\top \tilde \xi_{s}^\infty$ is a random variable whose conditional distribution given $\mathcal{F}_s$ is equal to that given $\mathcal{F}_0$, and its conditional distribution given $\mathcal{F}_0$ is absolutely continuous with respect to the Lebesgue measure due to $\Omega_s$ being positive definite by Assumption \ref{assump: limit var}. Hence we obtain the desired result. $\blacksquare$

\begin{lemma}
	\label{lemm: norm approx}
	For any vector $b \in \mathbf{R}^{2 d_Z}$ such that $b^\top b = 1$ and for $\tilde c \in \mathbf{R}$,
	\begin{align*}
		P\left\{ \sum_{s=1}^T b^\top \xi_s^* \le \tilde c \mid \mathcal{F}_0 \right\} - P\left\{  \sum_{s=1}^T b^\top \tilde \xi_s^\infty \le \tilde c \mid \mathcal{F}_0 \right\} 
		=  \sum_{s=1}^T \mathbf{E}\left[  \Delta_{s-1} \left(\tilde c - b^\top R_{s-1}\right) \mid \mathcal{F}_0\right],
	\end{align*}
	where
	\begin{align*}
		\Delta_{s-1}\left( \tilde c - b^\top R_{s-1} \right) &= P\left\{ b^\top \xi_s^* \le \tilde c - \sum_{t=s+1}^T b^\top \tilde \xi_t^\infty -  b^\top R_{s-1} \mid \mathcal{F}_s \right\}\\
		&\quad - P\left\{ b^\top \tilde \xi_{s}^\infty \le \tilde c - \sum_{t=s+1}^T b^\top \tilde \xi_t^\infty -  b^\top R_{s-1} \mid \mathcal{F}_s \right\}
	\end{align*}
	and $R_s = \sum_{t=1}^s \xi_t^*$, $R_0 = 0$, and $\tilde \xi_s^\infty$, with $s=1,...,T$, are defined in Lemma \ref{lemm: approx}.
\end{lemma}

\noindent \textbf{Proof: } First, we write
\begin{align*}
	P\left\{ \sum_{s=1}^T b^\top \xi_s^* \le \tilde c \mid \mathcal{F}_T \right\} &= P\left\{ b^\top \xi_T^* \le \tilde c - b^\top R_{T-1} \mid \mathcal{F}_T \right\}\\
	&= P\left\{ b^\top \tilde \xi_T^\infty \le \tilde c - b^\top R_{T-1} \mid \mathcal{F}_T \right\} + \Delta_{T-1}(\tilde c - b^\top R_{T-1}),
\end{align*}
where
\begin{align*}
	\Delta_{T-1}(\tilde c - b^\top R_{T-1}) &= P\left\{ b^\top \xi_T^* \le \tilde c - b^\top R_{T-1} \mid \mathcal{F}_T \right\}\\
	&\quad - P\left\{ b^\top \tilde \xi_T^\infty \le \tilde c - b^\top R_{T-1} \mid \mathcal{F}_T \right\}.
\end{align*}
(Note that $R_{T-1}$ is $\mathcal{F}_{T}$-measurable.) We write
\begin{align*}
	P\left\{ b^\top \tilde \xi_T^\infty \le \tilde c - b^\top R_{T-1} \mid \mathcal{F}_{T-1} \right\}
	&=P\left\{ b^\top \xi_{T-1}^* \le \tilde c - b^\top \tilde \xi_T^\infty - b^\top R_{T-2} \mid \mathcal{F}_{T-1} \right\} \\
	&=P\left\{ b^\top \tilde \xi_{T-1}^\infty + b^\top \tilde \xi_T^\infty \le \tilde c - b^\top R_{T-2} \mid \mathcal{F}_{T-1} \right\} \\
	&\quad \quad + \Delta_{T-2}(\tilde c - b^\top R_{T-2}).
\end{align*}
We continue this procedure until we have
\begin{align*}
	&P\left\{ b^\top \tilde \xi_2 \le \tilde c -b^\top \tilde \xi_T^\infty -  b^\top \tilde \xi_{T-1}^\infty \cdots- b^\top \tilde \xi_3^\infty - b^\top R_{1} \mid \mathcal{F}_1 \right\}\\
	&=P\left\{ b^\top \xi_1^* \le \tilde c -b^\top \tilde \xi_T^\infty -  b^\top \tilde \xi_{T-1}^\infty \cdots- b^\top \tilde \xi_2^\infty \mid \mathcal{F}_1 \right\}\\
	&=P\left\{ b^\top \tilde \xi_T^\infty +  b^\top \tilde \xi_{T-1}^\infty \cdots+ b^\top \tilde \xi_2 + b^\top \tilde \xi_1^\infty \le \tilde c  \mid \mathcal{F}_1 \right\} + \Delta_0(\tilde c - b^\top R_0),
\end{align*}
where $R_0 = 0$, and
\begin{align*}
	\Delta_0(\tilde c - b^\top R_0) &= P\left\{ b^\top \xi_1^* \le \tilde c -b^\top \tilde \xi_T^\infty -  b^\top \tilde \xi_{T-1}^\infty \cdots- b^\top \tilde \xi_2^\infty \mid \mathcal{F}_1 \right\}\\
	&\quad - P\left\{ b^\top \tilde \xi_1^\infty \le \tilde c -b^\top \tilde \xi_T^\infty -  b^\top \tilde \xi_{T-1}^\infty \cdots- b^\top \tilde \xi_2^\infty \mid \mathcal{F}_1 \right\}.
\end{align*}
By taking conditional expectations given $\mathcal{F}_0$ of all the conditional probabilities above, we obtain the desired result. $\blacksquare$

\begin{lemma}
	\label{lemm: normal approx}
	As $n_B,n_F \rightarrow \infty$, for any vector $b \in \mathbf{R}^{2d_Z}$ such that $b^\top b = 1$,
	\begin{align*}
		\sup_{\tilde c \in \mathbf{R}}	\left|P\left\{ b^\top \sum_{s=1}^T \xi_s^* \le \tilde c \mid \mathcal{F}_0\right\} - P\left\{ b^\top  \Omega^{1/2} \mathbb{Z} \le \tilde c \mid \mathcal{F}_0 \right\} \right| \rightarrow_P 0,
	\end{align*}
	where $\mathbb{Z} \in \mathbf{R}^{2d_Z}$ is a standard normal random vector independent of other random variables, and $\Omega$ is defined in (\ref{Omega 3}).
\end{lemma}

\noindent \textbf{Proof: } By Lemma \ref{lemm: norm approx},
\begin{align*}
	\sup_{\tilde c \in \mathbf{R}}  \left| P\left\{ \sum_{s=1}^T b^\top \xi_s^* \le \tilde c \mid \mathcal{F}_0 \right\} - P\left\{  \sum_{s=1}^T b^\top \tilde \xi_s^\infty \le \tilde c \mid \mathcal{F}_0 \right\} \right| 
	&\le \sum_{s=1}^T \mathbf{E}\left[\sup_{\tilde c \in \mathbf{R}} \Delta_{s-1}(\tilde c) \mid \mathcal{F}_0 \right],
\end{align*}
because $R_{s-1}$ is $\mathcal{F}_s$-measurable. Note that
\begin{align*}
	\sup_{\tilde c \in \mathbf{R}} \Delta_{s-1}(\tilde c) \le \sup_{\tilde c \in \mathbf{R}} \left| P\left\{ b^\top \xi_{s}^* \le \tilde c \mid \mathcal{F}_s \right\} - P\left\{ b^\top \tilde \xi_{s}^\infty \le \tilde c \mid \mathcal{F}_s \right\} \right|,
\end{align*}
because $\mathbb{Z}_t$'s that constitute $\tilde \xi_{t}^\infty$'s are independent of $\mathcal{F}_s$, and $\Omega_t$'s are all $\mathcal{F}_0$-measurable by Assumption \ref{assump: limit var}. The last supremum is $o_P(1)$ by Lemma \ref{lemm: approx}. Since $\sup_{\tilde c \in \mathbf{R}} \Delta_{s-1}(\tilde c)$ is bounded by 1, it is uniformly integrable. Hence we find that
\begin{align*}
	\mathbf{E}\left[\sup_{\tilde c \in \mathbf{R}} \Delta_{s-1}(\tilde c) \mid \mathcal{F}_0 \right] = o_P(1),
\end{align*}
for each $s=1,...,T$. Thus, we obtain the desired result. $\blacksquare$\medskip

\noindent \textbf{Proof of Theorem \ref{thm: asym norm}: } We write
\begin{align}
	\label{eq2}
	\sum_{s=1}^T \left[ \begin{array}{c}
		\displaystyle \frac{1}{\sqrt{n}}\sum_{i \in N} \tilde Z_{i,s}^* 1_{i,B} \varepsilon_{i,s} \\ \displaystyle \frac{1}{\sqrt{n}}\sum_{i \in N} \tilde Z_{i,s}^* 1_{i,F} \varepsilon_{i,s}
	\end{array}
	\right]
	=\sum_{s=1}^T \left[ \begin{array}{c}
		\displaystyle \frac{1}{\sqrt{n_B}}\sum_{i \in N_B} \tilde Z_{i,s}^*  \varepsilon_{i,s} \\ \displaystyle \frac{1}{\sqrt{n_F}}\sum_{i \in N_F} \tilde Z_{i,s}^*  \varepsilon_{i,s}
	\end{array}
	\right] =  \left[ \begin{array}{c}
		\displaystyle \sqrt{n_B} \hat U_B \\ \displaystyle \sqrt{n_F} \hat U_F,
	\end{array}
	\right] + o_P(1),
\end{align}
by Lemma \ref{lemm: repres}. Let $\hat U = [\sqrt{n_B}\hat U_B^\top/\sqrt{n}, \sqrt{n_F}\hat U_F^\top/\sqrt{n}]^\top$. Hence
\begin{align}
	\label{eq43}
	\left[ \begin{array}{c}
		\sqrt{n_B} \hat V_B^{-1/2} (\hat \delta_B - \delta_{B})\\ \\
		\sqrt{n_F} \hat V_F^{-1/2} (\hat \delta_F - \delta_{F})
	\end{array}
	\right] = \left[ \begin{array}{cc}
		\left(\hat B_B^\top \hat \Omega_B^{-1} \hat B_B\right)^{-1/2} \hat B_B^\top \hat \Omega_B^{-1} & 0 \\ \\
		0 &\left(\hat B_F^\top \hat \Omega_F^{-1} \hat B_F\right)^{-1/2} \hat B_F^\top \hat \Omega_F^{-1}
	\end{array}
	\right] \sqrt{n} \hat U.
\end{align}
By Lemma \ref{lemm: normal approx}, (\ref{eq2}), and Cram\'{e}r-Wold device, we find that
\begin{align*}
	\Omega^{-1/2} \sqrt{n} \hat U \rightarrow_d N(0,I),
\end{align*}
as $n_B, n_F \rightarrow \infty$, with the matrix $\Omega$ defined in (\ref{Omega 3}). This implies that $\sqrt{n} \hat U = O_P(1).$

Since $\sqrt{n} \hat U = O_P(1)$, we use Lemmas \ref{lemm: B} and \ref{lemm: Omega}, and Assumption \ref{assump: limit var} to rewrite the last term in (\ref{eq43}) as
\begin{align*}
	&\left[ \begin{array}{cc}
		\left(B_B^\top \Omega_B^{-1} B_B\right)^{-1/2} B_B^\top \Omega_B^{-1} & 0 \\ \\
		0 &\left(B_F^\top \Omega_F^{-1} B_F\right)^{-1/2} B_F^\top \Omega_F^{-1}
	\end{array}
	\right] \sqrt{n} \hat U +o_P(1)\\
	&=\left[ \begin{array}{cc}
		\left(B_B^\top \Omega_B^{-1} B_B\right)^{-1/2} B_B^\top \Omega_B^{-1/2} & 0 \\ \\
		0 &\left(B_F^\top \Omega_F^{-1} B_F\right)^{-1/2} B_F^\top \Omega_F^{-1/2}
	\end{array}
	\right] \Omega^{-1/2}\sqrt{n} \hat U +o_P(1).
\end{align*}
The leading term on the right hand side converges in distribution to $N(0,I)$, which is our desired result. $\blacksquare$\medskip

\noindent \textbf{Proof of Theorem \ref{thm: asym FWER control}: } In view of Theorem 2.1 of \cite{Romano/Shaikh:10:Eca}, it suffices to show the following three statements:\medskip

(a) $\lim_{n_B, n_F \rightarrow \infty}P\left\{ \hat Q_{FB} \le c_{1-\alpha}\right\} = 1 - \alpha$.

(b) $\lim_{n_B, n_F \rightarrow \infty}P\left\{ \hat Q_{BF} \le c_{1-\alpha}\right\} = 1 - \alpha$.

(c) $\lim_{n_B, n_F \rightarrow \infty}P\left\{ \max\{\hat Q_{FB}, \hat Q_{BF}\} \le c_{\sqrt{1-\alpha}}\right\} = 1 - \alpha$.\medskip

The statements (a) and (b) follow from Theorem \ref{thm: asym norm} immediately by the Continuous Mapping Theorem (CMT). We show (c). Again, by the CMT applied to Theorem \ref{thm: asym norm}, we find that as $n_B, n_F \rightarrow \infty$, $\left[ \hat Q_{FB}, \hat Q_{BF}\right]^\top \rightarrow_d [Q_1, Q_2]^\top$, where $Q_1$ and $Q_2$ are independent random variables that follow $\chi^2(1)$. Since the maximum is a continuous map, by the CMT, we have as $n_B, n_F \rightarrow \infty$,
\begin{align*}
	P\left\{ \max\{\hat Q_{FB}, \hat Q_{BF}\} \le c_{\sqrt{1-\alpha}}\right\} &= P\left\{ \max\{Q_1, Q_2\} \le c_{\sqrt{1-\alpha}}\right\} + o(1)\\
	&= P\left\{ Q_1 \le c_{\sqrt{1-\alpha}}\right\}P\left\{ Q_2 \le c_{\sqrt{1-\alpha}}\right\} + o(1) \\
	&= \left(\sqrt{1-\alpha}\right)^2 + o(1) = 1 - \alpha + o(1).
\end{align*}
Thus, we have the desired result. $\blacksquare$
 
\section{Further Results from the Empirical Application}

\begin{figure}[t]\centering
	\caption{Network characteristics (10th prc.)}
	\small
	\begin{subfigure}{.49\textwidth}
		\caption{Sector to bank}
		\includegraphics[width=\textwidth]{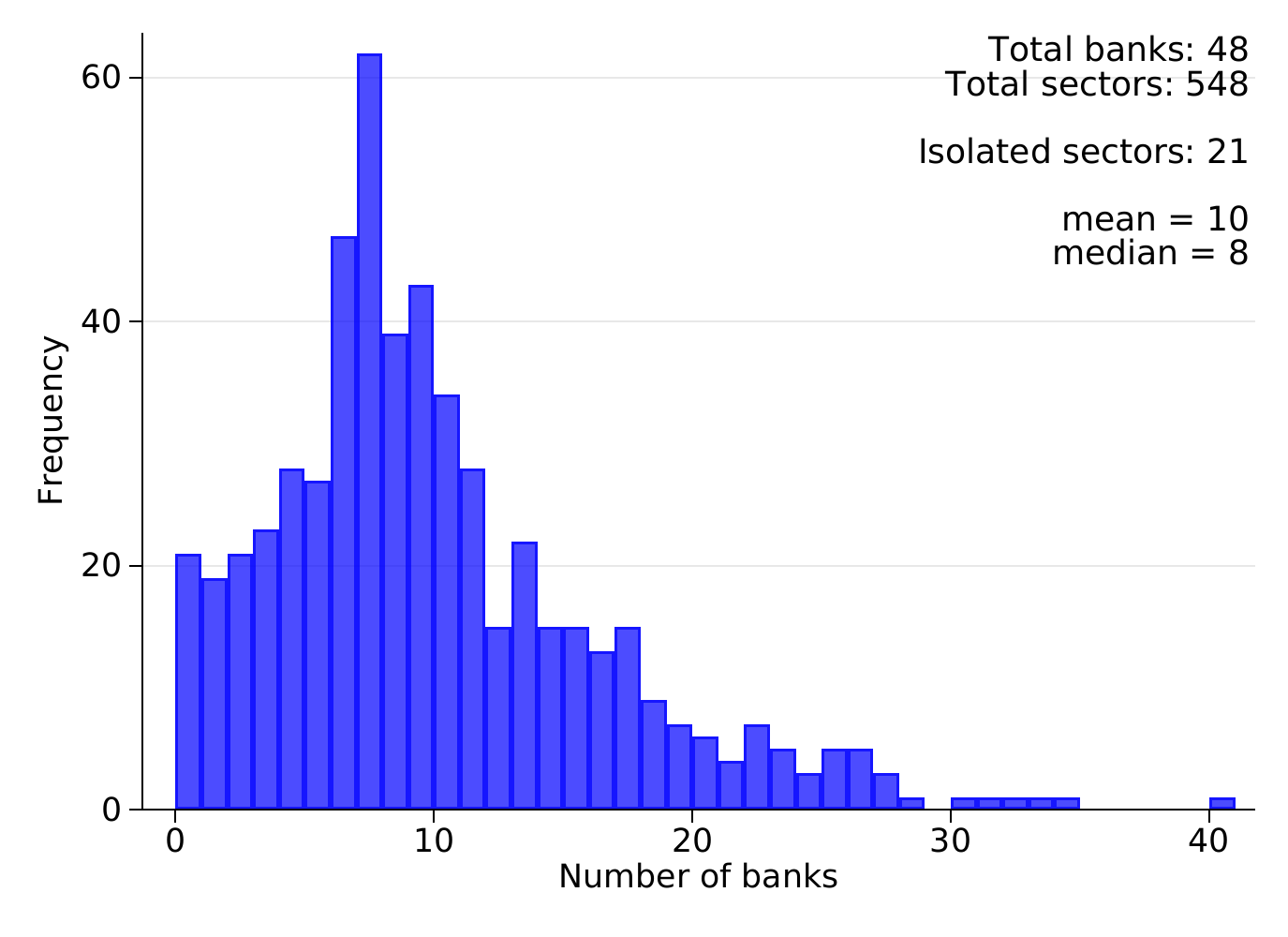}
	\end{subfigure}
	\hfill
	\begin{subfigure}{.49\textwidth}
		\caption{Bank to sector}
		\includegraphics[width=\textwidth]{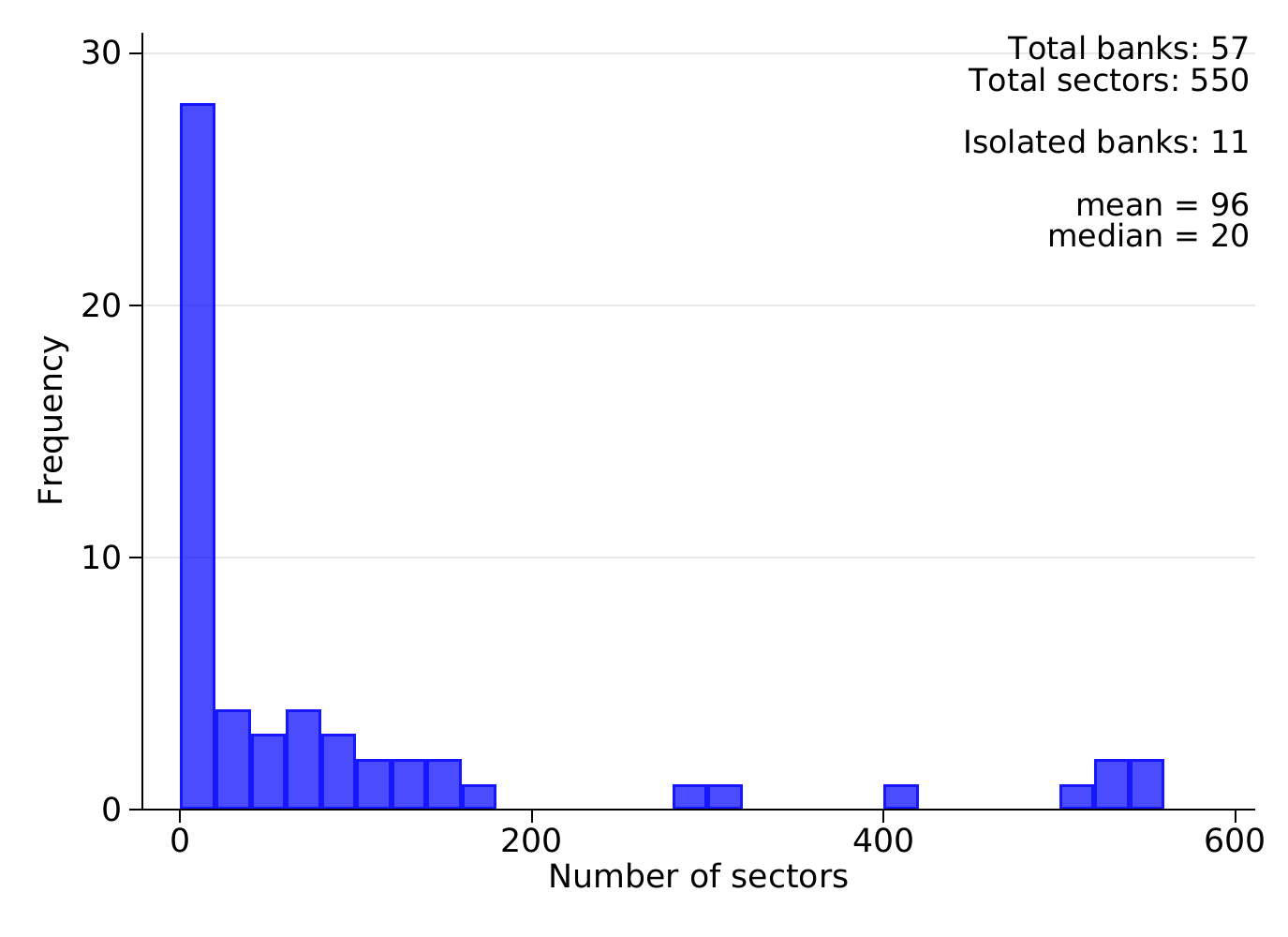}
	\end{subfigure}
	\floatfoot{Notes: This figure shows the network characteristics for the baseline specification. Panel (a) shows the distribution of the number of sectors per bank, while Panel (b) shows the distribution of the number of banks per sector. The network thresholds in Panels (a) and (b) are set to the 10th percentile of the relevant debt share and asset share, respectively.}
	\label{fig:net10}
\end{figure}

\begin{figure}[t]\centering
	\caption{Network characteristics (50th prc.)}
	\small
	\begin{subfigure}{.49\textwidth}
		\caption{Sector to bank}
		\includegraphics[width=\textwidth]{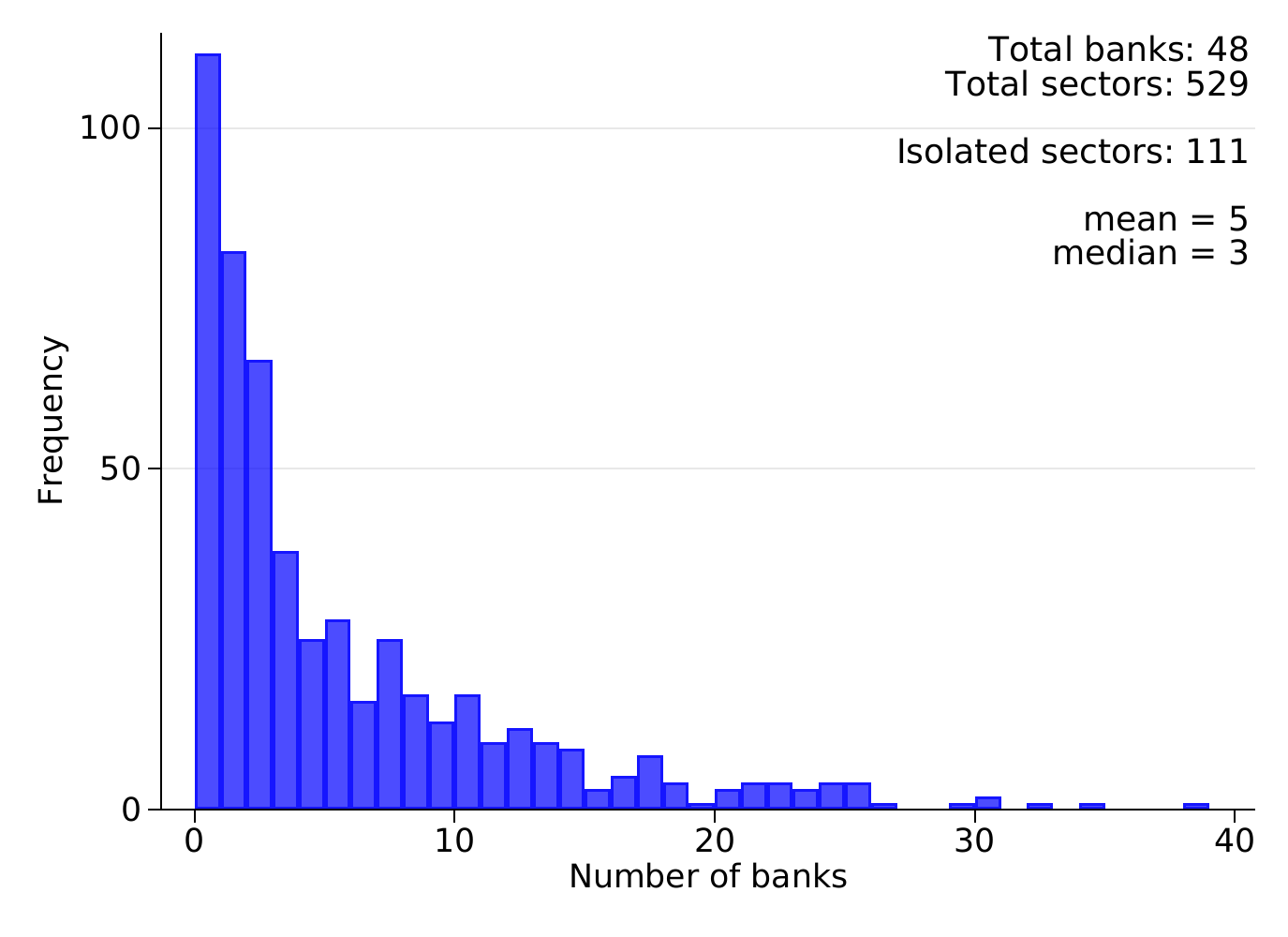}
	\end{subfigure}
	\hfill
	\begin{subfigure}{.49\textwidth}
		\caption{Bank to sector}
		\includegraphics[width=\textwidth]{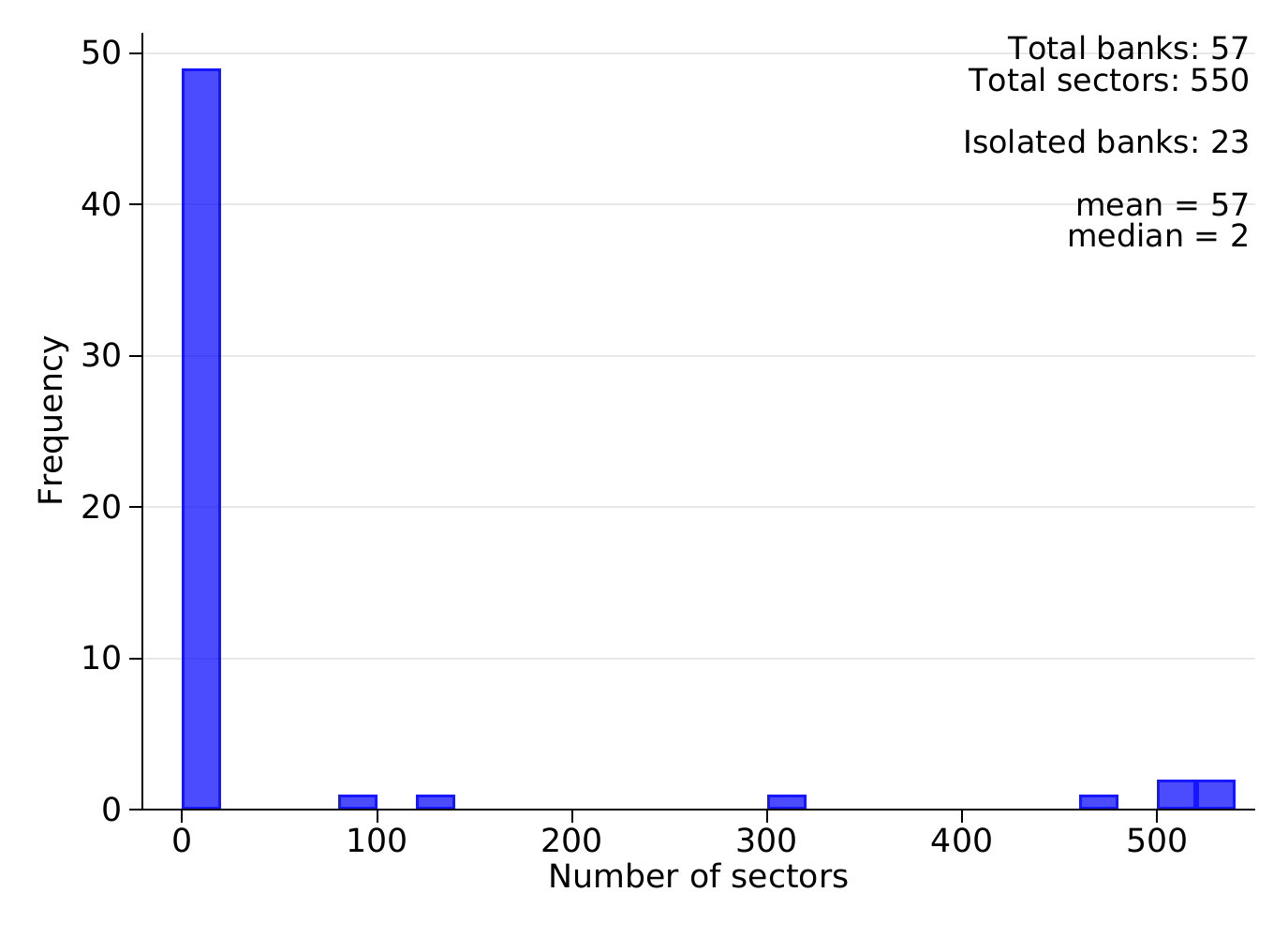}
	\end{subfigure}
	\floatfoot{Notes: This figure shows the network characteristics for the baseline specification. Panel (a) shows the distribution of the number of sectors per bank, while Panel (b) shows the distribution of the number of banks per sector. The network thresholds in Panels (a) and (b) are set to the 50th percentile of the relevant debt share and asset share, respectively.}
	\label{fig:net50}
\end{figure}

\begin{figure}[t]\centering
	\caption{Network characteristics (25th prc.)}
	\small
	\begin{subfigure}{.49\textwidth}
		\caption{Sector to bank}
		\includegraphics[width=\textwidth]{netgr_bpers_all_2005_2012_ltdebt_asymm_net25_spec12.pdf}
	\end{subfigure}
	\hfill
	\begin{subfigure}{.49\textwidth}
		\caption{Bank to sector}
		\includegraphics[width=\textwidth]{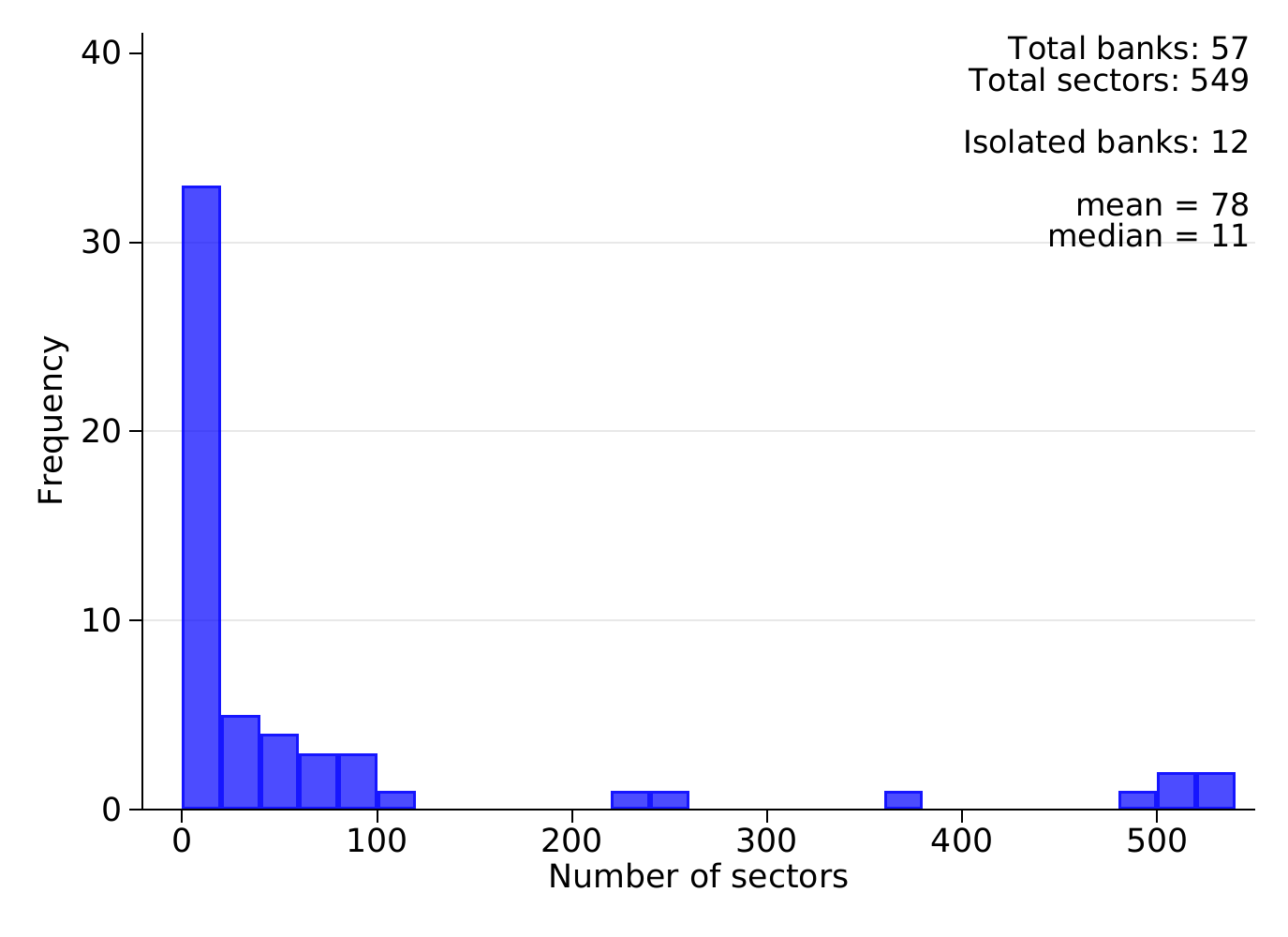}
	\end{subfigure}
	\floatfoot{Notes: This figure shows the network characteristics for the baseline specification. Panel (a) shows the distribution of the number of sectors per bank, while Panel (b) shows the distribution of the number of banks per sector. The network thresholds in Panels (a) and (b) are set to the 25th percentile of the relevant debt share.}
	\label{fig:net25_ltdebt}
\end{figure}

Here we report results cited from Section \ref{subsec: robustness} which discusses robustness of our results. In Figures \ref{fig:net10} and \ref{fig:net50}, we present the network characteristics based on the 10-th percentile thresholds and the 50-th percentile thresholds. Figure \ref{fig:net25_ltdebt} presents the characteristics of the networks based on the debt share of firms in the sectors. 

The estimation results are presented in Tables \ref{tab:results_all_2005_2012_asymm_optB_net50_spec12}-\ref{tab:results_all_2005_2012_asymm_optC_net25_spec12}. We refer the reader to Section \ref{subsec: robustness} for the discussion of these additional results.
\bigskip
\bigskip
\bigskip
\bigskip

\bigskip
\bigskip

\bigskip
\bigskip

\begin{table}[t]\centering
	\caption{Robustness: Sparser network (using IV based on LP with option B)}
	\small
	\begin{tabular}{l*{5}{SS}}
		\toprule
		&\multicolumn{1}{c}{Coef.}&\multicolumn{1}{c}{Std. err.}&\multicolumn{1}{c}{$p$-value}&\multicolumn{2}{c}{[95\% Conf. interval]}\\
		\hline \multicolumn{5}{l}{ \linebreak \textbf{\textit{Panel A: Sector to bank}}} \\
		Bank weakness $ _{t-1} $&       0.333&       0.002&       0.000&       0.329&       0.338\\
		Zombie congestion $ _{t-1} $&       0.963&       0.290&       0.001&       0.395&       1.531\\
		\hline
		Number of banks&          48&            &            &            &            \\
		Number of sectors&         529&            &            &            &            \\
		Observations&       23616&            &            &            &            \\
		\midrule
		\multicolumn{5}{l}{\linebreak \textbf{\textit{Panel B: Bank to sector}}} \\
		Zombie congestion $ _{t-1} $&       0.695&       0.049&       0.000&       0.599&       0.791\\
		Bank weakness $ _{t-1} $&       0.607&       3.485&       0.862&      -6.224&       7.438\\
		Market concentration $ _{t-1} $&      -0.033&       0.097&       0.737&      -0.224&       0.158\\
		Sales growth $ _{t-1} $&      -0.000&       0.008&       0.951&      -0.016&       0.015\\
		Capital intensity $ _{t-1} $&      -0.011&       0.007&       0.134&      -0.024&       0.003\\
		\hline
		Number of banks&          57&            &            &            &            \\
		Number of sectors&         550&            &            &            &            \\
		Observations&       26176&            &            &            &            \\
		\midrule
		Direction of spillover & & & & &\\
		\quad with FWER control & $S \rightarrow B$ & $\text{ at 1\%}$ & & & \\
		\midrule
	\end{tabular}
	\label{tab:results_all_2005_2012_asymm_optB_net50_spec12}
	\floatfoot{Notes: This table shows the results from estimating equation (\ref{bank firm model}). Panel A refers to the direction from sectors to banks while Panel B refers to the direction from banks to sectors. The network thresholds in Panels A and B are set to the 50th percentile of the relevant debt share and asset share, respectively. The sample contains all reported firm-bank relationships and covers the period between 2005 and 2012. Results are obtained using option B. Bank weakness is given by the ratio of loan loss provsions to total assets; zombie congestion is the asset-weighted share of zombie firms in a sector according to the baseline definition; market concentration is the HHI index of sector-level sales; sales growth is computed as the log difference in total sector sales from the year before; capital intensity is the share of a sector's fixed assets in total sales. }
\end{table}

\begin{table}[t]\centering
	\caption{Robustness: Denser network (using IV based on LP with option B)}
	\small
	\begin{tabular}{l*{5}{SS}}
		\toprule
		&\multicolumn{1}{c}{Coef.}&\multicolumn{1}{c}{Std. err.}&\multicolumn{1}{c}{$p$-value}&\multicolumn{2}{c}{[95\% Conf. interval]}\\
		\hline \multicolumn{5}{l}{ \linebreak \textbf{\textit{Panel A: Sector to bank}}} \\
		Bank weakness $ _{t-1} $&       0.330&       0.002&       0.000&       0.326&       0.334\\
		Zombie congestion $ _{t-1} $&       0.920&       0.298&       0.002&       0.335&       1.505\\
		\hline
		Number of banks&          48&            &            &            &            \\
		Number of sectors&         548&            &            &            &            \\
		Observations&       41840&            &            &            &            \\
		\midrule
		\multicolumn{5}{l}{\linebreak \textbf{\textit{Panel B: Bank to sector}}} \\
		Zombie congestion $ _{t-1} $&       0.696&       0.048&       0.000&       0.601&       0.791\\
		Bank weakness $ _{t-1} $&       0.150&       0.056&       0.007&       0.041&       0.260\\
		Market concentration $ _{t-1} $&      -0.027&       0.099&       0.786&      -0.220&       0.167\\
		Sales growth $ _{t-1} $&      -0.001&       0.008&       0.908&      -0.016&       0.014\\
		Capital intensity $ _{t-1} $&      -0.014&       0.009&       0.091&      -0.031&       0.002\\
		\hline
		Number of banks&          57&            &            &            &            \\
		Number of sectors&         550&            &            &            &            \\
		Observations&       43936&            &            &            &            \\
		\midrule
		Direction of spillover & & & & &\\
		\quad with FWER control & $S \leftrightarrow B$ & $\text{ at 1\%}$ & & & \\
		\midrule
	\end{tabular}
	\label{tab:results_all_2005_2012_asymm_optB_net10_spec12}
	\floatfoot{Notes: This table shows the results from estimating equation (\ref{bank firm model}). Panel A refers to the direction from sectors to banks while Panel B refers to the direction from banks to sectors. The network thresholds in Panels A and B are set to the 10th percentile of the relevant debt share and asset share, respectively. The sample contains all reported firm-bank relationships and covers the period between 2005 and 2012. Results are obtained using option B. Bank weakness is given by the ratio of loan loss provisions to total assets; zombie congestion is the asset-weighted share of zombie firms in a sector according to the baseline definition; market concentration is the HHI index of sector-level sales; sales growth is computed as the log difference in total sector sales from the year before; capital intensity is the share of a sector's fixed assets in total sales. }
\end{table}

\begin{table}[t]\centering
	\caption{Robustness: Networks based on debt only (using IV based on LP with option B)}
	\small
	\begin{tabular}{l*{5}{SS}}
		\toprule
		&\multicolumn{1}{c}{Coef.}&\multicolumn{1}{c}{Std. err.}&\multicolumn{1}{c}{$p$-value}&\multicolumn{2}{c}{[95\% Conf. interval]}\\
		\hline \multicolumn{5}{l}{ \linebreak \textbf{\textit{Panel A: Sector to bank}}} \\
		Bank weakness $ _{t-1} $&       0.330&       0.002&       0.000&       0.326&       0.334\\
		Zombie congestion $ _{t-1} $&       0.964&       0.322&       0.003&       0.333&       1.596\\
		\hline
		Number of banks&          48&            &            &            &            \\
		Number of sectors&         541&            &            &            &            \\
		Observations&       35144&            &            &            &            \\
		\midrule
		\multicolumn{5}{l}{\linebreak \textbf{\textit{Panel B: Bank to sector}}} \\
		Zombie congestion $ _{t-1} $&       0.695&       0.049&       0.000&       0.599&       0.791\\
		Bank weakness $ _{t-1} $&       0.140&       0.096&       0.146&      -0.049&       0.328\\
		Market concentration $ _{t-1} $&      -0.028&       0.099&       0.777&      -0.221&       0.165\\
		Sales growth $ _{t-1} $&      -0.001&       0.008&       0.898&      -0.016&       0.014\\
		Capital intensity $ _{t-1} $&      -0.015&       0.009&       0.090&      -0.032&       0.002\\
		\hline
		Number of banks&          57&            &            &            &            \\
		Number of sectors&         549&            &            &            &            \\
		Observations&       35680&            &            &            &            \\
		\midrule
		Direction of spillover & & & & &\\
		\quad with FWER control & $S \rightarrow B$ & $\text{ at 1\%}$ & & & \\
		\midrule
	\end{tabular}
	\label{tab:results_all_2005_2012_asymm_optB_net25_spec12_ltdebt}
	\floatfoot{Notes: This table shows the results from estimating equation (\ref{bank firm model}). Panel A refers to the direction from sectors to banks while Panel B refers to the direction from banks to sectors. The network thresholds in Panels A and B are set to the 25th percentile of the relevant debt share. The sample contains all reported firm-bank relationships and covers the period between 2005 and 2012. Results are obtained using option B. Bank weakness is given by the ratio of loan loss provsions to total assets; zombie congestion is the asset-weighted share of zombie firms in a sector according to the baseline definition; market concentration is the HHI index of sector-level sales; sales growth is computed as the log difference in total sector sales from the year before; capital intensity is the share of a sector's fixed assets in total sales. }
\end{table}

\begin{table}[t]\centering
	\caption{Robustness: First bank only (using IV based on LP with option B)}
	\small
	\begin{tabular}{l*{5}{SS}}
		\toprule
		&\multicolumn{1}{c}{Coef.}&\multicolumn{1}{c}{Std. err.}&\multicolumn{1}{c}{$p$-value}&\multicolumn{2}{c}{[95\% Conf. interval]}\\
		\hline \multicolumn{5}{l}{ \linebreak \textbf{\textit{Panel A: Sector to bank}}} \\
		Bank weakness $ _{t-1} $&       0.335&       0.002&       0.000&       0.331&       0.339\\
		Zombie congestion $ _{t-1} $&       0.751&       0.519&       0.148&      -0.266&       1.769\\
		\hline
		Number of banks&          46&            &            &            &            \\
		Number of sectors&         537&            &            &            &            \\
		Observations&       26552&            &            &            &            \\
		\midrule
		\multicolumn{5}{l}{\linebreak \textbf{\textit{Panel B: Bank to sector}}} \\
		Zombie congestion $ _{t-1} $&       0.696&       0.049&       0.000&       0.601&       0.792\\
		Bank weakness $ _{t-1} $&       0.133&       0.040&       0.001&       0.053&       0.212\\
		Market concentration $ _{t-1} $&      -0.027&       0.098&       0.784&      -0.219&       0.166\\
		Sales growth $ _{t-1} $&      -0.001&       0.008&       0.914&      -0.016&       0.014\\
		Capital intensity $ _{t-1} $&      -0.015&       0.009&       0.090&      -0.032&       0.002\\
		\hline
		Number of banks&          57&            &            &            &            \\
		Number of sectors&         550&            &            &            &            \\
		Observations&       28512&            &            &            &            \\
		\midrule
		Direction of spillover & & & & &\\
		\quad with FWER control & $B \rightarrow S$ & $\text{ at 1\%}$ & & & \\
		\midrule
	\end{tabular}
	\label{tab:results_first_2005_2012_asymm_optB_net25_spec12}
	\floatfoot{Notes: This table shows the results from estimating equation (\ref{bank firm model}). Panel A refers to the direction from sectors to banks while Panel B refers to the direction from banks to sectors. The network thresholds in Panels A and B are set to the 25th percentile of the relevant debt share and asset share, respectively. The sample only includes the first reported bank relationship for each firm and covers the period between 2005 and 2012. Results are obtained using option B. Bank weakness is given by the ratio of loan loss provsions to total assets; zombie congestion is the asset-weighted share of zombie firms in a sector according to the baseline definition; market concentration is the HHI index of sector-level sales; sales growth is computed as the log difference in total sector sales from the year before; capital intensity is the share of a sector's fixed assets in total sales. }
\end{table}

\begin{table}[t]\centering
	\caption{Baseline specification (using IV based on LP with option A)}
	\small
	\begin{tabular}{l*{5}{SS}}
		\toprule
		&\multicolumn{1}{c}{Coef.}&\multicolumn{1}{c}{Std. err.}&\multicolumn{1}{c}{$p$-value}&\multicolumn{2}{c}{[95\% Conf. interval]}\\
		\hline \multicolumn{5}{l}{ \linebreak \textbf{\textit{Panel A: Sector to bank}}} \\
		Bank weakness $ _{t-1} $&       0.281&       0.010&       0.000&       0.262&       0.300\\
		Zombie congestion $ _{t-1} $&       5.715&       3.801&       0.133&      -1.734&      13.164\\
		\hline
		Number of banks&          48&            &            &            &            \\
		Number of sectors&         541&            &            &            &            \\
		Observations&       35144&            &            &            &            \\
		\midrule
		\multicolumn{5}{l}{\linebreak \textbf{\textit{Panel B: Bank to sector}}} \\
		Zombie congestion $ _{t-1} $&       0.710&       0.059&       0.000&       0.595&       0.825\\
		Bank weakness $ _{t-1} $&       0.147&       0.050&       0.003&       0.050&       0.245\\
		Market concentration $ _{t-1} $&      -0.086&       0.110&       0.434&      -0.302&       0.130\\
		Sales growth $ _{t-1} $&       0.001&       0.008&       0.890&      -0.015&       0.017\\
		Capital intensity $ _{t-1} $&      -0.014&       0.011&       0.208&      -0.037&       0.008\\
		\hline
		Number of banks&          57&            &            &            &            \\
		Number of sectors&         550&            &            &            &            \\
		Observations&       37448&            &            &            &            \\
		\midrule
		Direction of spillover & & & & &\\
		\quad with FWER control & $B \rightarrow S$ & $\text{ at 1\%}$ & & & \\
		\midrule
	\end{tabular}
	\label{tab:results_all_2005_2012_asymm_optA_net25_spec12}
	\floatfoot{Notes: This table shows the results from estimating equation (\ref{bank firm model}). Panel A refers to the direction from sectors to banks while Panel B refers to the direction from banks to sectors. The network thresholds in Panels A and B are set to the 25th percentile of the relevant debt share and asset share, respectively. The sample contains all reported firm-bank relationships and covers the period between 2005 and 2012. Results are obtained using option A. Bank weakness is given by the ratio of loan loss provsions to total assets; zombie congestion is the asset-weighted share of zombie firms in a sector according to the baseline definition; market concentration is the HHI index of sector-level sales; sales growth is computed as the log difference in total sector sales from the year before; capital intensity is the share of a sector's fixed assets in total sales. }
\end{table}

\begin{table}[t]\centering
	\caption{Baseline specification (using IV based on LP with option C)}
	\small
	\begin{tabular}{l*{5}{SS}}
		\toprule
		&\multicolumn{1}{c}{Coef.}&\multicolumn{1}{c}{Std. err.}&\multicolumn{1}{c}{$p$-value}&\multicolumn{2}{c}{[95\% Conf. interval]}\\
		\hline \multicolumn{5}{l}{ \linebreak \textbf{\textit{Panel A: Sector to bank}}} \\
		Bank weakness $ _{t-1} $&       0.330&       0.002&       0.000&       0.326&       0.333\\
		Zombie congestion $ _{t-1} $&       0.712&       0.312&       0.023&       0.100&       1.325\\
		\hline
		Number of banks&          48&            &            &            &            \\
		Number of sectors&         541&            &            &            &            \\
		Observations&       35144&            &            &            &            \\
		\midrule
		\multicolumn{5}{l}{\linebreak \textbf{\textit{Panel B: Bank to sector}}} \\
		Zombie congestion $ _{t-1} $&       0.688&       0.050&       0.000&       0.589&       0.787\\
		Bank weakness $ _{t-1} $&       0.134&       0.047&       0.005&       0.041&       0.227\\
		Market concentration $ _{t-1} $&       0.014&       0.099&       0.891&      -0.181&       0.208\\
		Sales growth $ _{t-1} $&      -0.002&       0.008&       0.808&      -0.018&       0.014\\
		Capital intensity $ _{t-1} $&      -0.007&       0.005&       0.136&      -0.016&       0.002\\
		\hline
		Number of banks&          57&            &            &            &            \\
		Number of sectors&         550&            &            &            &            \\
		Observations&       37448&            &            &            &            \\
		\midrule
		Direction of spillover & & & & &\\
		\quad with FWER control & $B \leftrightarrow S$ & $\text{ at 1\%}$ & & & \\
		\midrule
	\end{tabular}
	\label{tab:results_all_2005_2012_asymm_optC_net25_spec12}
	\floatfoot{Notes: This table shows the results from estimating equation (\ref{bank firm model}). Panel A refers to the direction from sectors to banks while Panel B refers to the direction from banks to sectors. The network thresholds in Panels A and B are set to the 25th percentile of the relevant debt share and asset share, respectively. The sample contains all reported firm-bank relationships and covers the period between 2005 and 2012. Results are obtained using option C. Bank weakness is given by the ratio of loan loss provsions to total assets; zombie congestion is the asset-weighted share of zombie firms in a sector according to the baseline definition; market concentration is the HHI index of sector-level sales; sales growth is computed as the log difference in total sector sales from the year before; capital intensity is the share of a sector's fixed assets in total sales. }
\end{table}

\end{document}